\documentclass[11pt]{article}          
\usepackage[english]{babel}
\usepackage{latexsym}
\makeatletter
\def\@begintheorem#1#2{\trivlist%
 \item[\hskip \labelsep{\sffamily\bfseries #2\ #1}]\itshape}
\newtheorem{teo}{Theorem}[section]
\newtheorem{defi}[teo]{Definition}

\newtheorem{lem}[teo]{Lemma}
\newtheorem{pro}[teo]{Proposition}
\newtheorem{_rem}[teo]{Remark}
\newtheorem{_eje}[teo]{Example}
\newtheorem{claim}[teo]{Claim}
\newenvironment{rem}{\def\@begintheorem##1##2{\trivlist%
 \item[\hskip\labelsep{\sffamily\bfseries ##2\ ##1}]}\begin{_rem}}{\end{_rem}}

\makeatother

\newenvironment{beweis}{{\em Proof:}}{\hfill $\rule{2mm}{2mm}$
\vspace{3mm}

}

\DeclareMathAlphabet{\Ma}{U}{msa}{m}{n}
\DeclareMathAlphabet{\Mb}{U}{msb}{m}{n}
\DeclareMathAlphabet{\Meuf}{U}{euf}{m}{n}

\def\z#1{\Mb{#1}}
\def\got#1{\Meuf{#1}}

\DeclareSymbolFont{ASMa}{U}{msa}{m}{n}
\DeclareSymbolFont{ASMb}{U}{msb}{m}{n}
\DeclareMathSymbol{\hrist}{\mathord}{ASMa}{"16}
\DeclareMathSymbol{\varkappa}{\mathalpha}{ASMb}{"7B}
\DeclareMathSymbol{\CrPr}{\mathord}{ASMb}{"6F}

\newfont{\EinsFont}{cmr7 scaled 1070}
\def\EINS{{\mathchoice{% -> displaystyle
 \mbox{\unitlength1cm\begin{picture}(.25,.2)\put(0,0){$1$}%
 \put(0.105,0){{\mbox{\fontfamily{cmr}\upshape\small l}}}\end{picture}}}{%
 \mbox{\unitlength1cm\begin{picture}(.25,.2)\put(0,0){$1$}%
 \put(0.105,0){{\mbox{\fontfamily{cmr}\upshape\small l}}}\end{picture}}}{%
 \mbox{\unitlength1cm\begin{picture}(.18,.15)\put(0,0){$\scriptstyle 1$}%
 \put(0.07,0){{\mbox{\fontfamily{cmr}\upshape\EinsFont l}}}\end{picture}}}{%
 \mbox{\unitlength1cm\begin{picture}(.18,.15)\put(0,0){$\scriptstyle 1$}%
 \put(0.07,0){{\mbox{\fontfamily{cmr}\upshape\EinsFont l}}}\end{picture}}}}}

\def\restriction{{\mathchoice{%diplaystyle
 \mbox{\unitlength1cm\begin{picture}(.2,.4)%
  \bezier{5}(.07,.3)(.1,.27)(.13,.24)%
  \put(.07,.35){\line(0,-1){.5}}\end{picture}}}{%textstyle
 \mbox{\unitlength1cm\begin{picture}(.2,.4)%
  \bezier{5}(.07,.3)(.1,.27)(.13,.24)%
  \put(.07,.35){\line(0,-1){.5}}\end{picture}}}{%scriptstyle
  \hrist}{\hrist}}}

  \def\al #1.{{\mathcal{#1}}}
  \def\ot #1.{{\got{#1}}}
\def\CCR{\overline{\Delta({\got X},\,B)}}
\def\CCRy{\overline{\Delta({\got Y},\,B)}}
  \def\ccr #1,#2.{\overline{\Delta(#1,\,#2)}}
  \def\frak{\got} 
  \def\wp{\got S}
  \def\b #1.{{\bf #1}}
  \def\cross#1.{\mathrel{\mathop{\times}\limits_{#1}}}
  \def\B{\Theta}                        % The revolutionary Theta notation.
  \def\C{\Mb{C}}
  \def\N{\Mb{N}}
  
  \def\R{\Mb{R}}
  \def\Z{\Mb{Z}}
  \def\ff{\widehat{f}}

  \def\ww{\widehat{\omega}}
  \def\wh{\widehat}
  \def\wt{\widetilde}
\def\ilim{\mathop{{\rm lim}}\limits_{\longrightarrow}}
  \def\cross #1.{\mathrel{\raise 3pt\hbox{$\mathop\times\limits_{#1}$}}}
\def\d{\dagger}
\def\b #1.{{\bf #1}}
\def\ker{{\rm Ker}\,}
\def\aut{{\rm Aut}\,}
\def\chop{\hfill\break}
\def\rest{\restriction}
\title{\bf Local quantum constraints}

\author{
 {\sc Hendrik Grundling}                                            \\[1mm] 
 {\footnotesize Department of Mathematics,}                         \\ 
 {\footnotesize University of New South Wales,}                      \\
 {\footnotesize Sydney, NSW 2052, Australia.}                             \\ 
 {\footnotesize hendrik@maths.unsw.edu.au}                  \\ 
\and
  {\sc Fernando Lled\'o}\thanks{On leave from:             
   {\em Mathematical Institute, University of Potsdam,  
   Am Neuen Palais 10, Postfach 601~553,           
   D--14415 Potsdam, Germany.} 
   E-mail: lledo@rz.uni-potsdam.de}  \\[1mm]
 {\footnotesize Max--Planck--Institut f\"ur Gravitationsphysik,}\\
 {\footnotesize Albert--Einstein--Institut,}     \\   
 {\footnotesize Am M\"uhlenberg 1,}   \\      
 {\footnotesize D--14476 Golm, Germany.}                       \\
 {\footnotesize lledo@aei-potsdam.mpg.de}}

\date{RUNNING TITLE: Local Quantum Constraints.}
\date{KEYWORDS: constraints, locality, Haag--Kastler axioms, quantum field theory,
                Gupta--Bleuler electromagnetism, C*--algebra, CCR--algebra}
\date{AMS classification: 81T05, 81T10, 46L60, 46N50}

\begin{document}
\maketitle

%%%%%%%%%%%%%%%%%%%%%%%%%%%%%%%%%%%%%%%%%%%%%%%%%%%%%%%%%%%%%%%%%%%%%%%%%%%%%%
\begin{abstract}
We analyze the situation of a local quantum field theory
with constraints, both indexed by the same set of space--time regions.
In particular we find ``weak'' Haag--Kastler axioms
which will ensure that the 
 final constrained
theory satisfies the usual Haag--Kastler axioms.
Gupta--Bleuler electromagnetism is developed in detail as an
example of a theory which satisfies the ``weak''
Haag--Kastler axioms but not the usual ones.
This analysis is done by pure C*--algebraic means
without employing any indefinite metric representations, 
and we obtain the same physical algebra and positive
energy representation for it than by the usual means.
The price for avoiding the indefinite metric, is the use
of nonregular representations and complex valued test functions.
We also exhibit  the precise connection with the usual indefinite
metric representation.

We conclude the analysis by comparing the final physical
algebra produced by a system of local constrainings with
the one obtained from a single global constraining and also
consider the issue of reduction by stages.
For the usual spectral condition on the generators of
the translation group, we also find a ``weak'' version,
and show that the Gupta--Bleuler example satisfies it.
\end{abstract}
%%%%%%%%%%%%%%%%%%%%%%%%%%%%%%%%%%%%%%%%%%%%%%%%%%%%%%%%%%%%%%%%%%%%%%%%%%%%%%
\section{Introduction}

In many quantum field theories there are constraints consisting of local 
expressions of the quantum fields, generally written as a selection
condition for the physical subspace $\al H.^{(p)}$. In the physics literature
the selection condition usually takes the form:
\[
  \al H.^{(p)}:=\big\{\psi\,\mid\,\chi(x)\,\psi=0
\quad\forall\, x\in\R^4\big\}
\]
where $\chi$ is some operator--valued distribution
(so more accurately $\chi(x)\,\psi=0$ should be written
as $\chi(f)\,\psi=0$ for all test functions $f$).
Since the constraints $\chi$ are constructed from the
smeared quantum fields, one expects them to have the same net
structure in space--time as the smeared quantum fields.
The question then arises as to how
locality properties and constraining intertwines.
This question will be at the focus of our interest in this paper.

To properly study locality questions, we shall use 
algebraic quantum field theory, a well--developed 
theory built on a net of C*--algebras 
satisfying the Haag--Kastler axioms~\cite{Haag64,bHaag92},
but we shall assume in addition a local net
of constraints (to be defined in Section 3). To impose these constraints
at the algebraic level,
we use the method developed by Grundling and Hurst~\cite{Grundling85},
and this can be done either in each local algebra separately
or globally in the full field algebra. We will compare the results
of these two different routes, and will find conditions on the local
net of constraints to ensure that the net of algebras obtained
after constraining satisfies the Haag--Kastler axioms.
In fact one can weaken the Haag--Kastler axioms on the original
system, providing that after constraining the final net obtained
satisfies these axioms.
We characterise precisely what these ``weak''
Haag--Kastler axioms are. In our subsequent example (Gupta--Bleuler
electromagnetism) we find that this weakening is crucial,
since the original constraints violate the causality axiom.
In our example we will avoid the usual indefinite
metric representations, but will obtain by C*--algebraic
means both the correct physical algebra, and the same (positive energy)
representation than the one produced via the indefinite metric.
Thus we show that a gauge quantum field can be completely described
by C*--algebraic means, in a framework of Algebraic QFT, without 
the use of indefinite metric representations.
There is however a cost for avoiding the indefinite metric, and this
consists of the use of nonregular representations, and the use of
complex valued test functions (this is related to causality violation).
Fortunately, both of these pathologies only involve nonphysical
parts of the theory which are eliminated after constraining, thus the
final theory is again well--behaved.

Since local constraints are usually generators of gauge transformations
of the second kind, the theory developed here can be considered as
complementing the deep Doplicher--Haag--Roberts analysis of systems
with gauge transformations of the first kind~\cite{DHR69a}.
Our axioms will be slightly different (weakened
Haag--Kastler axioms), and we will work with an abstract net
of C*--algebras, whereas the DHR analysis is done concretely
in a positive energy representation.

We need to remark that there is a range of reduction
algorithms for quantum constraints available in the literature,
cf.~\cite{lands} at different degrees of rigour.
These involve either more structure and choices, or are
representation dependent, or the maps involved have pathologies
from the point of view of C*--algebras. That is why we chose
the method of~\cite{Grundling85}.
Furthermore, the Haag--Kastler axioms have not previously been included
in any of the reduction techniques in~\cite{lands}.
Locality has been examined in specific constrained theories
in the literature (cf. \cite {S+C}, \cite{Strocchi74})
 but not in the general terms we do here.

The architecture of the paper is as follows.
In Section 2 we collect general facts of the constraint
procedure of \cite{Grundling85} which we will need in the
subsequent sections. There is some new material in this section,
since we need to extend the previous method to cover the
current situation. In Section 3 we introduce our basic object,
a ``system of local quantum constraints'' as well as the 
``weak Haag--Kastler axioms'' and prove that after local
constraining of such a system,
 we obtain a system satisfying the Haag--Kastler
axioms. Section 4 consists of some preliminary material
necessary for the development of our example in 
Section 5, Gupta--Bleuler
electromagnetism, (C*--algebra version adapted from \cite{Grundling88b})
and we verify all the weak Haag--Kastler axioms for it. 
We concretely characterize the net of constrained algebras, but it turns
out that in order to obtain a simple global algebra we need to do a 
second stage of constraining (traditionally thought of as
imposing the Maxwell equations, but here it is slightly
stronger than that). We also verify the weak Haag--Kastler axioms for 
this second stage of constraints, and we work out in detail the
connection of the C*--theory with the usual indefinite metric
representation.
In Section 6 we consider miscellaneous topics raised by the previous
Sections. 
First, for a system of local constraints,
we consider the relation between the algebras obtained from a single
global constraining, and the inductive limit of the algebras
found from local constrainings.
We show that for the Gupta--Bleuler example these two algebras
are the same.
Secondly, we develop the theory of constraint reduction by stages
(i.e. impose the constraints sequentially along an increasing
chain of subsets instead of all at once). %Here we find a criterion
%when the final algebra obtained from an n--step reduction
%procedure will be the same as the one obtained from 
%enforcing all constraints at once. 
We then show that the
two--step reduction procedure of the Gupta--Bleuler example 
satisfies this criterion.
Thirdly, we consider the spectral condition (on the
 generators of translations)
which occur in Haag--Kastler theory, and find a ``weak''
version of it which will guarantee that the final constrained
theory satisfies the usual spectral condition. 
We show that the Gupta--Bleuler example satisfies it, by 
demonstrating that from the indefinite metric in the heuristic theory
we can define a (positive metric) representation of the constrained
algebra which satisfies the spectral condition.
There are two appendices containing additional constraining facts
needed in proofs,
and one long proof which would have interrupted the flow of the paper.

%%%%%%%%%%%%%%%%%%%%%%%%%%%%%%%%%%%%%%%%%%%%%%%%%%%%%%%%%%%%%%%%%%%%%%%%%%%%%%
%%%%%%%%%%%%%%%%%%%%%%%%%%%%%%%%%%%%%%%%%%%%%%%%%%%%%%%%%%%%%%%%%%%%%%%%%%%%%%
\section{Kinematics for Quantum Constraints.}
\label{TProcedure}

In this section we present the minimum preliminary material
necessary to define our primary problem.
The reader whose main interest is quantum constraints will find this
section interesting in its own right, as well as many general constraint results
scattered throughout the paper.
Here we present a small generalisation of the T--procedure of Grundling and 
Hurst \cite{Grundling85,Grundling88b}.
 All new results will be proven here and for 
other  proofs we refer to the literature.

In heuristic physics a set of constraints is a set $\{A_i\;\big|\;i\in I\}$ 
(with $I$ an index set) of operators on some Hilbert space together with  
a selection condition for the subspace of physical vectors:
\[
   \al H.^{(p)}:=\overline{\Big\{\psi \mid A_i\psi=0
                 \quad {\forall}\, i\in I\Big\}}. 
\]
The set of traditional observables is then the commutant $\{A_i\mid
i\in I\}'$ which one can enlarge to the set of all observables
which preserve $\al H.^{(p)}$. The final constrained system is the 
restriction of this algebra to the subspace $\al H.^{(p)}$. On abstraction of 
such a system into C*--algebra terms, one starts with a unital C*--algebra 
$\al F.$ (the {\it field algebra}) containing all physical relevant 
observables. This is an abstract C*--algebra, i.e.~we ignore the
initial representation in which the system may be defined.
 We need to decide in what form the constraints should appear in
$\al F.$ as a subset $\al C.$. We have the following possibilities:
\begin{itemize}
\item{} If all $A_i$ are bounded 
we can identify $\al C.$ directly with $\{A_i\mid i\in I\}\subset\al F.$.
\item{} If the $A_i$ are unbounded but essentially selfadjoint, we can
 take $\al C.:=\{U-\EINS\;|\;U\in\al U.\}=:\al U.-\EINS$, where
 the set of unitaries $\al U.\subset\al F.$ is identified with
 $\{\exp(it\bar A_j)\;|\;t\in\z{R},\;\; j\in I\}$. This is the form in
 which constraints were analyzed in \cite{Grundling85}, and also 
 the form which we will use here in the following sections.
\item{} If the $A_i$ are unbounded and normal, we can identify
$\al C.$ with $\{f(A_j)\;|\; j\in I\}$ where $f$ is a bounded real valued
 Borel function with $f^{-1}(0)=\{0\}$.
\item{} If the $A_i$ are unbounded, closable and not normal, then we can 
 replace each $A_i$ by the essentially selfadjoint operator $A_i^*A_i$
 which is justified because for any closed operator $A$
 we have ${\rm Ker}\,A={\rm Ker}\,A^*A$, reducing this case to 
 the one for essentially selfadjoint constraints.
\end{itemize}
Finally, notice that we can replace any constraint set $\al C.$ as above, 
by one which satisfies $\al C.^*=\al C.$ as a set and which selects the 
same physical subspace, using the fact that ${\rm Ker}\,A={\rm Ker}\,A^*A$.

%%%%%%%%%%%%%%%%%%%%%%%%%%%%%%%%%%%%%%%%%%%%%%%%%%%%%%%%%%%%%%%%%%%%%%%%%%%%%%
%\subsection{Imposing constraints: The T--procedure.}

Motivated from above, our starting point is:
\begin{defi}
A {\bf quantum system with constraints} is a
pair $(\al F.,\;\al C.)$ where the {\bf field algebra}
$\al F.$ is a unital {\rm C*}--algebra containing
the {\bf constraint set} $\al C.=\al C.^*.$ A
{\bf constraint condition} on $(\al F.,\,\al C.)$ consists of
the selection of the physical 
state space by:
\[
  {\got S}_D:=\Big\{ \omega\in{\got S}({\al F.})\mid\pi_\omega(C)
              \Omega_\omega=0\quad {\forall}\, C\in {\al C.}\Big\}\,,
\]
where ${\got S}({\al F.})$ denotes the state space of $\al F.$,
and $(\pi_\omega,\al H._\omega,\Omega_\omega)$ denotes the
GNS--data of $\omega$. The elements of ${\got S}_D$ are called 
{\bf Dirac states}.
The case of {\bf unitary constraints} means 
that $\al C.=\al U.-\EINS$, $\;\al U.\subset\al F._u$, and 
for this we will
also use the notation $(\al F.,\,\al U.)$.
\end{defi}
Thus in the GNS-representation of each Dirac state, the GNS cyclic vector 
$\Omega_\omega$ satisfies the physical selection condition above. The 
assumption is that all physical information is contained in the
pair $(\al F.,{\got S}_D)$.
 
For the case of unitary constraints we have the following equivalent 
characterizations of the Dirac states 
(cf.~\cite[Theorem~2.19~(ii)]{Grundling85}):
\begin{eqnarray}
  \label{DiracU1}{\got S}_{{D}}&=&\Big\{ \omega\in{\got S}({\al F.})\mid 
                   \omega(U)=1 \quad {\forall}\, U\in {\al U.}\Big\} \\[1mm]
  \label{DiracU2}              &=&\Big\{ \omega\in{\got S}({\al F.})\mid 
                   \omega(FU)=\omega(F)=\omega(UF) \quad {\forall}\,
                   F\in\al F.,\; U\in {\al U.}\Big\}.
\end{eqnarray}
Moreover, the set $\{\alpha_U:= {\rm Ad}(U)\mid U\in\al U.\}$ of 
automorphisms of $\al F.$ leaves every Dirac state invariant, 
i.e.~we have $\omega\circ\alpha_U=\omega$ for all 
$\omega\in {\got S}_{{D}}$, $U\in{\al U.}$. 

For a general constraint set $\al C.$, observe that we have:
\begin{eqnarray*}
 {\got S}_D &=& \Big\{ \omega\in{\got S}({\al F.})\mid\omega(C^*C)=0 
                \quad {\forall}\, C\in\al C.\Big\}                \\[1mm]
            &=& \Big\{ \omega\in{\got S}({\al F.})\mid \al C.\subseteq 
           N_\omega\Big\}\kern2mm=\kern2mm\al N.^\perp\cap{\got S}(\al F.)\;.
\end{eqnarray*}
Here $N_\omega:=\{F\in\al F.\mid\omega(F^*F)=0\}$ is the left kernel of
$\omega$ and 
$\al N.:=\cap\; \{N_{\omega}\mid\omega\in{\got S}_D \}$, 
and the superscript $\perp$ denotes the annihilator of the 
corresponding subset in the dual of $\al F.$. 
The equality $\al N.=[\al FC.]$
(where we use the notation $[\cdot]$ for the closed linear space 
generated by its argument), follows from the fact that every closed 
left ideal is the intersection of the left kernels which contains it 
(cf.~3.13.5 in \cite{bPedersen89}).
Thus $\al N.$ is the left ideal generated by $\al C.$. 
Since $\al C.$ is selfadjoint and contained in 
$\al N.$ we conclude $\al C.\subset {\rm C}^*(\al C.)\subset 
\al N.\cap\al N.^*=[\al FC.]\cap[\al CF.]$, where ${\rm C}^*(\cdot)$ 
denotes the C*--algebra in $\al F.$ generated by its argument.

\begin{teo}
\label{Teo.2.1}
Now for the Dirac states we have:
\begin{itemize}
\item[{\rm (i)}] ${\got S}_{{D}}\neq\emptyset\;$ iff 
   $\;\EINS\not\in {\rm C}^*(\al C.)$ iff 
$\;\EINS\not\in \al N.\cap\al N.^*=:\al D.$.
\item[{\rm (ii)}] $\omega\in {\got S}_D\;$ iff 
   $\; \pi_{\omega}({\al D.})\Omega_{\omega}=0$.
\item[{\rm (iii)}] An extreme Dirac state is pure.
\end{itemize}
\end{teo}
\begin{beweis} (i) The first equivalence is proven in Theorem~2.7 of 
\cite{Grundling85}. If $\EINS\in\al D.\subset\al N.$, then 
$\omega(\al N.)\not=0$ for all states $\omega$, i.e.~${\got S}_D=\emptyset$.
If $\EINS\not\in\al D.$, then $\EINS\not\in\al N.$ so $\al N.$ 
is a proper closed left ideal
and hence by 3.10.7 in \cite{bPedersen89} ${\got S}_D\not=\emptyset$.

(ii) If $\omega\in{\got S}_D$, then $\al D.\subset\al N.\subset N_\omega$,
hence $\pi_\omega(\al D.)\Omega_\omega=0$. Conversely, since
$\al C.\subset\al D.=[\al FC.]\cap[\al CF.]$ we have that 
$\pi_\omega(\al D.)\Omega_\omega=0$ implies $\pi_\omega(\al C.) 
\Omega_\omega=0$  hence $\omega\in{\got S}_D$.

(iii) Denote the quasi-state space of $\al F.$ by $Q$ \cite{bPedersen89}.
We can write the set of Dirac states as
\[
 {\got S}_D={\got S}(\al F.)\cap 
             \{ \phi\in Q\mid \phi(L)=0 \quad\forall\; L\in\al N.\}\,.
\]
Since $\al N.$ is a left ideal, if it is in ${\rm Ker}\;\omega$,
it must be in $N_\omega$.
By Theorem~3.10.7 in \cite{bPedersen89} the set $Q_0:=\{ 
\phi\in Q\mid \phi(L)=0 \quad\forall\; L\in\al N.\}$ is
a weak* closed face in $Q$. Now if we can decompose a
Dirac state $\omega$, since it is in $Q_0$, so are its components
by the facial property of $Q_0$. These components are multiples of Dirac 
states, dominated by $\omega$ so $\omega$ cannot be extreme. Thus 
extreme Dirac states must be pure.
\end{beweis}

We will call a constraint set $\al C.$ {\it first class} if
$\EINS\not\in {\rm C}^*(\al C.)$, and this is the nontriviality 
assumption which we henceforth make \cite[Section~3]{Grundling88a}.

Now define 
\[
 {\al O.} := \{ F\in {\al F.}\mid [F,\, D]:= FD-DF \in {\al D.}\quad 
               {\forall}\, D\in{\al D.}\}.
\]
Then ${\al O.}$ is the C$^*$--algebraic analogue of Dirac's observables 
(the weak commutant of the constraints) \cite{bDirac64}.

\begin{teo}
\label{Teo.2.2}
With the preceding notation we have:
\begin{itemize}
\item[{\rm(i)}] $\al D.=\al N.\cap \al N.^*$ is the unique maximal 
  {\rm C}$^*$--algebra in $\, \cap\; \{ {\rm Ker}\,\omega\mid \omega\in 
  {\got S}_{{D}} \}$. Moreover $\al D.$ is a hereditary 
  {\rm C}$^*$--subalgebra of $\al F.$.
\item[{\rm(ii)}] ${\al O.} = {M}_{\al F.}({\al D.})
  :=\{ F\in{\cal F}\mid FD\in{\cal D}\ni DF\quad\forall\, D\in{\cal D}\}$, 
  i.e.~it is the relative multiplier algebra of ${\al D.}$ in ${\al F.}$.
\item[{\rm(iii)}] $\al O.=\{F\in\al F.\mid\; [F,\,\al C.]\subset\al D.\}$.
\item[{\rm(iv)}] $\al D.=[\al OC.]=[\al CO.]$.
\item[{\rm(v)}] For the case of unitary constraints, i.e.
 $\al C.=\al U.-\EINS$, we have
 $\al U.\subset\al O.$ and $\al O.={\{F\in\al F.\mid\alpha_U(F)-F\in\al D.
\quad\forall\; U\in\al U.\}}$
  where $\alpha_U:={\rm Ad}\,U$.
\end{itemize}
\end{teo}
\begin{beweis}
(i) The proof of Theorem~2.13 in \cite{Grundling85} is still valid for the 
current more general constraints. To see that $\al D.$ is hereditary, use
Theorem~3.2.1 in \cite{bMurphy90} and the fact that 
$\al N.=[ {\al F.}\al C.]$ is a closed left ideal of $\al F.$.

(ii) Since $\al D.$ is a two--sided ideal for $M_{\al F.}(\al D.)$ it is
obvious that $M_{\al F.}(\al D.)\subset \al O.$. Conversely, consider
$B\in\al O.$, then for any $D\in\al D.$, we have
$BD=DB + D'\in\al N.$ with $D'$ some element of $\al D.$,
where we used $\al F.\al D.=\al F.(\al N.\cap\al N.^*)\subset\al N.$.
Similarly we see that $DB\in\al N.^*$. But then
$\al N.\ni BD=DB+D'\in\al N.^*$, so $BD\in\al N.\cap\al N.^*=\al D.$.
Likewise $DB\in\al D.$ and so $B\in M_{\al F.}(\al D.)$.

(iii) Since $\al C.\subset\al D.$ we see from the definition of $\al O.$
that $F\in\al O.$ implies that $[F,\,\al C.]\subset\al D.$.
Conversely, let $[F,\,\al C.]\subset\al D.$
for some $F\in\al F.$.
Now $F[\al FC.]\subset[\al FC.]$ and $F[\al CF.]=[F\al CF.]\subset
[(\al CF.+\al D.)\al F.]\subset[\al CF.]$ because
$\al C.F+\al D.\subset\al C.F+[\al CF.]\subset[\al CF.]$.
Thus $F\al D.=F\big([\al FC.]\cap[\al CF.]\big)
\subset[\al FC.]\cap[\al CF.]=\al D.$.
Similarly $\al D.F\subset\al D.$, and thus by (ii) we see
$F\in\al O.$.

(iv) $\al D.\subset\al O.$ so by (i) it is the unique maximal C*--algebra
annihilated by all the states $\omega\in\ot S._D(\al O.)=
\ot S._D\restriction\al O.$ (since $\al C.\subset\al O.$).
Thus $\al D.=[\al OC.]\cap[\al CO.]$. But $\al C.\subset\al D.$,
so by (ii) $[\al OC.]\subset\al D.\subset[\al OC.]$ and so
$\al D.=[\al OC.]=[\al CO.]$.

(v) $\al U.\subset\al O.$ because $\al U.-\EINS\subset\al D.\subset\al O.\ni
\EINS$, and so $[F,\,\al C.]\subset\al D.$ implies
$[U-\EINS,\, F]U^{-1}=\alpha_U(F)-F\in\al D.$ for $U\in\al U.$.
The converse is similar.
\end{beweis}

Thus $\al D.$ is a closed two-sided ideal of $\al O.$ and it
is proper when ${\got S}_D\not=\emptyset$ (which we assume here
by $\EINS\not\in {\rm C}^*(\al C.)$).
From (iii) above, we see that the traditional observables $\al C.'\subset\al O.$,
where $\al C.'$ denotes the relative commutant of $\al C.$ in $\al F.$.

Define the {\it maximal {\rm C}$^*$--algebra of physical observables} as
\[
 {\al R.}:={\al O.}/{\al D.}.
\]
The factoring procedure is the actual step of imposing constraints. 
This method of constructing $\al R.$  from $(\al F.,\,\al C.)$
is called the {\bf T--procedure}
in \cite{Grundling87}. 
We require that after the T--procedure all physical
information is contained in the pair $({\al R.}\kern.4mm ,{\got S}
({\al R.}))$, where ${\got S}({\al R.})$ denotes the set of 
states on $\al R.$. Now, it is possible that $\al R.$ may not be simple
\cite[Section~2]{Grundling87}, and this would not be acceptable for a 
physical algebra. So, using physical arguments, one would in practice
choose a C$^*$--subalgebra $\al O._c\subset \al O.$ containing 
the traditional observables $\al C.'$ such that
\[
 \al R._c :=\al O._c / (\al D.\cap\al O._c )\subset \al R.\,,
\]
is simple. The following result justifies the choice of $\al R.$ as the
algebra of physical observables (cf. Theorem 2.20 in \cite{Grundling85}):

\begin{teo}
\label{Teo.2.6}
 There exists a w$^*$--continuous isometric bijection 
         between the Dirac states on ${\al O.}$ and the states on ${\al R.}$. 
\end{teo}

The hereditary property of $\al D.$ can be further analyzed, and we
do this in Appendix 1 (it will be useful occasionally in proofs).

%%%%%%%%%%%%%%%%%%%%%%%%%%%%%%%%%%%%%%%%%%%%%%%%%%%%%%%%%%%%%%%%%%%%%%%%%%%%%%
\section{Local Quantum Constraints}
\label{LocalQuantumConstraints}

In this section we will introduce our main object of study,
viz a system of local quantum constraints.
In practice, a large class of constraint systems occur in quantum field theory
(henceforth denoted by QFT), for instance gauge theories.
A prominent property of a QFT, is space--time locality, and usually when
constraints occur, they also have this property. Heuristically
such constraints are written as 
\[
   \chi(x)\,\psi=0\qquad\hbox{for}\quad\psi\in\al H.^{(p)}\,,\;
   x\in M^4\,,
\]
where Minkowski space is $M^4=(\R^4,\,\eta)$ with metric 
$\eta:={\rm diag}(+,-,-,-)$, and this makes the locality explicit. 
Since $\chi$ is actually an operator--valued distribution, the 
correct expression should be of the form
\[
   \chi(f)\,\psi=0\qquad\hbox{for}\quad\psi\in\al H.^{(p)}\,,\;
   f\in C_c^\infty(\R^4)\;.
\]
In this section we now want to analyze how locality intertwines with the 
T--procedure of constraint reduction. To make this precise, recall
that the Haag--Kastler axioms \cite{Haag64,bHaag92} express locality
for a QFT as follows:
\begin{defi} 
\label{HK}
A {\bf Haag--Kastler QFT} (or HK--QFT for short) consists of
the following.
\begin{itemize}
\item A directed set $\Gamma\subseteq\{\B\subset M^4\mid\B\quad\hbox{open
 and bounded}\,\}$ partially ordered by set inclusion, such that
 $\R^4=\cup\{\B\mid\B\in\Gamma\}$ and under the
 action of the orthochronous Poincar\'e group $\al P._+^\uparrow$
 on $M^4$ we have $g\B\in\Gamma$ for all $\B\in\Gamma$, 
 $g\in\al P._+^\uparrow$. 
\item A directed set $\widetilde\Gamma$ of {\rm C*}--algebras with a common 
 identity $\EINS$, ordered by inclusion, with an inductive limit 
 {\rm C*}--algebra $\al F._0$ (over $\widetilde\Gamma$).
We will call the elements of $\widetilde\Gamma$ the {\bf local
field algebras} and $\al F._0$ the \b quasi--local algebra..
\item A surjection $\al F.\colon\ \Gamma\to \widetilde\Gamma$, satisfying:
\begin{itemize}
\item[{\rm (1)}] (Isotony) $\al F.$ is order preserving,
 i.e.~$\al F.(\B_1)\subseteq\al F.(\B_2)$ if $\B_1\subseteq\B_2$. 
\item[{\rm (2)}] (Causality) If $\B_1,\;\B_2\in\Gamma$ are spacelike separated, 
 (henceforth denoted $\B_1\perp\B_2$), then\hfill\break
 ${[\al F.(\B_1),\,\al F.(\B_2)]=0}$ in $\al F._0$.
\item[{\rm (3)}] (Covariance) There is an action $\alpha\colon\
 \al P._+^\uparrow\to{\rm Aut}\,\al F._0$ such that
 $\alpha_g(\al F.(\B))=\al F.(g\B)$, $\; g\in\al P._+^\uparrow$, $\,\B
 \in\Gamma$.
\end{itemize}
\end{itemize}
\end{defi}

\begin{rem}
\begin{itemize}
\item[(i)]
In the usual 
 algebraic approach to QFT
(cf.~\cite{bHaag92}),  
there are additional axioms, e.g. that $\al F._0$ must be primitive, 
that there is a vacuum state with GNS--representation in which 
the generators of the translations in the covariant representation
of $\al P._+^\uparrow$ have spectra in the forward light cone,
that there is a compact gauge group, local definiteness, local
normality etc. We will return to some of these
axioms later, but for now, we restrict our analysis to those
listed in Definition~\ref{HK}. The net $\Gamma$ is usually
taken to be the set of double cones in $M^4$, but here we will
keep it more general.
There is some redundancy in this definition
 e.g. for the results in this Section,
 we will not need the assumption that $\B\in\Gamma$
is bounded.
\item[(ii)]
In the literature, sometimes ``local'' is used synonymously with
``causal.'' Since we want to weaken the Haag--Kastler axioms below,
we will make a distinction between these terms, in particular, we will
call an algebra $\al F.(\B)$ in an isotone net as above {\it local,} (as well as 
its elements). If such an isotone net also satisfies causality,
it is called {\it causal.}
\end{itemize}
\end{rem}

In the context of the Haag--Kastler axioms, we would like to define a
system of ``local quantum constraints'' in such a way that it includes the
major examples from QFT. 

\begin{defi}
\label{loc.cons}
A system of {\bf local quantum constraints} consists of a surjection 
$\al F.\colon\ \Gamma\to\widetilde\Gamma$ as in Definition~$\ref{HK}$
satisfying isotony as well as
\begin{itemize}
\item[{\rm (4)}] (Local Constraints) There is a map $\al U.$ from $\Gamma$ to
the set of first class subsets of the unitaries in the
 local field algebras such that
\begin{itemize}
\item[] $\;\al U.(\B)\subset\al F.(\B)_u\;$ for all $\;\B\in\Gamma$, and
\item[] if $\B_1\subseteq\B_2$, then %$(\al F.(\B_1),\,\al U.(\B_1))\subseteq
%(\al F.(\B_2),\,\al U.(\B_2))$ where the inclusion notation means that
%$\al F.(\B_1)\subseteq\al F.(\B_2)$ and
$\al U.(\B_1)=\al U.(\B_2)\cap
 \al F.(\B_1)$.
\end{itemize}
\end{itemize}
\end{defi}

\begin{rem}
In this definition we have made the minimum assumptions to start the
analysis. We have omitted causality and covariance because these
are physical requirements which one should demand for the final
physical theory, not the initial (unconstrained) theory which contains 
nonphysical objects. There are examples of constrained QFTs satisfying 
these conditions, e.g.~\cite[Remark~4.3]{Lledo97},
\cite[Chapter~4]{tLledo98} and see also below in 
 Section~5 our example of Gupta--Bleuler electromagnetism.
\end{rem}

Given a system of local quantum constraints, $\B\to(\al F.(\B),\,
\al U.(\B))$, we can apply the T--procedure to each local system
$(\al F.(\B),\, \al U.(\B))$, to obtain the ``local'' objects:
\begin{eqnarray*}
{\got S}_D^\B &:=& \Big\{ \omega\in{\got S}({\al F.}(\B))\mid 
                \omega(U)=1\quad {\forall}\, U\in \al U.(\B)\Big\}
         \kern2mm=\kern2mm {\got S}_D(\al F.(\B))\,, \\[2mm]
\al D.(\B) &:=& [\al F.(\B)\,\al C.(\B)]\cap[\al C.(\B)\,\al F.(\B)]\,,\\[2mm]
\al O.(\B) &:=& \{ F\in \al F.(\B)\mid  FD-DF \in \al D.(\B) \quad 
                  {\forall}\, D\in\al D.(\B) \}\kern2mm = \kern2mm
                  M_{\al F.(\B)}(\al D.(\B))\,, \\[2mm]
\al R.(\B) &:=& \al O.(\B) / \al D.(\B)\,.
\end{eqnarray*}
We can be now more precise about what our task is in this section:
\begin{itemize}
\item[]{\bf Problem:} For a system of local quantum constraints
 $\B\to (\al F.(\B),\,\al U.(\B))$, find minimal conditions
 such that the net of local physical observables 
 $\B\to\al R.(\B)$ becomes a HK--QFT.
\end{itemize}

Now, to analyze the isotony property, we need to determine
what the inclusions in Definition~\ref{loc.cons}
% ${(\al F.(\B_1),\,\al U.(\B_1))}\subseteq
%{(\al F.(\B_2),\,\al U.(\B_2))}$ 
imply for the associated
objects ${\big({\got S}_D^{\B_i},\,\al D.(\B_i),\,\al O.(\B_i),
\al F.(\B_i)\big)}$, and this is the task of the next subsection.

\subsection{Inclusion structures}

For this  subsection we assume unitary 
first class constraints $\al C.=\al U.-\EINS$, 
indicated by a pair $(\al F.,\,\al U.)$.
Motivated by Definition~\ref{loc.cons}, we define:

\begin{defi}
\label{inclu}
A first class constrained system $({\al F.},\,{\al U.})$
is said to be {\bf included} in another one
$({\al F.}_e,\,{\al U.}_e)$, if the {\rm C*}--algebras
 ${\al F.}\subset{\al F.}_e$ have a common identity and
${\al U.}= {\al U.}_e \cap {\al F.}$. We denote this by
$(\al F.,\,\al U.)\subseteq(\al F._e,\,\al U._e)$.
\end{defi}
For the rest of this subsection we will assume that
$(\al F.,\,\al U.)\subseteq(\al F._e,\,\al U._e)$.
{}From the T--procedure sketched above we 
obtain the corresponding quadruples
\[
 ({\got S}_D,\,{\al D.},\,{\al O.},\,{\al R.})\qquad {\rm and} \qquad 
 ({\got S}_{De},\,{\al D.}_e,\,{\al O.}_e,\,{\al R.}_e)\,.
\]

\begin{lem}
\label{Lem.3.1}
Suppose that ${\al D.}= {\al D.}_e\cap \al O.$ and $\al O.\subset\al
O._e$, then there is a {*--isomorphism} from $\al R.$ to a {\rm C}$^*$--subalgebra 
of ${\al R.}_e$, which maps the identity of $\al R.$ to the 
identity of $\al R._e$.
\end{lem} 
\begin{beweis}
{}From
\[
 {\al R.}_e =\al O._e / {\al D.}_e = \Big(\al O. / {\al D.}_e \Big) \cup
             \Big( (\al O._e\setminus\al O.) / {\al D.}_e \Big)\,,
\]
 it is enough to show that $\al O. / {\al D.}_e\cong {\al R.}=
\al O. / \al D.$. Now, a
 ${\al D.}_e$--equivalence class consists of $A,B\in\al O.$ such that
 $A-B\in{\al D.}_e$ and therefore $A-B\in{\al D.}_e\cap\al O.={\al D.}$.
 This implies $\al O. / {\al D.}_e\cong \al O. / {\al D.}={\al R.}$.
Moreover, since $\EINS\in\al O.\subset\al O._e$,
and the $\al D.$--equivalence class of $\EINS$ is contained in
the $\al D._e$--equivalence class of $\EINS$, it follows that 
the identity maps to the identity.
\end{beweis}

\begin{rem}
\label{Rem.3.2} 
\begin{itemize}
\item[(i)] By a simple finite dimensional example one can verify
that the condition $\al D.=\al D._e\cap\al O.$ does not
imply $\al O.\subset\al O._e$.
\item[(ii)] Instead of the conditions in Lemma~\ref{Lem.3.1}
another natural set of restriction conditions one can also choose is 
$\al D.=\al D._e\cap\al F.$ and $\al O.=\al O._e
\cap\al F.$, but below we will see that these imply
those in Lemma~\ref{Lem.3.1}.
\end{itemize}
\end{rem}

The next result gives sufficient conditions for the equation
${\al D.}={\al D.}_e \cap\al O.$ to hold.
\begin{lem}
\label{Lem.3.3}
 ${\al D.}={\al D.}_e \cap {\al F.}$ if either
\begin{itemize}
 \item[{\rm (i)}] ${\got S}_D={\got S}_{De}\restriction {\al F.}$ 
                  \qquad{\it or}
 \item[{\rm (ii)}] $[{\al F.}\;{\rm C}^*({\al U.} - \EINS) ]
    =[{\al F.}_e\;{\rm C}^*({\al U.}_e - \EINS)] \;\cap\; {\al F.}\,$.
\end{itemize}
\end{lem} 
\begin{beweis}
 Take $\omega\in {\got S}_{De}$ and recall the definition for the left
 kernel $N_{\omega}$ given in the preceding subsection. Then 
 $N_{\omega}\cap {\al F.}=N_{\omega\restriction {\al F.}}$ and from (i)
 we get
\begin{eqnarray*}
 \al N.=
 \cap\; \{N_{\omega}\mid \omega\in {\got S}_D\}
   &=& \cap\;\{N_{\omega}\mid\omega\in{\got S}_{De}\restriction {\al F.}\} \\
   &=& \cap\;\{N_{\omega}\cap{\al F.}\mid\omega\in{\got S}_{De} \}        \\
   &=& \cap\;\{N_{\omega}\mid\omega\in{\got S}_{De} \} \cap {\al F.}
    =\al N._e\cap\al F.\,.
\end{eqnarray*}
This produces 
\[
 \al N.=[{\al F.}\;{\rm C}^*({\al U.} - \EINS) ]
    =[{\al F.}_e\;{\rm C}^*({\al U.}_e - \EINS)]\; \cap\; {\al F.}
\]
as well as 
\begin{eqnarray*}
 {\al D.}
   &=& [{\al F.}\;{\rm C}^*({\al U.} - \EINS)] \;\cap\;
       [{\rm C}^*({\al U.} - \EINS)\;{\al F.}]                        \\ 
   &=& [{\al F.}_e\;{\rm C}^*({\al U.}_e - \EINS)]\; \cap \;
       [{\rm C}^*({\al U.}_e - \EINS)\;{\al F.}_e]\; \cap \; {\al F.} \\
   &=& {\al D.}_e \cap {\al F.}
\end{eqnarray*}
and the proof is concluded.
\end{beweis}

\begin{lem}
\label{Lem.3.4}
 We have
\[
 \al O.\subset\al O._e \quad {\it iff} \quad \al O.\subset 
 \{F\in{\al F.}\mid UFU^{-1}-F\in{\al D.}_e 
 \quad{\forall}\,U \in ({\al U.}_e\setminus {\al U.}) \}\,.  
\]
\end{lem} 
\begin{beweis}
By Theorem~\ref{Teo.2.2}~(v) we have
\begin{eqnarray}
\lefteqn{\!\!\!\!\!\!\!\!\!\!\! \al O._e\cap{\al F.} } \quad
  &=& \{ F\in{\al F.}\mid UFU^{-1}-F\in{\al D.}_e 
         \quad{\forall}\,U \in{\al U.}_e \}         \nonumber\\[2mm]
  &=&  \{ F\in{\al F.}\mid UFU^{-1}-F\in{\al D.}_e\cap{\al F.} 
         \quad{\forall}\,U \in{\al U.} \}          \nonumber   \\ 
  & &  \;\cap \; \{ F\in{\al F.}\mid UFU^{-1}-F\in{\al D.}_e 
    \quad{\forall}\,U \in ({\al U.}_e\setminus {\al U.})\}\nonumber\\[2mm]
  &=& \Big( \al O. \;\cup\;\{ F\in{\al F.}\mid UFU^{-1}-F\in
      ({\al D.}_e\cap{\al F.} \setminus {\al D.})
       \quad{\forall}\,U \in{\al U.} \}  \Big)           \nonumber\\ 
  & & \;\cap \; \{ F\in{\al F.}\mid UFU^{-1}-F\in{\al D.}_e 
    \quad{\forall}\,U \in ({\al U.}_e\setminus {\al U.})\}\nonumber \\[2mm]
  &=& \Big(\al O. \;\cap\; \{ F\in{\al F.}\mid UFU^{-1}-F\in{\al D.}_e 
   \quad{\forall}\,U \in ({\al U.}_e\setminus {\al U.})\}\Big)
   \nonumber \\
  & &  \;\cup \; \Big\{ F\in{\al F.}\mid UFU^{-1}-F\in
      ({\al D.}_e\cap{\al F.} \setminus {\al D.}) \quad{\forall}\,
      U \in{\al U.}\;\;                                        \nonumber \\
  & & \qquad{\rm and} \;\; UFU^{-1}-F\in{\al D.}_e 
      \quad{\forall}\,U \in ({\al U.}_e\setminus {\al U.})\Big\}\,. 
  \label{LastEquation}
\end{eqnarray}
 Now if $\al O.\subset \{F\in{\al F.}\mid UFU^{-1}-F\in{\al D.}_e 
 \quad{\forall}\,U \in ({\al U.}_e\setminus {\al U.}) \}$, then 
 it is clear from the above equations that 
 $\al O.\subset\al O._e\cap{\al F.}\subset \al O._e$.

 To prove the other implication note that the second set in the union of
 Eqn.~(\ref{LastEquation}) 
 is contained in ${\al F.}\setminus\al O.$ and therefore from
 $\al O.\subset\al O._e$ we obtain 
\[
 \al O.=\al O.\cap\al O._e\cap{\al F.}
=\al O.\cap \{F\in{\al F.}\mid UFU^{-1}-F\in{\al D.}_e 
 \quad{\forall}\,U \in ({\al U.}_e\setminus {\al U.}) \}\,,
\]
 which implies the desired inclusion.
\end{beweis}

\begin{teo}
\label{Lem.3.5} 
Given an included pair of first class constrained 
systems $(\al F.,\,\al U.)\subseteq(\al F._e,\,\al U._e)$ and 
notation as above, the following statements
are in the relation $(iv)\Rightarrow(iii)\Rightarrow(ii)
\Rightarrow(i)$ where:
\begin{itemize}
\item[{\rm (i)}] $\ot S._D\restriction\al O.=
\ot S._{De}\restriction\al O.$.
\item[{\rm (ii)}] $\al D.=\al D._e\cap\al O.$ and
$\al O.\subseteq\al O._e$.
\item[{\rm (iii)}] $\al D.=\al D._e\cap\al F.$ and
$\al O.=\al O._e\cap\al F.$.
\item[{\rm (iv)}]  $\ot S._D=\ot S._{De}\restriction\al F.$
and $\al O.\subseteq\al O._e$.
\end{itemize}  
\end{teo} 
\begin{beweis}
We first prove the implication $(ii)\Rightarrow(i)$, so assume $(ii)$.
It suffices to show that all Dirac states on $\al O.$ extend to
Dirac states on $\al O._e$. 
 Denote by ${\got S}({\al R.})$ and ${\got S}_D(\al O.)$ respectively
 the set of states on $\al R.$ and
 the set of Dirac states on ${\al O.}$ (assume also corresponding 
 notation for ${\al R.}_e$ and $\al O._e$). Then from Theorem
 \ref{Teo.2.6} there exist $w^*$--continuous, isometric 
 bijections $\theta$ and $\theta_e$:
\[
 \theta\colon\ {\got S}_D(\al O.) \rightarrow {\got S}({\al R.}) \quad{\rm and} 
 \quad\theta_e\colon\ {\got S}_{De}(\al O._e) \rightarrow {\got S}({\al R.}_e)\,.
\]
 Now take $\omega\in{\got S}_D(\al O.)$, so that 
 $\theta (\omega)\in{\got S}({\al R.})$. From Lemma~\ref{Lem.3.1}, 
 ${\al R.}\subset{\al R.}_e$ so we can extend $\theta (\omega)$ to
 $\widetilde{\theta (\omega)}\in{\got S}({\al R.}_e)$. Finally, 
 $\theta_e^{-1}\Big( \widetilde{\theta (\omega)} \Big)\in{\got S}_{De}
  (\al O._e)$
 is an extension of $\omega$, since for any $A\in \al O.$ we have
\[
 \theta_e^{-1}\Big( \widetilde{\theta (\omega)} \Big)(A)
   = \widetilde{\theta(\omega)}(\xi_e(A))
   = \theta (\omega)(\xi(A))
   = \omega (A)\,,
\]
where $\xi\colon\al O.\to\al R.$ and $\xi_e\colon\ \al O._e\to\al R._e$ 
are the canonical factorization maps. This proves $(i)$.

Next we prove $(iii)\Rightarrow(ii)$. Obviously $\al O.=\al O._e
\cap\al F.$ implies that $\al O.\subseteq\al O._e$.
Since $\al D.\subset\al O.\subset\al F.$ we have
$\al D.=\al D._e\cap\al F.\subset\al O.$ and so
$\al D.=(\al D._e\cap\al F.)\cap\al O.=\al D._e\cap\al O.$.

Finally we prove $(iv)\Rightarrow(iii)$.
By Lemma~\ref{Lem.3.3}(i), $\ot S._D=\ot S._{De}\restriction
\al F.$ implies that $\al D.=\al D._e\cap\al F.$.
By Theorem~\ref{Teo.2.2}(iii) and the fact that $\al O.$ is a 
C*--algebra   (hence the span of its selfadjoint elements), we have
\begin{eqnarray}
\al O.
  &=& \big[\{F\in\al F._{sa}\;\mid\;[F,\,\al U.]\subset\al D.
         \}\big]         \nonumber\\[2mm]
  &=&\left[\{F\in\al F._{sa}\;\mid\;\omega\big([F,\, U^*]\cdot
[F,\, U])=0\quad\forall\;\omega\in\ot S._D,\; U\in\al U.\}\right]
\nonumber
\end{eqnarray}
where the last equality follows from
$\al D.=\al N.\cap\al N.^*$, $[F,\, U]^*=-[F,\, U^*]$ for $F=F^*$,
and $\al U.=\al U.^*$. Since $\ot S._D=\ot S._{De}\restriction\al F.$
we have 
\begin{eqnarray}
\al O.
  &=&\left[\{F\in\al F._{sa}\;\mid\;\omega\big([F,\, U^*]\cdot
[F,\, U])=0\quad\forall\;\omega\in\ot S._{De},\; U\in\al U.\}\right]
\nonumber\\[2mm]
  &\supseteq&
\left[\{F\in\al F._{sa}\;\mid\;\omega\big([F,\, U^*]\cdot
[F,\, U])=0\quad\forall\;\omega\in\ot S._{De},\; U\in\al U._e\}\right]
\nonumber\\[2mm]
&=&\al O._e\cap\al F.\;.\nonumber
\end{eqnarray}
Thus since $\al O.\subseteq\al O._e$ we conclude that
$\al O.=\al O._e\cap\al F.$.
\end{beweis}

\begin{rem}
\label{Rem.3.6}
 $\ot S._D\restriction\al O.=
\ot S._{De}\restriction\al O.$ does not seem to imply
that  $\ot S._D=\ot S._{De}\restriction\al F.$
since for a state in $\ot S._D$, if one restricts it to
$\al O.$ and then extends to a state in $\ot S._{De}$,
it seems nontrivial whether this extension can coincide
with the original state on $\al F.$.
\end{rem}

%%%%%%%%%%%%%%%%%%%%%%%%%%%%%%%%%%%%%%%%%%%%%%%%%%%%%%%%%%%%%%%%%%%%%%%%%%%%%%
\subsection{Isotony and causality weakened.}

Return now to the previous analysis of a system of local quantum constraints,
and define:
\begin{defi} 
\label{weak.con}
Fix a system of local quantum constraints 
$\B\to(\al F.(\B),\,\al U.(\B))$ (cf. Def.~\ref{loc.cons}), 
then we say that it satisfies:
\begin{itemize}
\item[\rm (5)] {\bf reduction isotony} if $\B_1\subseteq\B_2\;$
 implies $\;\al O.(\B_1)\subseteq\al O.(\B_2)\;$ and $\;\al D.(\B_1)
 ={\al D.(\B_2)\cap\al O.(\B_1)}$ (cf.~Lemma~$\ref{Lem.3.1}$ for motivation).
\item[\rm (6)] {\bf weak causality} if for
  $\B_1\perp\B_2$ there is some $\B_0\supset\B_1\cup\B_2$, $\B_0\in\Gamma$ such that 
\[
 \Big[\,\al O.(\B_1)\,,\,\al O.(\B_2)\,\Big] \subset \al D. (\B_0)\,.
\]
\end{itemize}
\end{defi}

\begin{rem}
\label{Rem.4.1}
\begin{itemize}
\item[(i)] Given a system with reduction isotony, we have by 
 Lemma~\ref{Lem.3.1} that when $\B_1\subseteq\B_2$, then
 $\al R.(\B_1)$ is isomorphic to a C*--subalgebra of $\al R.(\B_2)$,
 which we will denote as $\iota_{12}\colon\ \al R.(\B_1)\to\al R.(\B_2)$.
\item[(ii)]
 The weak causality condition is considerably weaker than requiring 
 causality (cf.~(2) in Definition \ref{HK}) for the field algebra, 
 and this will be crucial below for Gupta--Bleuler electromagnetism. 
\end{itemize}
\end{rem}

Now we state our first major claim.
\begin{teo}
\label{Teo.4.2}
Let $\Gamma\ni\B\to (\al F.(\B),\al U.(\B))$ be a system of local 
quantum constraints. 
\begin{itemize}
\item[{\rm (i)}] 
 If it satisfies reduction isotony, then $\B\to\al R.(\B)$ has isotony,
 i.e.~$\;\B_1\subset\B_2\,,\;$ implies $\;\al R.(\B_1)\subset\al R.(\B_2)\,.$
 In this case, the net $\B\to\al R.(\B)$ has an inductive limit, which we 
 denote by $\al R._0:=\mathop{{\rm lim}}\limits_{\longrightarrow} \al R.(\B)$, and call it the \b quasi--local physical algebra..
\item[{\rm (ii)}] 
 If it satisfies weak causality and reduction isotony, then 
 $\B\to\al R.(\B)$ has causality, i.e.
 $\;\B_1\perp\B_2\,,\;$ implies $\;[\al R.(\B_1),\,\al R.(\B_2)]=0$.
\end{itemize}
\end{teo}
\begin{beweis}
(i) By reduction isotony we obtain from Lemma \ref{Lem.3.1}
for $\B_1\subset\B_2$ a unital monomorphism $\iota_{12}\colon\ \al R.(\B_1)
\to\al R.(\B_2)$. To get isotony from these monomorphisms,
we have to verify that they  satisfy
Takeda's criterion: $\iota_{13}=\iota_{23}\circ\iota_{12}$
(cf.~\cite{Takeda55}), which will ensure the existence of
the inductive limit $\al R._0$, and in which case we can
write simply inclusion $\al R.(\B_1)\subset\al R.(\B_2)$
for $\iota_{12}$. Recall that $\iota_{12}
(A+\al D.(\B_1))=A+\al D.(\B_2)$ for $A\in\al O.(\B_1)$.
Let $\B_1\subset\B_2\subset\B_3$, then by reduction isotony,
$\al O.(\B_1) \subset\al O.(\B_2)\subset\al O.(\B_3)$, and so
for $A\in\al O.(\B_1)$, 
$\iota_{23}\big(\iota_{12}(A+\al D.(\B_1))\big)=\iota_{23}(
A+\al D.(\B_2))=A+\al D.(\B_3)=\iota_{13}(A+\al D.(\B_1))$.
This establishes Takeda's criterion.

(ii) Let $\B_1\perp\B_2$ with $\B_0\supset\B_1\cup\B_2$ such that
$[\al O.(\B_1),\al O.(\B_2)]\subset\al D.(\B_0)$ 
as in Definition~\ref{weak.con}~(6), then since $\al O.(\B_1)\cup
\al O.(\B_2)\subset\al O.(\B_0)$ (by reduction isotony), the commutation 
relation is in $\al O.(\B_0)$, so when we factor out by $\al D.(\B_0)$,
the right hand side vanishes and since factoring is a homomorphism,
we get $[\al R.(\B_1),\,\al R.(\B_2)]=0$ in $\al R.(\B_0)$ and
therefore in $\al R._0$.
\end{beweis}

Next we would like to analyze the covariance requirement for the net,
but here too, we need a preliminary subsection on equivalence
of constraints (i.e. when they select the same set of Dirac states),
since it will only be necessary for the constraint set to be covariant
up to equivalence to ensure that the net of physical algebras is covariant.

\subsection{Equivalent constraints.}

\begin{defi}
\label{Def.Equi1}
Two first class constraint sets $\al C._1,\;\al C._2$ for the
field algebra $\al F.$ are called {\bf equivalent} if they
select the same Dirac states, i.e.
if for any state $\omega\in{\got S}(\al F.)$ we have
\[
   \al C._1\subseteq N_\omega\quad{\rm iff}\quad 
   \al C._2\subseteq N_\omega.
\]
In this case we denote $\al C._1\sim \al C._2$, and for unitary constraints
situation $\al C._i=\al U._i-\EINS$, $i=1,2$, we also write  
$\al U._1\sim \al U._2$.
\end{defi}
 
\begin{rem}
\label{Rem.Equi2}
\begin{itemize}
\item[(i)]
 It is clear that the preceding definition introduces an equivalence 
 relation on the family of first class constraint sets for $\al F.$. Denote by 
 $\al D._i=[\al FC._i]\cap [\al C._i\al F.]$, $i=1,2$, the C$^*$--algebras
 of Theorem~$\ref{Teo.2.1}$~(i). Now $\,\al C._1\sim\al C._2$ iff 
 $\al D._1=\al D._2$ by Theorems~$\ref{Teo.2.2}$~(i) and $\ref{Teo.2.1}$~(ii),
 therefore the corresponding observables $\al O._i$ and physical
 observables $\al R._i$ given by the T--procedure will also coincide. 
 This justifies calling these constraint sets equivalent -- 
 the replacement of $\al C._1$ by $\al C._2$ leaves the physics unchanged.
 Also note that $\al C.\sim[\al C.]\sim {\rm C}^*(\al C.)$.
If $\al C._1\sim\al C._2$, one can have that $\al C._1'\not=
\al C._2'$, i.e. the traditional observables is more sensitive to
the choice of constraints than $\al O.$.
\item[(ii)] Whilst the definition of equivalence $\al C._1\sim\al C._2$
 as stated, depends on $\al F.$, it depends in fact only on the subalgebra
 ${\rm C}^*(\al C._1\cup\al C._2\cup\{\EINS\})=:\al A.$. This is because
 the extension (resp. restriction) of a Dirac state from (resp. to) a
 unital C*--subalgebra containing the constraints, is again a Dirac state.
 Explicitly, the condition: $\omega(C^*C)=0$ for all $C\in\al C._1$
 iff $\omega(C^*C)=0$ for all $C\in\al C._2$, clearly depends only
 on the behaviour of $\omega$ on $\al A.$.
\end{itemize} 
\end{rem}

Next we give an algebraic characterization of equivalent constraints,
and introduce a maximal constraint set associated to 
an equivalence class of constraint sets. In the case of
unitary constraints, $\al C._i=\al U._i-\EINS$, we obtain
a unitary group in $\al F.$.

\begin{teo}
\label{Teo.Equi3}
Let $\al C._i$, $i=1,2$, be two first class constraint sets for $\al F.$, 
with associated algebras $\al D._i$ as above. Then
\begin{itemize}
\item[{\rm (i)}] $\al C._1\sim\al C._2\quad$ iff $\quad\al C._1-\al C._2\subset
 \al D._1\cap\al D._2$.\hfill\break
 In the case when $\al C._i=\al U._i-\EINS$, we have
 $\al U._1\sim\al U._2\quad$ iff $\quad\al U._1-\al U._2\subset\al D._1\cap\al D._2$.
\item[{\rm (ii)}] The maximal constraint set
 which is equivalent to $\al C._1$ is $\al D._1$.
 In the case when $\al C._1=\al U._1-\EINS$, the set of unitaries
\[
 \al U._m:=\mathop{\bigcup}\limits_{\al U.\,\sim\, \al U._1}\, \al U. 
 \subset \al F._u \,,
\]
is the maximal set of unitaries equivalent to $\al U._1$, and it is a group
with respect to multiplication in $\al F.$. 
\end{itemize}
\end{teo}
\begin{beweis}
(i) Suppose that $\al C._1\sim\al C._2$ so that by the 
Remark~\ref{Rem.Equi2}~(i), $\al D._1=\al D._2$. Then, we have that 
$\al C._1\subset\al D._1=\al D._2\supset\al C._2$, and hence
$\al C._1-\al C._2\subset\al D._1=\al D._1\cap\al D._2$.

Conversely, assume $\,\al C._1-\al C._2\subset \al D._1\cap\al D._2$. If 
$\omega\in{\got S}(\al F.)$ satisfies $\pi_\omega(\al C._1)\Omega_\omega=0$,
then by assumption, $\pi_\omega(\al C._1)\Omega_\omega-
\pi_\omega(\al C._2)\Omega_\omega\subseteq{\pi_\omega(\al D._1\cap\al D._2)
\Omega_\omega}\subset\pi_\omega(\al D._1)\Omega_\omega=0$
using Theorem~\ref{Teo.2.1}~(ii). Thus $\pi_\omega(\al
C._2)\Omega_\omega=0$, i.e.~$\al C._1\subset N_\omega$ implies 
that $\al C._2\subset N_\omega$. 
Interchanging the roles of $\al C._1$ and $\al C._2$, we conclude that
$\al C._1\sim\al C._2$. The second claim for $\al C._i=\al U._i-\EINS$ 
follows from $\al C._1-\al C._2=\al U._1-\al U._2$.

(ii) That $\al D._1\sim\al C._1$ is just the content of 
Theorem~$\ref{Teo.2.1}$~(ii). That it is maximal follows from
the implication $\al C._2\sim\al C._1\;\Rightarrow\;\al C._2\subseteq
\al D._2=\al D._1$. 

For unitary constraints, since 
$\al U._m:=\mathop{\bigcup}\limits_{\al U.\,\sim\, \al U._1}\, 
\al U.$ it follows from part (i) that $\al U._1-\al U._m\subset\al D._1$.
Further $\al U._1\subset\al U._m$ implies also $\al D._1\subset\al D._m$, 
so that $\al U._1-\al U._m\subset\al D._1\cap\al D._m=\al D._1$ and
$\al U._m\sim\al U._1$. By construction it is also clear that $\al U._m$
is the maximal unitary constraint set in $\al F.$ equivalent to $\al U._1$.
We only have to prove that $\al U._m$ is a group.
Let $\al U._0$ be the group generated in $\al F.$ by $\al U._m$.
If $\omega\in{\got S}(\al F.)$ satisfies $\omega(\al U._m)=1$, we have
$1=\overline{\omega(U)}=\omega(U^*)=\omega(U^{-1})$, 
$U\in\al U._m$, and also $\omega(UV)=\omega(U)=1$, $U,V\in\al U._m$,
i.e.~$\omega(\al U._0)=1$. Thus
$\al U._0\sim\al U._m\sim\al U.$ and maximality implies
$\al U._0=\al U._m$. Hence $\al U._m$ is a group.
\end{beweis}

\begin{rem}
\label{maxU}
Observe that for a given unitary constraint system
$(\al F.,\,\al U.)$ we have that 
\[
 \al U._m={\{U\in\al F._u\mid
 \omega(U)=1\;\,\forall\;\omega\in{\got S}_D\}}
 ={\{U\in\al O._u\mid\omega(U)=1\;\,\forall\;\omega\in{\got S}_D\}}
\]
since $\al U._m\subset\al O.$, cf.~Theorem~\ref{Teo.2.2}~(v).
\end{rem}
 Next we show that for a large class of first class 
constraint systems $(\al F.,\al C.)$ we can find a single constraint 
in $\al F.$ which is equivalent to $\al C.$, and hence can replace it.

\begin{teo}
\label{Teo.2.8}
If $[\al C.]$ is separable, there exists a 
positive element $C\in\al D._+$ such that $\{C\}\sim\al C.$.
\end{teo}
\begin{beweis}
Let $\{C_n\}_{n=1}^\infty$ be a denumerable basis of $[\al C.]$ such 
that $\|C_n\|<1$, $n\in\z{N}$. Then
\[
 {\got S}_D= \Big\{ \omega\in{\got S}({\al F.})\mid 
             \omega(C_n^*C_n)=0 \quad {\forall}\, n\in \z{N}\Big\}\,.
\]
Define
\[
 C:=\sum_{n=1}^\infty \frac{C_n^*C_n}{2^n}\in\al D._+\,.
\]
Then $\omega(C)=0$ iff $\omega(C_n^*C_n)=0$ for all $n\in \z{N}$,
which proves that ${\got S}_D=\{ \omega\in{\got S}(\al F.)\mid
\omega(C)=0\}$. Thus $C^{\frac12}\in\al N.$, but since for any positive
operator $A$ we have ${\rm Ker}\, A={\rm Ker}\, A^n$, for all $n\in\z{N}$,
we conclude ${\got S}_D=\{ \omega\in{\got S}(\al F.)\mid
\omega(C^2)=0\}$, so $\{C\}\sim\al C.$.
\end{beweis}

\begin{rem}
\label{Rem.2.9}
Note that from \cite[p.~85]{bMurphy90} the preceding statement is not true if
the separability condition is relaxed.
From Remark~$\ref{Rem.2.12}$(i) we see that if we are willing to enlarge
$\al F.$ to
${\rm C}^*(\al F.\cup\{P\})$ for a certain projection $P$, then
$\{P\}\sim\al C.$ so that for this larger algebra, the separability
assumption can be omitted.
\end{rem}

\begin{teo}
\label{Teo.2.9}
Let $(\al F.,\,\al C.)$ be a first class constraint system, then there
is a set of unitaries $\al U.\subset\al F._u$ such that 
$\al C.\,\sim\,\al U.-\EINS$ and $\al U.=\al U.^*$.
\end{teo}
\begin{beweis}
Define the unitaries $\al U.:=\{\exp(itD)\,\mid\,t\in\R,\; D\in\al D._+\}$.  
Then for $\omega\in{\got S}(\al F.)$ we have that
$1=\omega(\exp\; itD)= 1+\sum\limits_{k=1}^\infty(it)^k\omega(D^k)/k!\;$
for all $t\in\R$, $D\in\al D._+\;$ iff $\;\omega(D)=0$ for all $D\in\al D._+\;$
iff $\;\omega(\al D.)=0\;$ iff $\;\omega\in{\got S}_D$.
Thus $\al U.-\EINS\;\sim\;\al C.$.
It is obvious that $\al U.=\al U.^*$.
\end{beweis}
Hence no constraint system is excluded by the assumption of unitary
constraints. Moreover, by Theorem~\ref{Teo.Equi3}
there is a canonical unitary group $\al U._m$ associated 
with each first class constraint system, and hence a group
of inner automorphisms ${\rm Ad}\,\al U._m$, which one can take
as a gauge group in the absence of any further physical
restrictions.

\subsection{Weak covariance.}

We define:
\begin{defi} 
\label{weak.cov}
Fix a system of local quantum constraints 
$\B\to(\al F.(\B),\,\al U.(\B))$, then we say that it satisfies:
\begin{itemize}
\item[\rm (7)] {\bf weak covariance} if there is an action 
 $\alpha\colon\ \al P._+^\uparrow\to{\rm Aut}\,\al F._0$ 
 such that $\alpha_g(\al O.(\B))=\al O.(g\B)$ and
 $\alpha_g(\al U.(\B))\;\sim\;\al U.(g\B)$, for all $g\in\al P._+^\uparrow$,
 $\B\in\Gamma$ (cf.~Definition~$\ref{Def.Equi1}$).
\end{itemize}
\end{defi}

\begin{rem}
\label{WCov}
\begin{itemize}
\item[(i)]
 For the weak covariance condition, we do not need to state in which algebra
 the equivalence of constraints holds, since this only depends
 on the unital C*--algebra generated by the two constraint sets involved
 (cf.~Remark~\ref{Rem.Equi2}~(ii)).
 Note that if the net $\B\to\al F.(\B)$ is already covariant, then
weak covariance follows from the covariance
 condition $\alpha_g(\al F.(\B))=\al F.(g\B)$ and
 $\alpha_g(\al U.(\B))\;\sim\;\al U.(g\B)$, for all $g\in\al P._+^\uparrow$,
 using the fact that equivalent constraint sets produce the same
 observable algebra. 
\item[(ii)]
 It is instructive to compare the conditions in Definitions \ref{weak.con}
and \ref{weak.cov}
 with those of the Doplicher-Haag--Roberts analysis (DHR for short 
 \cite{DHR69a}), given that both are intended for application
 to gauge QFTs. First, in DHR analysis one assumes that the actions
 of the gauge group and the Poincar\'e group commute, which limits
 the analysis to gauge transformations of the first kind (and hence
 excludes quantum electromagnetism). In contrast, we assume weak
 covariance, hence include gauge transformations of the second kind
 (and also QEM). The DHR analysis also assumes field algebra
 covariance, which we omit. Second, the DHR analysis is done concretely
 in a positive energy representation, whereas we assume an abstract
 C*--system, hence we can avoid the usual clash between regularity
 and constraints, which appears as continuous spectrum problems
 for the constraints (cf.~Subsection~\ref{Manuceau})
 and which generally leads to indefinite metric representations.
 At the concrete level this problem manifests itself in the inability
 of constructing the vector potential satisfying Maxwell's equations
 as a covariant or causal quantum field on a space with an invariant 
 vacuum, cf.~\cite{Strocchi67,Strocchi70,Barut72} and
 \cite[Eq.~8.1.2]{bWeinberg95}.
\item[(iii)]
 In the next sections we will construct an example (Gupta--Bleuler 
 electromagnetism) which satisfies the conditions \ref{weak.con}
and \ref{weak.cov}.
\end{itemize}
\end{rem}

Now we show that the conditions in Definition \ref{weak.con}
and \ref{weak.cov} are
sufficient to guarantee that the net of local physical observables
$\B\to\al R.(\B)$ is a HK--QFT. This is a central result for this paper.

\begin{teo}
\label{Teo.4.4}
Let $\Gamma\ni\B\to (\al F.(\B),\al U.(\B))$ be a system of local 
quantum constraints. 
 If it satisfies weak covariance, then for each $\B$ we have 
 $\widetilde{\alpha}_g (\al R.(\B))=\al R.(g\B)$, $g\in\al P._+^\uparrow$, 
 where $\widetilde{\alpha}_g$ is the factoring through of $\alpha_g$ (cf.~$(7)$ in 
 Definition~$\ref{weak.cov}$) to the local factor algebra $\al R.(\B)$.
 If the system of constraints in addition satisfies 
 reduction isotony, then the isomorphisms 
 $\widetilde{\alpha}_g\colon\ \al R.(\B)\to\al R.(g\B)$, 
 $\B\in\Gamma$, are the restrictions of an automorphism $\widetilde{\alpha}_g
 \in{\rm Aut}\,\al R._0$, and moreover,
 $\widetilde{\alpha}\colon\ \al P._+^\uparrow\to{\rm Aut}\,\al R._0$ 
 is an action, i.e.~the net $\B\to\al R.(\B)$ satisfies covariance.
\end{teo}
\begin{beweis}
 Let $\alpha\colon\ \al P._+^\uparrow\to{\rm Aut}\,\al F._0$ be the 
action introduced by the weak covariance assumption in 
Definition~\ref{weak.cov}~(7).
Now $\alpha_g(\al O.(\B))$ are the observables of the constraint
system $(\alpha_g(\al F.(\B)),\,\alpha_g(\al U.(\B)))$
with maximal constraint algebra
$\alpha_g(\al D.(\B))\subset\alpha_g(\al O.(\B))=
\al O.(g\B)\supset\al D.(g\B)$.
Since $\alpha_g(\al U.(\B))\sim \al U.(g\B)$, they have the same Dirac states
and so on $\al O.(g\B)$ the same maximal {\rm C*}--algebra contained in the
kernels of all Dirac states. Thus $\alpha_g(\al D.(\B))=\al D.(g\B)$.
Denote the factoring map $\xi_\B\colon\ \al O.(\B)\to\al R.(\B)$, i.e.
$\xi_\B(A)=A+\al D.(\B)$ for all $A\in\al O.(\B)$.
Then we factor through $\alpha_g\colon\ \al O.(\B)\to\al O.(g\B)$ to a map
$\widetilde{\alpha}_g\colon\ \al R.(\B)\rightarrow \al R.(g\B)$ by
\[
\widetilde{\alpha}_g(\xi_{\B}(A))
    := \alpha_g(A)+\alpha_g(\al D.(\B))=
         \alpha_g(A)+\al D.(g\B) 
     =  \xi_{g\B}(\alpha_g(A))\,,
\]
and this is obviously an isomorphism.

Next assume in addition reduction isotony, then we show that
the isomorphisms $\widetilde{\alpha}_g$ defined on the net
$\B\to\al R.(\B)$ are the restrictions of an automorphism
$\widetilde\alpha_g\in{\rm Aut}\,\al R._0$.
Indeed, for $\B_1\subset\B_2$ and any $A\in\al O.(\B_1)$
we have using equation $\al D.(g\B_1)= \al D.(g\B_2) \cap \al O.(g\B_1)$
and the monomorphisms $\iota_{12}\colon\ \al R.(\B_1)\to\al R.(\B_2)$,
$\;\iota^g_{12}\colon\ \al R.(g\B_1)\to\al R.(g\B_2)$ that
\begin{eqnarray*}
 \iota^g_{12}\Big(
\widetilde{\alpha}_g\big(\xi_{\B_1}(A)\big)\Big) 
  &=& \iota^g_{12}\Big(\alpha_g(A)+\al D.(g\B_1)\Big)\kern2mm=\kern2mm 
      \alpha_g(A)+\al D.(g\B_2) \\
  &=& \widetilde{\alpha}_g\Big(\xi_{\B_2}(A)\Big) 
   \kern2mm=\kern2mm\widetilde\alpha_g\big(\iota_{12}(\xi_{\B_1}(A))\big)\,.
\end{eqnarray*}
This shows that the diagram
\begin{eqnarray*}
  \al R.(\B_1)\kern3mm &\mathop{\longrightarrow}\limits^{\iota_{12}^{\phantom{g}}}
     & \kern3mm \al R.(\B_2)  \\[3mm]
        \Big\downarrow\, \widetilde{\alpha}_g\kern3mm
                       & & 
       \kern6mm \Big\downarrow\, \widetilde{\alpha}_g \\[3mm]
  \al R.(g\B_1)\kern3mm &\mathop{\longrightarrow}\limits^{\iota^g_{12}}
   & \kern3mm  \al R.(g\B_2)  
\end{eqnarray*}
commutes.
Therefore by the uniqueness property of the inductive
limit \cite[Section~11.4]{bKadisonII} the isomorphisms 
$\widetilde{\alpha}_g$ of the local observable algebras characterize an
automorphism of $\al R._0$ which we also denote by $\widetilde{\alpha}_g$.
Since $\alpha\colon\ \al P._+^\uparrow\rightarrow {\rm Aut}\,\al F.$ 
is a group homomorphism, we see for the local isomorphisms
that the composition of
 $\widetilde\alpha_g\colon\ \al R.(\B)\to\al R.(g\B)$  with
 $\widetilde\alpha_h\colon\ \al R.(g\B)\to\al R.(hg\B)$  is
 \break $\widetilde\alpha_h\circ\widetilde\alpha_g=
  \widetilde\alpha_{hg}\colon\ \al R.(\B)\to\al R.(hg\B)$. 
{}From this it follows that $\widetilde\alpha\colon\
\al P._+^\uparrow\to{\rm Aut}\, \al R._0$ is a group homomorphism.
\end{beweis}

So combining Theorems~\ref{Teo.4.2} and \ref{Teo.4.4} we
obtain our main claim:
\begin{teo}
\label{Teo.4.5}
If the system of local constraints satisfies all three
conditions in Definitions~$\ref{weak.con}$ and $\ref{weak.cov}$,
 then $\B\to\al R.(\B)$ is a HK--QFT.
\end{teo}
In the following sections we will construct field theory examples
of local systems of quantum constraints which satisfy the weak conditions
of Definition \ref{weak.con} and \ref{weak.cov}, hence 
define HK--QFTs for their
net of physical algebras.

\begin{pro}
\label{Pro.4.3}
Given a system of local quantum constraints, 
$\B\to (\al F.(\B),\,\al U.(\B))$, which satisfies reduction isotony 
and weak covariance, then the net
$\B\to(\al F.(\B),\,\al U._m(\B))$ $($where $\al U._m(\B)$ is the maximal
constraint group of $\al U.(\B)$ in $\al F.(\B)$, 
cf.~Theorem~$\ref{Teo.Equi3}~(ii))$ is a system of local quantum 
constraints satisfying reduction isotony and covariance, i.e.
\[
 \alpha_g(\al O.(\B))=\al O.(g\B)\quad\hbox{and}\quad
 \alpha_g(\al U._m(\B))=\al U._m(g\B)\,,\quad g\in\al P._+^\uparrow
  \,,\;\B\in\Gamma\,.
\] 
The system $\B\to(\al F.(\B),\,\al U._m(\B))$  is clearly locally 
equivalent to $\B\to (\al F.(\B),\,\al U.(\B))$, in the sense that
$\al U._m(\B)\sim\al U.(\B)$ for all $\B\in\Gamma$, from which it follows
that if one of these two systems has weak causality, so has the other one.
\end{pro}
\begin{beweis}
Let $\B_1\subseteq\B_2$, then by $(\al F.(\B_1),\,\al U.(\B_1))\subseteq
(\al F.(\B_2),\,\al U.(\B_2))$ and reduction isotony, we conclude from
Theorem~\ref{Lem.3.5} that all Dirac states on $\al O.(\B_1)\subset
\al O.(\B_2)$ extend to Dirac states on $\al O.(\B_2)$. Thus
by Remark~\ref{maxU}, 
\begin{eqnarray*}
\al U._m(\B_2)\cap\al F.(\B_1)&=&
    \{U\in\al O._u(\B_2)\mid\omega(U)=1\quad\forall\;\omega\in
{\got S}_D^{\B_2}\}\cap\al F.(\B_1)  \\[2mm]
 &=&   \{U\in\al F._u(\B_1)\cap\al O.(\B_2)
\mid\omega(U)=1\quad\forall\;\omega\in
{\got S}_D^{\B_1}\} \\[2mm]
 &=&   \{U\in\al O._u(\B_1)
\mid\omega(U) = 1\quad\forall\;\omega\in{\got S}_D^{\B_1}\} \\[2mm]
 &=&  \al U._m(\B_1)\,,
\end{eqnarray*}
where we used the fact that if for a unitary $U$ we have
$\omega(U)=1$ for all Dirac states $\omega$, then $U\in\al O.$.
Thus $(\al F.(\B_1),\,\al U._m(\B_1))\subseteq
(\al F.(\B_2),\,\al U._m(\B_2))$ and so the system
$\B\to(\al F.(\B),\,\al U._m(\B))$  is a system of local
quantum constraints. Reduction isotony follows from that
of the original system and the equivalences
$\al U._m(\B)\sim\al U.(\B)$ for all $\B\in\Gamma$.

To prove the covariance property of $\al U._m(\B)$
recall that from weak covariance we have 
$\alpha_g(\al U.(\B))\sim \al U.(g\B)$ for all $\B$. We show first 
that if $\al U._1(\B)\sim \al U.(\B)$, then 
$\alpha_g(\al U._1(\B))\subset \al U._m(g\B)$. We have 
\begin{eqnarray*}
\alpha_g\Big( \al U._1(\B) \Big)-\al U.(g\B)
  &=&\alpha_g\Big( \al U._1(\B)-\EINS\Big)+\EINS-\al U.(g\B)\\[2mm]
  &\subset& \alpha_g\Big( \al D.(\B) \Big)
            +  \al D.(g\B)                                  \\[2mm]
  &=&  \al D.(g\B)\kern2mm =\kern2mm \alpha_g(\al D.(\B))\,,
\end{eqnarray*}
where the last equality follows from the proof of the previous
theorem, and we used also that $\al D._1(\B)=\al D.(\B)$. Since
$\alpha_g(\al D.(\B))$ is the $\al D.$--algebra of 
$\alpha_g(\al U._1(\B))$ in $\alpha_g(\al O.(\B))=\al O.(g\B)$, this
implies by Theorem~\ref{Teo.Equi3}~(i) that $\alpha_g(\al U._1(\B))\sim 
\al U.(g\B)$ and therefore it must be contained in $\al U._m(g\B)$. Thus
$\alpha_g(\al U._m(\B))\subset \al U._m(g\B)$, $g\in\al P.$, and finally
the inclusion $\alpha_{g^{-1}}(\al U._m(g\B))\subset \al U._m(\B)$ proves
covariance for $\B\to\al U._m(\B)$.
\end{beweis}

%%%%%%%%%%%%%%%%%%%%%%%%%%%%%%%%%%%%%%%%%%%%%%%%%%%%%%%%%%%%%%%%%%%%%%%%%%%%%%

\section{Preliminaries for the Example.}

In this section we collect the relevant material we need to develop
our Gupta--Bleuler example in the next section.

\subsection{Outer constraints.}
\label{OuterConstraints}

We will need the 
following variant where the constraints are defined through a group 
action which is not necessarily inner. One assumes, following 
\cite{Grundling88b} that:
\begin{itemize}
\item{} There is a distinguished group action
$\beta\colon\ \al G.\to{\rm Aut}\,\al F.$ on the field algebra $\al F.$, 
and all physical information is contained in $\al F.$ and its set of 
invariant states:
\[
  \wp^\al G.(\al F.):=\{\omega\in\wp(\al F.)\mid{\omega\bigl(\beta_g(A)\bigr)=
                 \omega(A)\quad\forall\,g\in \al G.,\;A\in\al F.}\}\,.
\]
\end{itemize}
If $\al G.$ is locally compact, we can construct the (abstract) multiplier
algebra of the crossed product $\al F._e={M(\al G.\cross\beta.\al F.)}$ and 
otherwise we will just take the discrete crossed product. In either case 
we obtain a C*--algebra $\al F._e\supset\al F.$ which contains unitaries 
$U_g\;$ for all $\,{g\in \al G.}$ that implement ${\beta\colon\
\al G.\to{\rm Aut}\,\al F.}$. Then this situation is reduced to the 
previous one by the following theorem \cite[Section~3]{Grundling88b}:
\begin{teo}
\label{Outer}
$\wp^\al G.(\al F.)$ is precisely the restriction to $\al F.$ of
the Dirac states on $\al F._e$ with respect to $\al C.=U_\al G.-\EINS$, 
i.e.~${\wp^\al G.(\al F.)}={\wp_D(\al F._e)\restriction\al F.}$ where
\[
 \wp_D(\al F._e):=\{\omega\in\wp(\al F._e)\mid\omega(U_g)=1\quad
 \forall\,g\in \al G.\}\;.
\]
\end{teo}
Hence we can apply the T--procedure to $U_\al G.-\EINS$ in $\al F._e$, 
and intersect the resulting algebraic structures with $\al F.$.
This is called the {\it outer} constraint situation.

%%%%%%%%%%%%%%%%%%%%%%%%%%%%%%%%%%%%%%%%%%%%%%%%%%%%%%%%%%%%%%%%%%%%%%%%%%%%%%
\subsection{Bosonic constraints.}
\label{Manuceau}

For free bosons, one takes for $\al F.$ the {\rm C*}--algebra of the CCRs,
which we now define following Manuceau
\cite{Manuceau68a,Manuceau73}. Let ${\got X}$ be a linear space and 
$B$ a (possibly degenerate) symplectic form on it. Denote by 
$\Delta({\got X},\,B)$ the linear space of complex--valued
functions on ${\got X}$ with finite support. It has as linear basis the set
$\{\delta_{f}\mid f\in {\got X}\}$, where
\[
 \delta_{f}(h):=\left\{\begin{array}{l}
                1  \quad     {\rm if} \kern3mm f=h \\
                0  \quad     {\rm if} \kern3mm f\neq h.
                         \end{array} \right.
\]
Make $\Delta({\got X},\,B)$ into a *--algebra, by defining the
product ${\; \delta_{f} \cdot \delta_{h}}:= e^{\frac{i}{2}B(f,\, h)}\,
\delta_{f+h}\;$ and involution $(\delta_{f})^*:=\delta_{-f}$, where 
$f,\; h\in S$ and the identity is $\delta_0$. Let 
$\Delta_1({\got X},\,B)$ be the closure of
$\Delta({\got X},\,B)$ w.r.t. the norm $\Big\|\sum\limits_{i=1}^m 
\alpha_i \delta_{f_i}\Big\|_1:=\sum\limits_{i=1}^m |\alpha_i|$, 
$\alpha_i\in\z{C}$, then the CCR--algebra 
$\overline{\Delta({\got X},\,B)}$ is defined as the enveloping
{\rm C*}--algebra of the the Banach *-algebra $\Delta_1({\got X},\,B)$. 
That is, it is the closure with respect to the enveloping C*-norm:
\[
 \| A\| := \mathop{{\rm sup}}_{{\omega\in {\got
           S}(\Delta_1({\got X},\,B))}} \sqrt{\omega(A^*A)}\,.
\]
It is well--known (cf.~\cite{Manuceau73}) that:
\begin{teo}
\label{Teo.2.13}
$\CCR $ is simple iff $B$ is nondegenerate. 
\end{teo}
An important state on $\CCR$ is the central state defined by
\begin{equation}\label{central}
\omega_0\Big( \delta_{f}\Big) :=\left\{\begin{array}{l}
                1  \quad     {\rm if} \kern3mm f=0 \\
                0  \quad     {\rm otherwise} \;.
                         \end{array} \right.
\end{equation}
Using it, we make the following useful observations. The
relation between the norms on $\Delta(\ot X.,\, B)$ is
\[
\|F\|_2:=\Big(\sum_{i=1}^n|\lambda_i|^2\Big)^{1/2}=
\omega_0(F^*F)^{1/2}\leq\|F\|\leq\|F\|_1\;\;
\hbox{for}\; F=\sum_{i=1}^n\lambda_i\delta_{f_i}
\]
and hence
$\Delta_1(\ot X.,\, B)\subset\CCR\subset\ell^2(\ot X.)$, so we can write
an $A\in\CCR$ as $A=\sum\limits_{i=1}^\infty\lambda_i\delta_{f_i}$
where $\{\lambda_i\}\in\ell^2$ and $f_i=f_j$ iff $i=j$.
Let $A_n:=\sum\limits_{i=1}^{M_n}\gamma^{(n)}_i\,\delta_{f_i^{(n)}}\subset
\Delta(\ot X.,\,B)$ converge to $A\in\CCR$ in C*--norm
then  since the family of $f_i^{(n)}$ is denumerable
we can arrange it into a single sequence $f_i$ and thus write 
 $A_n:=\sum\limits_{i=1}^{N_n}\lambda^{(n)}_i\,\delta_{f_i}$.
We shall frequently use this way of writing a Cauchy sequence 
in $\Delta(\ot X.,\, B)$.

Now to define a constrained system corresponding to linear selfadjoint
constraints in $\al F.=\CCR$, we choose a set $\al C.=\al U.-\EINS$
where $\al U.=\{\delta_f\mid f\in{\frak s}\}$ and ${\frak s}\subset
{\got X}$ is a subspace corresponding to the ``test functions'' of the heuristic
constraints.
\begin{teo}
\label{Teo.2.131}
Define the symplectic commutant ${\frak s}':=
\{f\in {\got X}\mid B(f,\,{\frak s})=0\}$, then 
$\al C.=\al U.-\EINS$ is first class iff ${\frak s}\subseteq {\frak s}'$.
\end{teo}
For the proof, see Lemma~6.1 in
\cite{Grundling88a}.

We saw after Theorem~\ref{Teo.2.2}
that for the observable algebra we sometimes need to 
choose a smaller algebra $\al O._c\subset\al O.$ in order to
ensure that the physical algebra $\al R._c$ is simple. For
bosonic constraints with nondegenerate $B$, such an algebra is 
$\al O._c={\rm C}^*(\delta_{{\frak s}'})=\al C.'$ 
(in which case $\al D.\cap\al O._c={\rm C}^*(\delta_{{\got s}'})\cdot 
 {\rm C}^*(\delta_{\got s}-\EINS)$), 
which is what was chosen in \cite{Grundling85,Grundling88a}.
However, we now show that with this choice we have in fact
$\al R._c=\al R.$, i.e.~we obtain the same physical algebra
than with the full T--procedure, so nothing was lost
by this choice of $\al O._c$.

\begin{teo}
\label{FixingBlemish}
Given nondegenerate $({\got X},\,B)$ and ${\frak s}\subset 
{\got X}$ as above, where ${\frak s}\subset{\frak s}'$ and 
${\frak s}={\frak s}''$, then
\[
 \al O.= {\rm C}^*(\delta_{{\frak s}'}\cup \al D.)=\Big[ 
          {\rm C}^*(\delta_{{\frak s}'}) \cup \al D. \Big]\,.
\]
\end{teo}
\begin{beweis} The proof of this is new but long, so we put it in 
Appendix~2.
\end{beweis}

\begin{teo}
\label{Teo.2.14}
Consider a nondegenerate symplectic space $({\got X},\,B)$ and a 
first class set ${\frak s}\subset{\got X} $.
Denote by $\widetilde{B}$ the factoring through of $B$ to the factor
space ${\frak s}'/ {\frak s}$. Then we have the following isomorphism:
\[
 {\rm C}^*(\delta_{{\frak s}'})\; / \;{\rm C}^*(\delta_{\frak s}-\EINS)\,
 {\rm C}^*(\delta_{{\frak s}'})\cong 
  \overline{\Delta({\frak s}'/ {\frak s}\,,\,\widetilde{B})}\,.
\]
In particular, if ${\frak s}={\frak s}''$, then 
  $({\frak s}'/ {\frak s}\,,\,\widetilde{B})$
is nondegenerate, so the above CCR--algebra is simple, and using
Theorem~$\ref{FixingBlemish}$ we have
\[
 \al R.\cong 
  \overline{\Delta({\frak s}'/ {\frak s}\,,\,\widetilde{B})}\,.
\]
\end{teo}
For proofs and further details see 
\cite[Theorem~5.2 and 5.3, as well as Corollary~5.4 and 5.5]{Grundling88b}.
The surprise is that for linear bosonic constrained systems, the choice of
traditional observables $\al O._c=\al C.'$ produces the same physical algebra 
$\al R.$ than the T--procedure, which is not true in general.

A typical pathology which occurs for bosonic constraints, is that
the Dirac states are not regular, i.e.~the one parameter groups
$\R\ni t\mapsto\pi_\omega(\delta_{tf})$ for $\omega\in{\got S}_D$ 
will not be strong operator continuous for all $f\in{\got X}$, and so 
the corresponding generators (which are the smeared quantum 
fields in many models of bosonic fields), 
will not exist for some $f\in{\got X}$, cf.~\cite{Grundling88c}.
The resolution of this, is that the pathology only occurs on
nonphysical elements, i.e.~on $\delta_f\not\in\al O.$, with the result
that a Dirac state when restricted to $\al O.$ and factored to
$\al R.$ (i.e.~taken through the bijection in Theorem~$\ref{Teo.2.6}$)
can be regular again on the physical algebra $\al R.$. This is also
obvious from Theorem~$\ref{Teo.2.14}$, since a nontrivial $\al R.$ clearly has 
regular states. Thus for the physical algebra, quantum fields can exist.

%%%%%%%%%%%%%%%%%%%%%%%%%%%%%%%%%%%%%%%%%%%%%%%%%%%%%%%%%%%%%%%%%%%%%%%%%%%%%%
\section{Example: Gupta--Bleuler electromagnetism}
\label{Examples}

Quantum electromagnetism, in the heuristic Gupta--Bleuler
formulation, has a number of special features, cf.~\cite{Gupta50,Bleuler50}.
First, it is represented on an indefinite inner product space,
second, gauge invariance is imposed by the noncausal constraint
\[
  \chi(x):=\Big(\partial^\mu A_\mu\Big)^{\mbox{\tiny{(+)}}}(x)\,,
\]
and third, Maxwell's equations (in terms of the vector potential)
are imposed as state conditions instead of as operator identities.
This is necessary, because from the work of Strocchi 
(e.g.~\cite{StrocchiIn73,Strocchi74}), we know that
Maxwell's equations are incompatible
with the Lorentz covariance of the vector potential.
Gupta--Bleuler electromagnetism has been  rigorously reconstructed in
a C*--algebra context \cite{Grundling88b}, in a way which allows one
to avoid indefinite inner product representations (using instead
representations which are nonregular on nonphysical objects).
Here we will refine that approach in order to include the local 
constraint structure and to make contact with Haag--Kastler QFT.
Our aim is to define Gupta--Bleuler electromagnetism as a local
system of constraints as in Definition \ref{loc.cons} and subsequently
to show that it has reduction isotony, weak causality and covariance.
Our starting point for defining this system, is
\cite[Sections~4 and 5]{Grundling88b}
where motivation and further results can be found.

\subsection{Gupta--Bleuler electromagnetism -- the heuristic theory.}
\label{GBh}

Heuristically the field is
\def\f #1,#2.{\mathsurround=0pt \hbox{${#1\over #2}$}\mathsurround=5pt}
   \def\hlf{{\f 1,2.}}
\def\set #1,#2.{\left\{\,#1\;\bigm|\;#2\,\right\}}
\[
F_{\mu\nu}(x):=A_{\nu,\,\mu}(x)-A_{\mu,\,\nu}(x)
\]
where the vector potential, constructed on a Fock--Krein space  $\al H.$ is:
\[
A_\mu(x)=\bigl(2(2\pi)^3\bigr)^{-\hlf}\int_{C_+}\Bigl(a_\mu(\b p.)\,e^{-ip\cdot
x}+a^\d_\mu(\b p.)\,e^{ip\cdot x}\Bigr){d^3p\over p_0}
\]
where 
$C_+:=\{p\in\R^4\mid p_\mu p^\mu=0,\; p_0\geq 0\}$ is the 
mantle of the positive light cone:
$V_+:=\{p\in\R^4\mid p_\mu p^\mu\geq 0,\; p_0\geq 0\}$. 
Note that the adjoints $a^\d$ are w.r.t. the indefinite inner product,
and that the latter comes from the indefinite inner product
on the one particle space:
\[
K( f,h):=-2\pi \int_{C_+}{d^3p\over p_0}\overline{f}_\mu(p)
h^\mu(p).
\]
Then  the CCR's are
\[
\bigl[A_\mu(x),\,A_\nu(x')\bigr]=-i\eta_{\mu\nu}D(x-x')\;,\qquad
D(x):=-(2\pi)^{-3}\int_{C_+}\, e^{i\b p.\cdot\b x.}\,\sin (p_0x_0)\,
{d^3p\over p_0}
\]
using $\bigl[a_\mu(\b p.),\,a^\d_\nu(\b p.')\bigr]=-\eta_{\mu\nu}\|\b p.\|\,
\delta^3(\b p.-\b p.')$ and the other commutators involving $a$ are zero.
At this point ${A_\mu(x)}$ does not yet satisfy the field equations
${F_{\mu\nu}}^{,\nu}(x)=0$.
On smearing we obtain:
\begin{eqnarray}
A(\hat f)&:=&
\int d^4x\,A_\mu(x)\,f^\mu(x)  \label{smearI} \\[2mm]
&=&\sqrt\pi\int_{C_+}\Bigl(a_\mu(
\b p.)\,\hat f^\mu(p)+a_\mu^\d(\b p.)\,\overline{\hat f^\mu(p)}\Bigr){d^3p\over p_0}
\label{smearII} \\[2mm]
&=&\big(a(f)+a(f)^\d\big)\Big/\sqrt{2}\nonumber\\[2mm] \hbox{with}\qquad
a(f)&:=&\sqrt{2\pi}\int_{C_+}a_\mu(
\b p.)\,\hat f^\mu(p){d^3p\over p_0}
\nonumber
\end{eqnarray}
 where $f\in\al S.({\R}^4,\,{\R}^4)$ and 
   $\hat f_\mu(p):=(2\pi)^{-2}\int d^4x\,e^{-ip\cdot x}f_\mu(x)
\in\widehat\al S.$ and the latter means
\[
 \widehat\al S.:=\{\widehat{f}\mid f\in
\al S.(\R^4,\R^4)\}={\{f\in\al S.(\R^4,\C^4)\mid
\overline{f(p)}=f(-p)\}}
\]
and as usual $\al S.$ denotes Schwartz functions.
The operators $A(\hat f)$ are Krein symmetric, but not
selfadjoint.
Then the smeared CCRs are
\begin{equation} \label{symp}
[A(f),\, A(h)]=iD(f,\,h):=
-\pi\int_{C_+}\Big(f_\mu(p)\,\overline{h^\mu}(p)
-\overline{f_\mu}(p)\,h^\mu(p)\Big){d^3p\over p_0}\;\;.
\end{equation}
Note that the distribution $D$ is actually the Fourier transform
of the usual Pauli--Jordan distribution, i.e.
\[
\widehat{D}(f,\, h):=D(\widehat{f},\,\widehat{h})=
\int\int f_\mu(x)\,h^\mu(y)\, D(x-y)\,d^4x\, d^4y
\]
in heuristic form.
The supplementary condition 
\[
\chi(x):=\partial^\mu A_\mu^{(+)}(x)={-i\bigl(2
(2\pi)^3\bigr)^{-\hlf}\int_{C_+} p^\mu a_\mu(\b p.)\,e^{-ip\cdot x}}
{d^3p\over p_0}
\]
 selects
the physical subspace $\al H.':=\set\psi\in\al H.,\chi(h)\psi=0,h\in{\al S.
(\R^4,\R)}.$
(to make this well--defined, we need to specify the domain of $\chi(h)$,-
this will be done in Subsection~5.6).
The Poincar\'e transformations are defined
in the natural way: $(\Lambda,\, a)f(p)=e^{ia\cdot p}\Lambda f(\Lambda^{-1}p)$,
and the given Krein inner product on $\al H.$ is invariant w.r.t. the Poincar\'e 
transformations, but not the Hilbert inner product.
Moreover $\al H.'$ is positive semidefinite w.r.t. the Krein inner
product $\langle\cdot,\cdot\rangle$, so the heuristic theory constructs 
the physical Hilbert space $\al H._{\rm phys}$
 as the closure of $\al H.'/\al H.''$ equipped with inner product
 $\langle\cdot,\cdot\rangle$ where $\al H.''$ is the zero norm part
of it. At the one particle level, $\al H.'$ consists of functions satisfying
$p_\mu f^\mu(p)=0,$ and  $\al H.''$ consists of gradients 
$f_\mu(p)=ip_\mu h(p).$
The physical observables consist of operators which can 
factor to $\al H._{\rm phys}$, and in particular contains
the field operators $F_{\mu\nu}.$ These satisfy the Maxwell
equations on $\al H._{\rm phys}$, because ${F_{\mu\nu}}^{,\mu}$ maps 
 $\al H.'$ to $\al H.''$.

Note  that since the Krein inner product becomes the Hilbert inner product
on $\al H._{\rm phys},$ the Krein adjoint becomes the Hilbert space
adjoint for physical observables. With this in mind, we will below
do a reconstruction in C*--algebraic terms where the C*--involution
corresponds to the Krein involution.

%For the C*--theory below, we first look at how the preceding fields smear.
% After smearing  with ${h_\mu(p)}={\bigl(p_\mu
%p^\nu-p^2\delta_{\mu\nu}\bigr)\,f_\nu(p)}$, ${A(h)}$ will correspond to
%${{F_{\mu\nu}}^{,\nu}(f)}$, and since the integral for $D$ 
%involves restriction to $C_+$,
%the set of test functions ${\set p_\mu p^\nu f_\nu(p)
%%\restriction C_+,{f\in\widehat\al S.}.}$
% will represent the field equations. 
%On smearing $F_{\mu\nu}(p)$ with an antisymmetric
%tensor function $f_{\mu\nu}$ to obtain ${F(f)}$, we note that the latter
%corresponds to the smearing of $A_\mu$ with ${2p_\nu f^{\mu\nu}}$.

\subsection{Gupta--Bleuler electromagnetism as a local constraint system.}
\label{GBi}

To model this in rigorous field theory, we start with
 the CCR algebra
$\al A.:=\CCR$
where the symplectic space $({\got X}, B)$ is
constructed as follows. Consider the real linear space
$\widehat\al S.$ from above, and equip it with
the presymplectic form $D$ obtained from the CCRs before.
Now define ${\got X}:=\widehat\al S.\big/{\rm Ker}(D)$ which is a
symplectic space with symplectic form $B$ defined as the factoring of 
$D$ to the factor space ${\got X}$.

Now since the constraints $\chi(x)$ are not (Krein) selfadjoint, there is
no space of test functions in ${\got X}$ which represent them, so
we want to
define them as outer constraints through the gauge transformations which they generate.
A heuristic calculation (cf. \cite{Grundling88b})
produces:
\[
  {\rm Ad}\Big( {\rm exp}(-it\,\chi(h)^\d\chi(h))\Big)
 \, {\rm exp}(i\,A(f))={\rm exp}(i\,A(T^t_hf))\,,
\]
where formally 
$\chi(h):=\int\chi(x)h(x)\,d^4x$, $h\in{\al S.(\R^4,\R)}$ and
\begin{equation}
\label{gauge}
  \big(T_h^tf)_\mu(p)=f_\mu(p)-it\pi\, p_\mu\widehat{h}(p)\,\int_{C_+}
f^\nu(p')\, p'_\nu\overline{\widehat{h}(p')}
\,{d^3p'\over p_0'}\;,
\end{equation}
and we used the smearing formula Eqn~(\ref{smearII}).
(Note that since the operators $A(f)$ are not selfadjoint, the
operators ${\rm exp}(i\,A(f))$ can be unbounded).
At this point a problem occurs (pointed out to us by Prof. D. 
Buchholz)~\footnote{This was also an error in  \cite{Grundling88b}.}.
Whilst the function $ip_\mu\widehat{h}(p)$ is the Fourier transformation
of the gradient of $h,$ hence in the allowed class of functions,
the coefficient $c(f,h):=\int_{C_+}
f^\nu(p')\, p'_\nu\overline{\widehat{h}(p')}
\,{d^3p'\over p_0'}$ need not be real, so $T_h^t$ will not preserve
${\got X}.$ The reason for this difficulty, is because 
$\chi(h)^\d\chi(h)$ is a product of noncausal operator--valued distributions, 
and so its commutator
with the causal $A(f)$ is unlikely to be causal.
So, since the gauge transformations can take an $f\in{\got X}$
out of ${\got X}$, we enlarge the space  ${\got X}$
by including complex valued Schwartz functions, i.e. we set 
\[
   {\got Y}:= \al S.(\R^4,\C^4)\widehat{\vphantom{)}}\big/{\rm Ker}(D)
   = \al S.(\R^4,\C^4)\big/{\rm Ker}(D)
\]
where $D$ is given by the same formula~(\ref{symp}) than before;-- it is still
real on ${\got Y}$ because it is  the imaginary part of 
$K.$
We will discuss in  Remark~\ref{comp.supp}(ii)
 below what this enlargement of symplectic space
corresponds to in terms of the
original heuristic smearing formulii.
(Note however, that the symplectic form $D$ given by Eqn~(\ref{symp})
for arbitrary complex Schwartz functions, does not satisfy
causality.)
 Since ${\rm Ker}(D\rest\widehat\al S.)=\widehat\al S.\,\cap\,
{\rm Ker}(D\rest
\al S.(\R^4,\C^4)),$ we have that
 ${\got X}\subset{\got Y}.$
Thus $\CCR\subset\CCRy,$ and moreover the gauge transformations
$T_h^t$ are well defined on ${\got Y}.$
The transformations $T_h^t$ are symplectic, in fact, if we define
$G_h(f):= T^1_h(f)-f,$ then 
\begin{itemize}
\item[(i)] $B(G_h(f),\,k)=-B(f,\,G_h(k)),$
\item[(ii)] $G_g\circ G_h=0,$
\item[(iii)] $T_h^t(T_k^s(f))=f+tG_h(f)+sG_k(f)$.
\end{itemize}
For each $h\in{\al S.(\R^4,\R)}$ we have a one--parameter group
of gauge transformations $T_h^t: \al S.(\R^4,\C^4)\to \al S.(\R^4,\C^4)$ 
(cf. \cite{Grundling88b})
and $\{T_h^t\,\mid\, t\in\R,\,h\in\al S.(\R^4,\R)\}$ is a commutative set of
symplectic transformations, hence preserve ${\rm Ker}(D)$ and so factor
to the space ${\got Y}$. Each $T^t_h$ is a one--parameter group in
$t$, but due to the nonlinearity in $h$, the map $h\to T^1_h=:T_h$ is
not a group homomorphism of $\al S.(\R^4,\R)$.

We let our group of gauge transformations 
$\al G.,$ be the discrete group generated in ${\rm Sp}({\got Y},B)$
by all $T_h^t,$
and define as usual the action ${\beta:\al G.\to{\rm Aut}\left(\CCRy\right)}$ by
$\beta_\gamma(\delta_f)=\delta_{\gamma(f)},$ $\gamma\in\al G.,$ $f\in\al Y..$
Our field algebra will be the discrete crossed product
$\al F._e:=\al G.\cross\beta.\CCRy.$
As a C*--algebra $\al F._e$
is generated by $\CCRy$ and a set of commuting unitaries $U_{\al G.}:=
\{U_\gamma\mid \gamma\in \al G.\}$ such that
$\gamma(F)=U_\gamma\,F\,U_\gamma^*$,
$F\in\al A.$, $U_{\gamma^{-1}}=U_\gamma^*$ and
$U_{\gamma\gamma'}=U_\gamma U_{\gamma'}$, $\gamma,\gamma'\in \al G.$.

\begin{rem}
\label{spaces}
\begin{itemize}
\item[(i)]
Sometimes we need a more concrete characterization of the
space $\ot X.$.
Now $\ot X.=\wh\al S./{\rm Ker}(D)$ and $\wh\al S.={\{f\in\al S.(\R^4,\,\C^4)
\mid\overline{f(p)}=f(-p)\}}=\al S._++i\al S._-$ where
$\al S._\pm:={\{u\in\al S.(\R^4,\,\R^4)\mid u(p)=\pm u(-p)\}}$.
{}From Eqn.~(\ref{symp}) we see that ${\rm Ker}(D)={\{f\in\wh\al S.
\mid f\restriction C_+=0\}}$, and hence factoring by
${\rm Ker}(D)$ is the same as restriction to $C_+$, i.e.
$\ot X.=\wh\al S.\restriction C_+$, and since
$f(p)\restriction C_+=f(\|\b p.\|,\,\b p.)$ we can identify these
functions with a subspace of $C(\R^3,\,\C^4)$.
Since we are restricting Schwartz functions, we note that
these functions on $\R^3$ are smooth except at the origin,
and Schwartz on the complement of any open neighbourhood
of the origin.
The conditions $u(p)=\pm u(-p)$ for $u\in\al S._\pm$
involve points outside $C_+$, so through smoothness they will 
influence the behaviour of $u\restriction C_+$ near the origin.
Specifically if $u\in\al S._+$ (resp. $u\in\al S._-$), then on each line
through the origin in $C_+$, $\{ta\mid
t\in\R\}$, $a\in C_+\backslash 0$, the function
$u_a(t):=u(ta)$ is smooth and even (resp. odd), hence
all its derivatives of odd degree must be odd (resp. even)
and its derivatives of even degree must be even (resp. odd).
Thus the derivatives of $u_a$ of odd (resp. even) degree
are zero at the origin. This is a property which does
restrict to $C_+$, and distinguishes between $\al S._+\restriction
C_+$ and $\al S._-\restriction C_+$.
Note from the above discussion, that
$\ot X.=\wh\al S.\restriction C_+$ contains all smooth functions
with compact support away from zero.
\item[(ii)]
The space to which we will next restrict our constructions,
  is the real span
of the orbit of ${\got X}$ under the gauge group $\al G.$, i.e.
$\ot Z.:={\rm Span}_{\R}(\al G.({\got X})).$
Denote the real space of gradients by 
\[
{\got G}:=\set{f\in\widehat{\al S.}},
{f_\mu(p)=ip_\mu \widehat{h}(p),\,\; h\in\al S.(\R^4,\R)}.
\]
(which is not in ${\rm Ker}(D)$).
Now we want to show that $\ot Z.={\got X}+\C\cdot{\got G}$
where we use the same symbol for $\ot G.$ and its image in $\ot Y.$
under factoring by ${\rm Ker}(D),$ and $\C\cdot{\got G}$
is a shorthand for the complex span ${\rm Span}_{\C}({\got G})$.
Note that a general element of $\ot Z.$ is of the form
\begin{eqnarray*}
&&\sum_{n=1}^N\lambda_n\left(f^{(n)}_\mu(p)-it_n\pi\, p_\mu\widehat{h}^{(n)}(p)
\cdot c(h^{(n)},f^{(n)}) \right)\\
&&\qquad= \sum_{n=1}^N\lambda_nf^{(n)}_\mu(p) - ip_\mu\sum_{n=1}^N\pi t_n\lambda_n
c(h^{(n)},f^{(n)})\cdot\widehat{h}^{(n)}(p)
\end{eqnarray*}
where $\lambda_n,\; t_n\in\R,$ $f^{(n)}\in\ot X.,$
${h}^{(n)}\in\al S.(\R^4,\R)$ and $c(h,f)\in\C$ as 
in Eqn.~(\ref{gauge}). Clearly this shows that $\ot Z.\subseteq
{\got X}+\C\cdot{\got G}.$
For the reverse inclusion, we have that ${\got X}$ is in $\ot Z.$
and to see that $\C\cdot{\got G}$ is in $\ot Z.$, note that it contains
$\pi^{-1}(T_h^tf-T_h^{t+1}f)(p)= ip_\mu\wh{h}(p)\cdot c(f,h)$
for all $f$ and $h.$ From the discussion in the previous  remark,
it is clear that we may choose the real and imaginary parts of 
$f$ and $\widehat{h}$ independently,
and so $ c(f,h)$ can be any complex number.
Thus  $\ot Z.={\got X}+\C\cdot{\got G}$.
From a physical point of view, one can justify the inclusion of
complex smearing functions in $\ot Z.$ by the fact that the constraints $\chi(f)$
are already noncausal, and that below for the final physical theory we will
eliminate these, retaining only the real valued smearing functions.
\end{itemize}
\end{rem}
   
To construct the net of local field algebras $\al F.:\Gamma\to
\widetilde\Gamma$ as in Definition~\ref{HK},
let $\B$ be any open set in $\R^4$
and define
\begin{eqnarray*}
\al S.(\B)&:=& \{f\in\al S.(\R^4,\C^4)\,\mid\,{\rm supp}(f)
\subset\B\} \\
\ot X.(\B)&:=&(\widehat{\al S.(\B)}\cap\wh{\al S.})\big/{\rm Ker}(D) \\
\ot Z.(\B)&:=&\big(\widehat{\al S.(\B)}\big/{\rm Ker}(D)\big)\cap\ot Z. \\
\al U.(\B)   &:=& \{U_{T_h}\mid h\in\al S.(\R^4,\R),\; {\rm supp}(h)\subset\B\} \\
 \al F.(\B)&:=&{\rm C}^*\Big( \delta_{{\got X}(\B)} \cup\al U.(\B) \Big)
  \subset\al F._e. 
\end{eqnarray*}
Note that if $\B$ is bounded, then $\al S.(\B)=C_c^\infty(\B,\,\C^4)$.
Moreover
$T_h\ot Z.(\B)\subset\ot Z.(\B)$ when ${\rm supp}(h)\subset\B$.
Thus if we let $\al G.(\B)$ be the discrete group generated in ${\rm Sp}(
\ot Y.,B)$ by
$\{T^t_h\,\mid\,{\rm supp}(h)\subset\B\,,\;t\in\R\}$, then it preserves
${\rm C}^*(\delta_{\ot Z.(\B)})$  %=\overline{\Delta(\ot X.(\B),\, B)}$, so
so that it makes sense to define 
$\al G.(\B)\cross\beta.{\rm C}^*(\delta_{\ot Z.(\B)})$.

\begin{lem}
\label{shorten}
We have: 
\[
\al F.(\B)=\al G.(\B)\cross\beta.{\rm C}^*(\delta_{\ot Z.(\B)})
              =[U_{\al G.(\B)}\,\delta_{{\got Z}(\B)}]
              =[\delta_{{\got Z}(\B)}\,U_{\al G. (\B)}]\,.
\]
\end{lem}
\begin{beweis}
We start with the proof of the first equality.
From $\delta_{{\got X}(\B)}$ and $\al U.(\B)$
we can produce $\delta_{\al G.(\B)(\ot X.(\B))}$ in $\al F.(\B)$.
Let $g={T_{h_1}^{t_1}\cdots T_{h_n}^{t_n}}\in\al G.(\B),$
then $g(f)=f+\sum_i^nt_iG_{h_i}(f)\in\ot Z.(\B)$
where $f\in\ot X.(\B),$ ${\rm supp}(h_i)\subset\B,$ 
and $G_h(f):=T_h(f)-f\in\C\cdot\ot G.\cap\widehat{\al S.(\B)}$.
By varying the $h_i$ we can get all possible complex multiples of
the gradients in $\ot G.\cap\widehat{\al S.(\B)}$, hence
$\al G.(\B)(\ot X.(\B))=\ot Z.(\B).$
Thus $\al F.(\B)={\rm C}^*\Big( \delta_{{\got Z}(\B)} \cup\al U.(\B) \Big).$
Now recall the fact that the crossed product
           ${\al G.(\B)\cross\beta.{\rm C}^*(\delta_{\ot Z.(\B)})}$
is constructed from the  twisted (by $\beta$) convolution algebra of functions
 $f:\al G.(\B)\to {\rm C}^*(\delta_{\ot Z.(\B)})$
of finite support, and these form a subalgebra of $\al F._e$.
The enveloping C*--norm on this convolution algebra coincides with
the C*--norm of $\al F._e$, and now the equality follows from
the fact that the *--algebra $\al A.(\B)$ generated by 
 $\{ \delta_{{\got Z}(\B)} \cup\al U.(\B) \}$
is dense in this convolution algebra.
For the next two equalities note that $\al A.(\B)$ consists of
linear combinations of products of unitaries in $\delta_{\ot Z.(\B)}$
and unitaries in $\al U.(\B)$. Each such a product of unitaries can be
written as a constant times a product of the form  $U_\gamma\cdot \delta_f$,
$\gamma\in\al G.$, $f\in\ot Z.(\B)$ as well as a product of the form
$\delta_{f'}\cdot U_{\gamma'}$, using the Weyl relation together with
the implementing relation $U_\gamma\delta_f=\delta_{\gamma(f)}U_{\gamma}$
to rearrange the order of the products.
Clearly now the last two relations follow from this.
\end{beweis}
By setting $\B=\R^4$, the global objects are included in this lemma.
Also observe that whilst $U_{\al G.(\B)}$ is clearly an equivalent
set of constraints to $\al U.(\B)$, in general it is strictly larger as a set.
Now to define a system of local quantum constraints (cf. Def. \ref{loc.cons})
let $\Gamma$ be any directed set of open bounded sets of $\R^4$
which covers $\R^4$, and such that orthochronous Poincar\'e transformations
map elements of $\Gamma$ to elements of $\Gamma$.
Then the map $\al F.$ from $\Gamma$ to subalgebras of $\al F._e$ by
$\B\to\al F.(\B)$ satisfies isotony.
The main result of this subsection is:

\begin{teo}
\label{GBcentral}
The map
$\Gamma\ni\B\to(\al F.(\B),\,\al U.(\B))$ defines a system of
local quantum constraints.
\end{teo}
\begin{beweis}
The net $\B\to \al F.(\B)$ is isotone and by construction
of the cross product we also have $\al U.(\B_1)=\al U.(\B_2)\cap
\al F.(\B_1)$ if $\B_1\subseteq\B_2$. It remains to show that $\al U.(\B)$
is first class in $\al F.(\B)$, $\B\in\Gamma$. 
Consider the central state $\omega_0$ on ${\rm C}^*(\delta_{\ot Z.})$ (cf.
Eqn.~(\ref{central})). 
This is
$\al G.$--invariant, and its restriction to ${\rm C}^*(\delta_{\ot Z.(\B)})$
is clearly $\al G.(\B)$--invariant.
By Theorem~\ref{Outer} 
it extends to a nontrivial Dirac state on $\al F.(\B)$, hence 
$\al U.(\B)$ is first--class.
\end{beweis}

\begin{rem}
\label{comp.supp}
\begin{itemize}
\item[(i)]
Observe that as $\Gamma$ is preserved by translations 
(cf. Definition~\ref{HK}), we can cover each compact set
in $\R^4$ by a finite number of elements in $\Gamma$.
Hence, since $\Gamma$ is a directed set, each
compact set in $\R^4$ is contained in an element of $\Gamma$.
Thus ${\cup\{\al S.(\B)\mid\B\in\Gamma\}}
={C_c^\infty(\R^4,\,\C^4)}$ and so
\[
  \al F._0=\lim_{\longrightarrow}\al F.(\B)
={\rm C}^*\left(\delta_{\ot Z._{(0)}}\cup\al U._{(0)}\right)
\subset\al F._e
\]
where $\ot Z._{(0)}:=\ot Z.\cap
C_c^\infty(\R^4,\,\C^4)\wh{\phantom{I}}\big/{\rm Ker}\, D$ and
$\al U._{(0)}:={\left\{U_{T_h}\mid h\in
C_c^\infty(\R^4,\,\R)\right\}}$.
\item[(ii)]
Having now constructed the proposed algebraic framework for Gupta--Bleuler
electromagnetism, we still need to justify the extension of our
symplectic space by complex test  functions.
From the heuristic smearing formulii, it seems that there are two inequivalent
ways of extending the smearing to complex functions, depending on whether
one generalises Eqn~(\ref{smearI}) or Eqn~(\ref{smearII}).
Specifically, for a complex--valued test function $f,$ one has the choice of
\begin{eqnarray*}
A_1(\hat f)&:=&
\int d^4x\,A_\mu(x)\,f^\mu(x)
=\big(a(f)+a(\overline{f})^\d\big)\Big/\sqrt{2}   \\[2mm]
\hbox{or:}\qquad A_2(\hat f)
&:=&\sqrt\pi\int_{C_+}\Bigl(a_\mu(
\b p.)\,\hat f^\mu(p)+a_\mu^\d(\b p.)\,\overline{\hat f^\mu(p)}\Bigr){d^3p\over p_0}
 \\[2mm]
&=&\big(a(f)+a(f)^\d\big)\Big/\sqrt{2}\\[2mm]
\hbox{with}\qquad
a(f)&:=&\sqrt{2\pi}\int_{C_+}a_\mu(
\b p.)\,\hat f^\mu(p){d^3p\over p_0}
\end{eqnarray*}
Now $A_1(f)$ is complex linear in $f,$ hence is not Krein--symmetric
if $f$ is not real, and it produces a complex--valued symplectic form:
\[
\big[A_1(\hat f),\,A_1(\hat h)\big]=i\int\int f_\mu(x)\, h^\mu(y)
\,D(x-y)\,d^4x\,d^4y
\]
hence it is not possible to define a CCR--algebra with this form.
It is causal though.

The choice which we use in this paper, is $A_2(f),$ and the reason for this
is that it is the smearing which was necessary to define our gauge transformations
Eqn~(\ref{gauge}). Furthermore, $A_2(f)$ is always Krein symmetric (and real linear), and it
defines a real valued symplectic form
\[
\big[A_2(\hat f),\,A_2(\hat h)\big]=iD(\hat f,\,\hat h)=i\,{\rm Im}\, K(\hat f,\,\hat h)
\]
which we can (and did) use to define a CCR--algebra.
$D$ is not causal for complex--valued functions,
but we have compensated for this by only extending the real space
$\got X$ by complex multiples of gradients $\got G.$ These gradients will
be eliminated by the subsequent  constrainings below. Their purpose is to
select the physical subalgebras.
\end{itemize}
\end{rem}
  
%%%%%%%%%%%%%%%%%%%%%%%%%%%%%%%%%%%%%%%%%%%%%%%%%%%%%%%%
\subsection{Reduction isotony and weak causality.}
\label{GBii}

In this subsection we establish reduction isotony and weak causality
for our example in Theorems~\ref{GBred.iso} and \ref{wcaus}.
We first enforce the
T--procedure locally as in Section~3, to obtain the objects:
\begin{eqnarray*}
{\got S}_D^\B 
 \!\!\!&:=&\!\!\!\{\omega\in\wp(\al F.(\B))\mid\omega(U_{T_h})=1\quad
 \forall\,h\in\al S.(\R^4,\R),\quad{\rm supp}(h)\subset\B \} \\[2mm]
\al D.(\B) 
 \!\!\!&:=&\!\!\!
   [\al F.(\B)(\al U.(\B)-\EINS)]\cap[(\al U.(\B)-\EINS)\al F.(\B)]\,,\\[2mm]
\al O.(\B) 
  \!\!\!&:=&\!\!\! \{ F\in \al F.(\B)\mid  FD-DF \in \al D.(\B) \; 
               {\forall}\, D\in\al D.(\B) \} = 
               M_{\al F.(\B)}(\al D.(\B))\,, \\[2mm]
\al R.(\B) 
  \!\!\!&:=&\!\!\! \al O.(\B) / \al D.(\B)\,\quad
\hbox{where $\B$ is any open set in $\R^4$.}
\end{eqnarray*}
For reduction isotony we need to prove that if $\B_1\subseteq\B_2$ then
$\al D.(\B_1)=\al D.(\B_2)\cap\al O.(\B_1)$ and $\al O.(\B_1)\subseteq
\al O.(\B_2)$, and this requires more explicit characterization
of the local algebras involved.

\begin{teo}
\label{Observ}
We have:
\begin{eqnarray*}
\al O.(\B)&=&
 {\rm C}^*(\delta_{\ot p.(\B)}\cup \al D.(\B))=[
          \delta_{\ot p.(\B)} \cup \al D.(\B) ]
     ={\rm C}^*(\delta_{\ot p.(\B)})+\al D.(\B)\,
      \;\hbox{where}\\[2mm]
  \ot p.(\B)&:=&\big\{f\in\ot Z.(\B)\,\mid\, T_h(f)=f\;\forall\,
   h\in\al S.(\R^4,\R),\;\;{\rm supp}(h)\subset\B\big\} \\[2mm]
  &=&\big\{f\in\ot Z.(\B)\,\mid\, B(f,\, G_h(f))=0\;\forall\,
   h\in\al S.(\R^4,\R),\;\;{\rm supp}(h)\subset\B\big\} \\[2mm]
  &=&\big\{f\in\ot Z.(\B)\,\mid\, p_\mu f^\mu\restriction C_+=0\big\}
\end{eqnarray*}
with $G_h(f):=T_h(f)-f$. Moreover $\al R.(\B)\cong{\rm C}^*(\delta_{\ot p.(\B)})$.
\end{teo}
\begin{beweis} 
For any $f\in\ot p.(\B)$ one has $\delta_{f}=\beta_{T_h}(\delta_{f})=
U_{T_h}\,\delta_{f}\,U_{T_h}^*$, ${\rm supp}(h)\subset\B$, so that
$\delta_{\ot p.(\B)}\subset \al U.(\B)'$ and Theorem~\ref{Teo.2.2}~(v) 
implies $\delta_{\ot p.(\B)}\subset\al O.(\B)$. Further $\al D.(\B)\subset\al O.(\B)$
proves the inclusion ${\rm C}^*(\delta_{\ot p.(\B)}\cup\al D.(\B))\subseteq\al O.(\B)$. 
To show the reverse inclusion take $A\in\al O.(\B)\subset \al F.(\B)$
and from Lemma~\ref{shorten} there is a sequence $\{A_n\}_{n\in\N}
\subset{\rm span}\big\{\delta_{\ot Z.(\B)}U_{\al G.(\B)}\big\}$ 
converging in the C*--norm to $A$. 
Put $A_n:=\sum_{i=1}^{k_n}\lambda_i^{n}\,\delta_{f_i^{n}}    
\,U_{\gamma_i^{n}}$, $\;\lambda_i^{n}\in\C$, $f_i^{n}\in\ot Z.(\B)$, 
$\gamma_i^{n}\in\al G.(\B)$, and since $\{f_i^{n}\mid i=1,\ldots,k_n\,,
\;n\in\N\}$ is a denumerable set we can rearrange it into a single
sequence $\{f_i\}_{i\in\N}$, where $f_i\neq f_j$ if
$i\neq j$. Thus we can rewrite
\begin{equation}
\label{A_n}
A_n=\sum_{i=1}^{N_n}\delta_{f_i}\sum_{j=1}^{L_n}\lambda_{ij}^{(n)}\, U_{\gamma_{ij}^{(n)}}.
\end{equation}
Observe that for $\omega\in\ot S._D^\B$ we have 
$\pi_\omega(A_n)\Omega_\omega=\sum_i\zeta_i^{(n)}\pi_\omega(
\delta_{f_i})\Omega_\omega$ where $\zeta_i^{(n)}:=\sum_j\lambda_{ij}^{(n)}$.
 Now $A\in\al O.(\B)$ and therefore we have for any 
$\omega\in\ot S._{D}^\B$, 
\begin{eqnarray*}
0 &=&
   \pi_\omega((U_{T_h}-\EINS)\cdot A)\Omega_\omega 
  =\lim_{n\to\infty}
   \pi_\omega((U_{T_h}-\EINS)\cdot A_n)\Omega_\omega 
          \\[2mm]
  &=&\lim_{n\to\infty}
   \pi_\omega(\beta_{T_h}
(A_n)- A_n)\Omega_\omega 
 =\lim_{n\to\infty}\sum_{i=1}^{N_n}\zeta_i^{(n)}
\pi_\omega(\delta_{T_h(f_i)}-\delta_{f_i})\Omega_\omega
\end{eqnarray*}
for $ h\in\al S.(\R^4,\R),\;\;{\rm supp}(h)\subset\B$,
where we made use of $(\al U.(\B)-\EINS)\al O.(\B)\subset
\al D.(\B)\subset N_\omega$ at the start.
In particular let $\omega$ be an extension
of the central state $\omega_0$ defined in 
Eq.~(\ref{central})  (which has Dirac state 
extensions by Theorem~\ref{Outer}).
Then for all $h$:
\begin{eqnarray}
0 &=&
   \omega(A^*(U_{T_h}-\EINS)^*(U_{T_h}-\EINS)
    A)\nonumber\\[2mm]
  &=&
 \lim_{n\to\infty}\Big\{2\omega(A_n^*A_n)  \nonumber \\[-1mm]
  & &\kern1.1cm -2\,{\rm Re}\Big(\sum_{i,j}^{N_n}\overline\zeta_i^{(n)}
\zeta_j^{(n)}\exp[i\,B(f_j,\, T_h(f_i))/2]\;
\omega(\delta_{f_j-T_h(f_i)})\Big)\Big\} \nonumber \\[2mm]
  &=&
 2\lim_{n\to\infty}\Big\{\sum_i^{N_n}\big|\zeta_i^{(n)}\big|^2
-{\rm Re}\Big(\sum_{(i,j)\in P_h(n)}\overline\zeta_i^{(n)}
\zeta_j^{(n)}
\Big)\Big\}
\label{zetas}
\end{eqnarray}
where $P_h(n):=\{(i,\,j)\in\{1,\ldots,\,N_n\}^2\,\mid\, f_j=T_h(f_i)\}
\subset P_h(n+m)$.
Observe that if $f_i\in\ot p.(\B)$, then $(i,\,i)\in P_h(n)$ for all
$h$, and that these terms cancel in Eqn.~(\ref{zetas}), i.e.
we may assume that $f_i\not\in\ot p.(\B)$ in (\ref{zetas}).
Furthermore by the Cauchy--Schwarz inequality
\begin{equation}
\Big|\sum_{(i,j)\in P_h(n)}\overline\zeta_i^{(n)}\zeta_j^{(n)}\Big|
\leq\Big(\sum_{i\in D_h(n)}\big|\zeta_i^{(n)}\big|^2\Big)^{\frac12}
\Big(\sum_{j\in R_h(n)}\big|\zeta_j^{(n)}\big|^2\Big)^{\frac12}
\leq\sum_{i=1}^{N_n}\big|\zeta_i^{(n)}\big|^2
\end{equation}
where $D_h(n):=\{i\mid(i,j)\in P_h(n)\}$ and
 $R_h(n):=\{j\mid(i,j)\in P_h(n)\}$ i.e. the domain and range of the relation
defined by $P_h(n)$.
If $D_h(n)$ or $R_h(n)$ is not $\{1,\, 2,\ldots,\, N_n\}$, then the last inequality
is strict and Eqn.~(\ref{zetas}) cannot hold unless $\lim\limits_{n\to\infty}\zeta_i^{(n)}=0$
for all $i\leq N_n$ not in $D_h(n)$ or $R_h(n)$.
Given that $f_i\not\in\ot p.(\B)$ in the surviving terms of the sum,
for each $i$, choose an $h$ such that $f_i\not=T_h(f_i)$, then by 
Eqn.~(\ref{gauge}) $T_{th}(f_i)=f_i+t^2(T_h(f_i)-f_i)$ for $t\in\R_+$,
and so $\{T_{th}(f_i)\mid t\in\R_+\}$ is a continuous family of distinct
elements of $\ot Z.(\B)$. Since $\{f_j\mid j\in \N\}$ is denumerable, there 
exists a $t_0$ such that $T_{t_0h}(f_i)\not= f_j$ for all $j\in\N$,
i.e. $i\not\in D_{t_0h}(n)$ for all $n$, and so Eqn.~(\ref{zetas})
can only hold if $\lim\limits_{n\to\infty}\zeta_i^{(n)}=0$.
We conclude that in the original expression for $A_n$, if
$f_i\not\in\ot p.(\B)$, then 
$\lim\limits_{n\to\infty}\zeta_i^{(n)}=0$.

Recall by Remark~\ref{Rem.2.12} in Appendix~1 that the factorization map
$\al O.(\B)\to\al R.(\B)$ is precisely the restriction of
$\al O.(\B)$ in the universal representation $\pi_u$ to the 
subspace $\al H.^{(p)}_u:=\{\psi\in\al H._u\,\mid\,\pi_u(\al U.(\B))\psi
=\psi\}$, so if we can show that $\pi_u(\al O.(\B))\restriction\al H._u^{(p)}
=\pi_u({\rm C}^*(\delta_{\ot p.(\B)}))\restriction\al H._u^{(p)}$, that
suffices to prove that $\al O.(\B)={\rm C}^*(\delta_{\ot p.(\B)})+\al D.(\B)$.
Let $\psi\in\al H._u^{(p)}$, and $f_j\not\in\ot p.(\B)$:
\begin{eqnarray}
\pi_u(A)\psi &=&
  \lim_{n\to\infty}
   \pi_u(A_n)\psi=\lim_{n\to\infty}\sum_{i=1}^{N_n}\zeta_i^{(n)}\pi_u(\delta_{f_i})\psi
  \nonumber          \\[2mm]
  &=&\lim_{n\to\infty}\Big(
    \sum_{i\not=j}^{N_n} \zeta_i^{(n)}\pi_u(\delta_{f_i})
    + \zeta_j^{(n)}\pi_u(\delta_{f_j})\Big)\psi
            \nonumber\\[2mm]
 &=&\lim_{n\to\infty}\sum_{i\not=j}^{N_n}\zeta_i^{(n)}
\pi_u(\delta_{f_i})\psi
\label{later}
\end{eqnarray}
so we can omit all contributions where $f_j\not\in\ot p.(\B)$ from the sum:
\[
  \pi_u(A)\psi=\lim_{n\to\infty}\sum_{f_i\in\ot p.(\B)}^{N_n}\zeta_i^{(n)}\pi_u
   (\delta_{f_i})\psi  %\in\pi_u\left({\rm C}^*(\delta_{\ot p.(\B)})\right)\psi
   \qquad\quad\forall\;\psi\in\al H._u^{(p)}\,.
\]
The latter will be in $\pi_u\left({\rm C}^*(\delta_{\ot p.(\B)})\right)\psi,$
providing we can show that $\lim_{n\to\infty}\sum_{f_i\in\ot p.(\B)}^{N_n}\zeta_i^{(n)}
   \delta_{f_i}$ converges in the C*--norm. This is easy to see, because from the
convergence of $A_n$ in Eqn.~(\ref{A_n}), we get the convergence of the subseries
\[
\sum_{f_i\in\ot p.(\B)}^{N_n}\delta_{f_i}
\sum_{j=1}^{L_n}\lambda_{ij}^{(n)}\, U_{\gamma_{ij}^{(n)}}
\in {\rm C}^*(\delta_{\ot p.(\B)}\cup\al U.(\B))
\]
and since $\delta_{\ot p.(\B)}$ commutes with $\al U.(\B)$
there is a *--homomorphism
$\varphi: {\rm C}^*(\delta_{\ot p.(\B)}\cup\al U.(\B))\to
{\rm C}^*(\delta_{\ot p.(\B)})$ by $\varphi(\al U.(\B))=\EINS$
(just apply the  T--procedure), hence the image of the preceding sequence
converges, i.e. $\lim_{n\to\infty}\sum_{f_i\in\ot p.(\B)}^{N_n}\zeta_i^{(n)}
   \delta_{f_i}$ converges in the C*--norm.
Thus $\pi_u(\al O.(\B))\psi\subseteq\pi_u\left({\rm C}^*(\delta_{\ot p.(\B)})\right)\psi$
for all $\psi\in\al H._u^{(p)}.$
Hence 
 $\al O.(\B)={\rm C}^*(\delta_{\ot p.(\B)})+\al D.(\B)$, using 
$\delta_{\ot p.(\B)}\subset\al O.(\B)$, (cf.~\cite{Grundling88b}).

For the last two equivalent characterisations of $\ot p.(\B)$,
observe first that if $T_h(f)=f$, then $B(f,\,G_h(f))=B(f,\,T_h(f)-f)=0$
and conversely
\[
0=B(f,\,G_h(f))=-2\pi\,\Big|\int_{C_+}{d^3p\over p_0}p_\mu\,f^\mu(p)\,
\overline{\widehat h(p)}\Big|^2
\]
which by Eqn.~(\ref{gauge}) implies that $G_h(f)=0$.
Choose $\widehat h=ip_\mu f^\mu$ (which is in the allowed class of
functions) to see the equivalence  
 with $p_\mu f^\mu\restriction C_+=0$.

Finally, to prove that $\al R.(\B)=\big({\rm C}^*(\delta_{\ot p.(\B)})
+\al D.(\B)\big)\big/\al D.(\B)\cong
{\rm C}^*(\delta_{\ot p.(\B)})$, it suffices to show that
the ideal
${\rm C}^*(\delta_{\ot p.(\B)})\cap\al D.(\B)=\{0\}$.
Consider a sequence
\[
  A_n=\sum_{i=1}^{N_n}\lambda_i^{(n)}\delta_{f_i}\in
\Delta(\ot p.(\B),\, B),\;\;\lambda_i^{(n)}\in\C,\qquad
  \hbox{converging to}\quad A\in\al D.(\B)
\]
then we show that it converges to zero.
Now $\delta_{f_j}\in\al O.(\B)$ for $f_j\in\ot p.(\B)$,
so for $N_n>j$ we have
\[
  \delta_{-f_j}\cdot A_n=\sum_{i\not=j}^{N_n}\lambda^{(n)}_i
\delta_{f_i-f_j}\exp\big(iB(f_i,\, f_j)/2\big)+
\lambda_j^{(n)}\EINS\longrightarrow\delta_{-f_j}\cdot A\in
\al D.(\B)\;.
\]
Since the central state $\omega_0$ (cf.~Eqn.~(\ref{central}))
extends to a Dirac state
we have
\[
  0=\omega_0(\delta_{-f_j}\cdot A)=\lim_{n\to\infty}
\omega_0(\delta_{-f_j}A_n)=\lim_{n\to\infty}
\lambda_j^{(n)}\quad\forall\, j\,.
\]
This implies that $A=0$, because
we can realize $A$ as an $\ell^2\hbox{--sequence}$ over
$\ot p.(\B)$ (recall discussion in Section~4.1), 
and the evaluation map at a point in $\ot p.(\B)$
is $\ell^2\hbox{--continuous}$, hence C*--continuous
so by the previous equation  evaluation of
$A$ at each point is zero.
\end{beweis}

One can now set $\B=\R^4$ to get the global version of this theorem.
An important physical observation, is that $\ot p.=\cup\ot p.(\B)$
contains the functions corresponding to the field operators
$F_{\mu\nu}$. To see this, smear
 $F_{\mu\nu}(p)$ with an antisymmetric
tensor function $f_{\mu\nu}$ to obtain ${F(f)}$, and note that the latter
corresponds to the smearing of $A_\mu$ with ${2p_\nu f^{\mu\nu}}\in\ot p.$.
\begin{teo}
\label{GBred.iso}
The system of local constraints defined here satisfies reduction isotony.
\end{teo}
\begin{beweis}
Let $\B_1\subseteq\B_2$, then we start by showing
that $\al O.(\B_1)\subseteq\al O.(\B_2)$, i.e. by Theorem
\ref{Observ} we show that ${\rm C}^*(\delta_{\ot p.(\B_1)})+\al D.(\B_1)
\subseteq 
{\rm C}^*(\delta_{\ot p.(\B_2)})+\al D.(\B_2)$. This follows directly
from $\al D.(\B_1)\subseteq\al D.(\B_2)$ and $\ot p.(\B_1)\subseteq
\ot p.(\B_2)$ where the last inclusion comes from the last
characterisation of $\ot p.(\B)$ in Theorem~\ref{Observ}.

It only remains to show that
 $\al D.(\B_1)
=\al D.(\B_2)\cap\al O.(\B_1)$. Recall that $\al D.(\B)\subset\al O.(\B)$,
and that $\al D.(\B)$ is the largest C*--algebra in $\al F.(\B)$
(hence in $\al O.(\B)$) which is annihilated by all $\omega\in\wp_D^\B$.
Since $\al O.(\B_1)\subseteq\al O.(\B_2)$, it suffices to show 
by Lemma~\ref{Lem.3.3} that
every Dirac state on $\al O.(\B_1)$ extends to a Dirac state on
$\al O.(\B_2)$. 

Recall that $\al O.(\B_1)={\rm C}^*(\delta_{\ot p.(\B_1)})
+\al D.(\B_1)$, so a Dirac state on $\al O.(\B_1)$ is uniquely
determined by its values on $\delta_{\ot p.(\B_1)}$.
 Moreover, from the fact that $f\in\ot p.(\B_1)$
implies $p_\mu f^\mu\restriction C_+=0$ and Eqn.~(\ref{gauge}),
we see that $\al U.(\B_2)$ commutes with ${\rm C}^*(\delta_{\ot p.(\B_1)})$:
\begin{eqnarray*}
U_{T_h}\delta_f U^{-1}_{T_h}&=&\delta_{T_hf}=\delta_f
  \quad\hbox{since}\\
  \big(T_hf)_\mu(p)&=&f_\mu(p)-i\pi\, p_\mu\widehat{h}(p)\,\int_{C_+}
f^\nu(p')\, p'_\nu\overline{\widehat{h}(p')}
\,{d^3p'\over p_0'}=f_\mu(p)
\end{eqnarray*}
for $f\in\ot p.(\B_1)$.
Next define $\wt\al O.:={\rm C}^*\big(\delta_{\ot p.(\B_1)}\cup\al U.(\B_2)
\big)\subset\al O.(\B_2)$.
Now $\wt\al O.$ is generated by the two mutually commuting C*--algebras
${\rm C}^*\big(\delta_{\ot p.(\B_1)}\big)$ and
${\rm C}^*\big(\al U.(\B_2)\big)\cong{\rm C}^*(\al G.(\B_2))$
where the latter is Abelian. If $AB=0$ for $A\in
{\rm C}^*\big(\delta_{\ot p.(\B_1)}\big)$ and $B\in
{\rm C}^*\big(\al U.(\B_2)\big)$, then either $A=0$
or $B=0$. This we can see from the realisation of
${\rm C}^*\big(\al U.(\B_2)\big)$ as scalar valued functions of denumerable
support in $\al G.(\B_2)$, so (pointwise) multiplication
by a nonzero $A\in
{\rm C}^*\big(\delta_{\ot p.(\B_1)}\big)$
cannot change support. Then by an application of the result in
\cite[Exercise~2, p.~220]{Takesaki}, we conclude that the map
$\varphi(A\otimes B):=AB$, 
 $A\in
{\rm C}^*\big(\delta_{\ot p.(\B_1)}\big)$, $B\in
{\rm C}^*\big(\al U.(\B_2)\big)$
extends to an isomorphism
$\varphi:
{\rm C}^*\big(\delta_{\ot p.(\B_1)}\big)\otimes
{\rm C}^*\big(\al U.(\B_2)\big)\to\wt\al O.$.
(Note that since 
${\rm C}^*\big(\al U.(\B_2)\big)$
is commutative it is nuclear, hence the
tensor norm is unique).

 Let $\omega\in\wp_D^{\B_1}\restriction\al O.(\B_1)$
and define a product state $\wt\omega$ on $\wt\al O.$ by
$\wt\omega:=\omega\otimes\hat\omega$, 
where $\hat\omega$ is  the state
$\hat\omega(U_\theta)=1$ for all $\theta\in\al G.(\B_2)$.
 Now extend $\wt\omega$ arbitrarily
to $\al O.(\B_2)$, then since it
coincides with $\omega$ on $\delta_{\ot p.(\B_1)}$ and
$\wt\omega(\al U.(\B_2))=1$, it is a Dirac state on 
$\al O.(\B_2)$ which extends $\omega\restriction
\al O.(\B_1)$. 
\end{beweis}

\begin{teo}\label{wcaus}
The system of local quantum constraints 
$(\al F.(\B),\,\al U.(\B))$ satisfies weak causality,
i.e.~if $\B_1\perp\B_2$ then $[\,\al O.(\B_1)\,,\,\al O.(\B_2)\,] 
\subset \al D.(\B_0)$ for some $\B_0\supset\B_1\cup\B_2$, 
 $\B_i\in\Gamma$. 
\end{teo}
\begin{beweis}
Since $\al O.(\B)={\rm C}^*(\delta_{\ot p.(\B)})+\al D.(\B)$
it is sufficient to consider commutants of $A_i\in\al O.(\B_i)$
being generating elements:
$A_i=\delta_{f_i}+D_i$ for  $f_i\in\ot p.(\B_i)$, 
$D_i\in\al D.(\B_i)$, $i=1,2$. Now 
\begin{eqnarray*}
[A_1,A_2] &=& [\delta_{f_1},\delta_{f_2}] +[\delta_{f_1},D_2]+
              [D_1,\delta_{f_2}]+[D_1,D_2] \\
          &=& \Big( e^{\frac{i}{2}B(f_1,\,f_2 )}
             -e^{-\frac{i}{2}B(f_1,\,f_2 )} \Big)\delta_{f_1+f_2} \\
          & &+[\delta_{f_1},\,D_2]+[D_1,\,\delta_{f_2}]+[D_1,\,D_2]\,.
\end{eqnarray*}
The first term vanishes because  $\B_1\perp\B_2$ implies the supports of
$f_1$ and $f_2$ are spacelike separated, so
\[
  B(\widehat{f}_1,\,\widehat{f}_2)=\int dx\, dx'\, D(x-x')\,
f_1^\mu(x)\,f_{2\mu}(x')=0
\]
because the Pauli--Jordan distribution $D$ has support  inside the
closed forward and backward light cones
\cite[p.~214]{bReedII}. Further for any $\B_0\supset\B_1\cup\B_2$
reduction isotony implies $\al D.(\B_1)\subset\al D.(\B_0)\supset\al D.(\B_2)$
and $\al O.(\B_1)\subset\al O.(\B_0)\supset\al O.(\B_2)$. But 
$\al D.(\B_0)$ is a closed 2--sided ideal in $\al O.(\B_0)$
and therefore the last 3 terms of the sum above are contained in 
$\al D.(\B_0)$ and the proof is concluded.
\end{beweis}

\begin{rem}
Note that the net $\B\to\al F.(\B)$ does not satisfy the causality
property, as we expect from the choice of 
noncausal constraints 
$(\partial^\mu A_\mu)^{\mbox{\tiny{(+)}}}(x)$. 
To see this, let $\B_1\perp\B_2$, and let
$\delta_{f}\in\al F.(\B_1)$ and $U_{T_h}\in\al F.(\B_2)$,
then the commutator
$[\delta_{f},\, U_{T_h}]$ need not vanish because
in Eqn.~(\ref{gauge}) we can have that
\[
  c(f,h)=\int_{C_+}\ff^\nu(p)\,p_\nu\,\overline{\widehat{h}(p)}{d^3p\over p_0}
  \not=0
\]
for ${\rm supp}(f)\subset\B_1$ and ${\rm supp}(h)\subset\B_2$.
\end{rem}

\subsection{Covariance.}
\label{covariance}

In order to examine weak covariance for this system of local constraints,
we first need to define the action of $\al P._+^\uparrow$ on $\al F._e$.
We start with the usual action of
 $\al P.^\uparrow_+$ on $\al S.(\R^4,\C^4)$. Define
\begin{equation}
\label{Lorenz}
(V_gf)(p):=e^{-ipa}\,\Lambda\,f(\Lambda^{-1}p)
\qquad\forall\; f\in\al S.(\R^4,\C^4),\;
g=(\Lambda,a)\in \al P.^\uparrow_+.
\end{equation}
Then $V_g$ is symplectic, hence factors through to a symplectic transformation on
$\ot Y.$, and this defines an action $\alpha:\al P._+^\uparrow\to
{\rm Aut}(\CCRy)$ by $\alpha_g(\delta_f):=\delta_{V_gf}$, $f\in\ot Y.$.

\begin{lem} 
\label{action}
Define $\alpha_g(U_{T_h}):=U_{T_{W_gh}}$,
where $(\wh{W_gh})(p):=e^{-ia\cdot p}\wh{h}(\Lambda^{-1}p)$, 
and we chose $g=(\Lambda,a)\in\al P._+^\uparrow$.
Then this extends
$\alpha_g$ from $\CCRy$ to $\al F._e$, producing a
consistent action $\alpha:\al P._+^\uparrow\to{\rm Aut}(\al F._e)$.
\end{lem}
\begin{beweis}
We need to show that if we extend $\alpha_g$ from the set
$\CCRy\cup\{U_{T_h}\mid h\}$ to the *-algebra generated
by it using the homomorphism property of $\alpha_g$, then
this is consistent with all relations of the $U_{T_h}$
amongst themselves, and between them and $\CCRy$.
First we need to establish how $V_g$ and $T_h$ intertwines.
\begin{eqnarray*}
\Big(T_{W_gh}V_gf\Big)_\mu(p)&=& (V_gf)_\mu(p)-i\pi\,p_\mu\widehat{W_gh}(p)
\int_{C_+}{d^3p'\over p_0'}\big(V_gf\big)^\nu(p')p'_\nu
\overline{\widehat{W_gh}(p')}   \\
&=& (V_gf)_\mu(p)-i\pi\,p_\mu e^{-ip\cdot a}\widehat{h}(\Lambda^{-1}p)
\int_{C_+}{d^3p'\over p_0'}\big(\Lambda f)^\nu(\Lambda^{-1}p')
p'_\nu\overline{\widehat{h}(p')}\\
&=& (V_gT_hf)_\mu(p)\;.
\end{eqnarray*}
Thus $V_gT_h=T_{W_gh}V_g$. Now the basic relation between $\CCRy$
and ${\{U_{T_h}\mid h\}}$ is the implementing relation, so 
\begin{eqnarray*}
  \alpha_g(U_{T_h}\delta_f U_{T_h}^*)
 &=& \alpha_g(\delta_{T_hf})=\delta_{V_gT_hf}
     =\delta_{T_{W_gh}V_gf}   \\
 &=& U_{T_{W_gh}}\delta_{V_gf}U_{T_{W_gh}}^*
     =\alpha_g(U_{T_h})\alpha_g(\delta_f)\alpha_g(U_{T_h})^*
\end{eqnarray*}
thus $\alpha_g$ is consistent with this. Finally we need to show that
$\alpha_g$ respects any group identities in $\beta(\al G.)
\subset{\rm Aut}\CCRy$. 
Recalling that $\al G.$ consists of finite products of $T_h$,
let $\gamma=T_{h_1}\ldots T_{h_n}\in\al G.$, then
$\gamma\to
T_{W_gh_1}\ldots T_{W_gh_n}$ defines a consistent
group homomorphism because
\begin{eqnarray*}
\alpha_g\big(\beta_\gamma(\delta_f)\big) &=& 
             \alpha_g(\delta_{T_{h_1}\cdots T_{h_n}f})
             =\delta_{V_gT_{h_1}\ldots T_{h_n}f}\\
             &=&\delta_{T_{W_gh_1}\ldots T_{W_gh_n}V_gf}
             =\beta\big( T_{W_gh_1}\cdots T_{W_gh_n}\big)
             \alpha_g( \delta_f)
\end{eqnarray*}
i.e. $ \beta(T_{W_gh_1}\cdots T_{W_gh_n})=\alpha_g\circ\beta_\gamma
\circ\alpha_g^{-1}$.
Thus $\alpha_g(U_{T_h})=U_{T_{W_gh}}$ extends consistently
to $U_{\al G.}$.
\end{beweis}
Observe that the action $V_g$ preserves the reality condition 
$\overline{f(p)}=f(-p)$ which defines $\ot X.$, hence it
preserves $\ot X.$ and in fact $V_g\ot X.(\B)=\ot X.(g\B).$ 

\begin{teo}\label{GBcov}
Consider the action $\alpha\colon\ \al P._+^\uparrow\to{\rm Aut}\,
\al F._e$ defined above. Then the system of local
quantum constraints $\Gamma\ni\B\to(\al F.(\B),\,\al U.(\B))$
satisfies $\alpha_g(\al U.(\B))=\al U.(g\B)$ and the
net $\Gamma\ni\B\to\al F.(\B)$ transforms covariantly, 
i.e.~$\alpha_g(\al F.(\B))=\al F.(g\B)$, $\B\in\Gamma$. Therefore
the local observables define a covariant net, 
i.e.~$\alpha_g(\al O.(\B))=\al O.(g\B)$.
\end{teo}
\begin{beweis} We have:
\begin{eqnarray*}
  \alpha_g(\al U.(\B))
  &=& \alpha_g(\{U_{T_h}\mid{\rm supp}(h)\subset\B\})
      =\{U_{T_{V_gh}}\mid{\rm supp}(h)\subset\B\} \\
  &\subseteq& \{U_{T_h}\mid{\rm supp}(h)\subset g\B\}=\al U.(g\B)
\end{eqnarray*}
and replacing $g$ by $g^{-1}$ gives the reverse inclusion.

For covariance of the net $\al F.(\B)$, recall that each $\al F.(\B)$
is generated by $\al U.(\B)$ and $\delta_{\ot X.(\B)}$, so since
\[
  \alpha_g(\delta_{\ot X.(\B)})=\delta_{V_g\ot X.(\B)}=\delta_{\ot X.(g\B)}\,,
\]
it follows that $\alpha_g(\al F.(\B))=\al F.(g\B)$.
 The covariance property for the net
of local observables follows from Remark~\ref{WCov}~(i)
\end{beweis}

Finally putting together Theorems~\ref{GBred.iso}, \ref{wcaus}, 
\ref{GBcov}, \ref{Teo.4.5}
 we have proved for the Gupta--Bleuler
model a major claim:
\begin{teo}
The system of local quantum constraints 
$\Gamma\ni\B\to(\al F.(\B),\,\al U.(\B))$ satisfies 
reduction isotony, weak causality and covariance and therefore the
corresponding net of local physical observables 
$\Gamma\ni\B\to\al R.(\B)$ is a HK--QFT. 
\end{teo}

%%%%%%%%%%%%%%%%%%%%%%%%%%%%%%%%%%%%%%%%%%%%%%%%%%%%%%%%%%%%%%%%%%%%%%%%%%%%%%
\subsection{A simple physical observable algebra}

The net $\B\to\al R.(\B)={\rm C}^*(\delta_{\ot p.(\B)})$
produces a quasi--local physical algebra 
\[
 \al R._0=\ilim\al R.(\B)=\ilim{\rm C}^*(\delta_{\ot p.(\B)})
         ={\rm C}^*(\delta_{\ot p.})=\ccr\ot p.,B.\,,
\] 
where $\ot p.:={\rm Span}\{\ot p.(\B)\mid\B\in\Gamma\}
=\mathop{\cup}\limits_{\B\in\Gamma}\ot p.(\B)$
since $\Gamma$ is a directed set.
Since $B$ is degenerate on $\ot p.$ (see below), $\al R._0$
is not simple and thus cannot be the final physical algebra.
This is also evident from the fact that 
$\ot p.$ contains complex multiples of gradients,
so $\ot p.$ is not in $\ot X..$
Moreover since we have not enforced Maxwell's equations,
from a physical point of view $\al R._0$ cannot be considered as
representing the observables of an electromagnetic field as yet.
To solve these problems, we now do a second stage of constraining
where we choose for our constraint system ${(\al R._0,\,\wt\al U.)}$
where $\wt\al U.:=\delta_{\ot p._0}$ and $\ot p._0$ is the kernel
of $B\restriction\ot p.$. 
The
T--procedure applied to this pair 
will result in a simple algebra via Corollary 5.4 in
\cite{Grundling88b}. For the connection with the Maxwell
equations, we need the following proposition:
\begin{pro} \label{Flemma}
We have:
\begin{eqnarray*}
 \ot p._0 &:=& \big\{f\in\ot p.\mid B(f,\,k)=0\quad\forall\; 
               k\in\ot p.\big\}\\
          &=&\big\{f\in\ot p.\mid f_\mu(p)=p_\mu\, h(p)\;\;
             \hbox{for}\;\; p\in C_+,\;\;\hbox{where}\;\;
             h:C_+\to\C \\
          && \qquad\qquad\!\hbox{is any function such that}
            \quad p\to p_\mu h(p) \;\;\hbox{is in}\;\;\ot Z._{(0)}\;\big\} %\\
%&=&\C\cdot\ot G.\cap\ot Z._{(0)}\,,
\end{eqnarray*}
where $\ot  Z._{(0)}=\ot Z.\cap C_c^\infty(\R^4,\C^4)\wh{\vphantom{|}}\big/
{\rm Ker}(D).$ 
\end{pro}
\begin{beweis}
Recall that $\ot Z.=\ot X.+\C\cdot\ot G.$, then it is easy to see
from Theorem~\ref{Observ} that the gradients $\C\cdot\ot G.\cap\ot Z._{(0)}$
are in $\ot p.$. Moreover, we have in fact that
$\C\cdot\ot G.\cap\ot Z._{(0)}\subset{\rm Ker}(B\rest\ot p.)$ since if we
take $h_\mu=p_\mu k\in\C\cdot\ot G.$ and $f\in\ot p.$ (hence $p_\mu f^\mu=0$),
then
\[
B(h,f)=i\pi\int_{C_+}\left(f_\mu(p)\, p^\mu\overline{k(p)}-
\overline{f_\mu(p)}\, p^\mu k(p)\right){d^3p\over p_0}=0\,.
\]
Thus $\ot p._0=\C\cdot\ot G.\cap\ot Z._{(0)}+{\rm Ker}(B\rest(\ot p.\cap\ot X.)).$
Now, to examine ${\rm Ker}(B\rest(\ot p.\cap\ot X.))$ we first 
want to extend to a larger class of functions, since $\ot p.$
consists of Fourier transforms of functions of compact support,
hence cannot have compact support, which we will want to use below.
Now $\ot p.\cap\ot X.\subset\ot X._{(0)}=\rho\big(C_c^\infty
(\R^4,\R^4)\big)$ where $\rho(f):=\wh{f}\rest C_+$ and by definition
$D(\widehat{f},\, \widehat{k})=:
B(\rho(f),\,\rho(k))$. Moreover, by Theorem~\ref{Observ},
$\ot p.\cap\ot X.=\rho(\ot P.)$ where
$\ot P.:=
{\{f\in C_c^\infty(\R^4,\,\R^4)\mid\partial^\mu f_\mu=0\}}$.
Since the smooth functions of compact
support are dense with respect to the Schwartz topology in the
Schwartz space, and the divergence operator is continuous for
the Schwartz topology, the closure of $\ot P.$ in the 
Schwartz topology is
$\wt{\ot P.}:=
{\{f\in\al S.(\R^4,\,\R^4)\mid\partial^\mu f_\mu=0\}}$.
It is well--known that
$\widehat{D}$ is a tempered distribution (it is the two--point function for the
free electromagnetic field), hence it is continuous with respect
to the Schwartz topology on $\al S.(\R^4,\,\R^4)$ in
each entry.
Thus $\widehat{D}(f,\, k)=0$ for all $k\in\ot P.$ iff
$\widehat{D}(f,\, k)=0$ for all $k\in\wt{\ot P.}$ and hence
${\rm Ker}(B\rest\ot p.\cap\ot X.)
={\{f\in\ot p.\cap\ot X.\mid\, B(f,\, k)=0\;\;\forall\, k
\in\wt{\ot p.}\}}$ where $\wt{\ot p.}:=\rho\big(
\wt{\ot P.}\big)$. We will need this below.

Let $f\in\ot p.\cap\ot X.$, so $p_\mu f^\mu(p)=0$ for $p\in C_+$,
i.e. $\b p.\cdot\b f.(p)=\|\b p.\|\, f_0(p)$, so for
$p\in C_+\backslash 0$, we have $f_0(p)=\b p.\cdot\b f.(p)\big/\|\b p.\|
=\b e.(p)\cdot\b f.(p)$ where $\b e.(p):=\b p./\|\b p.\|$. Now 
in terms of real and imaginary parts 
$f=u+iv\in{\rm Ker}(B\rest\ot p.\cap\ot X.)$ 
iff for all $k=w+ir\in\wt{\ot p.}$ we have that
\begin{eqnarray*}
 0 \!\!\!&=&\!\!\! D(f,\, k)=2i\int\limits_{\R^3\backslash 0}
       \Big(v_\mu w^\mu-u_\mu r^\mu\Big){d^3p\over\|\b p.\|} 
       \qquad\hbox{(using Eqn.~(\ref{symp}))}\\[1mm]
   \!\!\!&=&\!\!\! 2i\int\limits_{\R^3\backslash 0}
         \Big(\b u\cdot r.-\b v\cdot w.
         +\big(\b e.(p)\cdot\b v.\big)\big(\b e.(p)\cdot\b w.\big)
         -\big(\b e.(p)\cdot\b u.\big)\big(\b e.(p)\cdot\b r.\big)
          \Big){d^3p\over\|\b p.\|}\;.
\end{eqnarray*}
Choose $w=0$  (which is possible in $\wt{\ot p.}$)
to get that for all $r\in\wt{\ot p.}\cap(\al S._-\restriction C_+)$
(recall Remark~\ref{spaces}):
\begin{eqnarray}
 0 &=& D(f,\, k)=
     2i\int_{\R^3\backslash 0}\Big(\b u\cdot r.
         -\big(\b e.(p)\cdot\b u.\big)\big(\b e.(p)\cdot\b r.\big)
          \Big){d^3p\over\|\b p.\|}
           \nonumber           \\[1mm]
   &=&  2i\int_{\R^3\backslash 0}\;\b r.\cdot\Big(\b u.
         -\b e.(p)\big(\b e.(p)\cdot\b u.\big)
          \Big){d^3p\over\|\b p.\|}\;.\label{integ}
\end{eqnarray}
Now let $m:C_+\to\R_+$ be a smooth bump function with compact support
away from zero, then we know that
that the function $s$  given by
\[
  \b s.(p):= \big(\b u.(p)-\b e.(p)\big(\b e.(p)\cdot\b u.(p)\big)\big)\,
   m(p)\qquad\hbox{and}\qquad s_0(p):=\b e.(p)\cdot\b s.(p)
\]
is in $\ot X.$ by the characterisation of $\ot X.$ given 
in Remark~\ref{spaces}(i), that it contains all smooth functions with
compact support away from zero. Moreover, since
$p_\mu s^\mu=0$, we conclude $s\in\wt{\ot p.}$.
(Note that $s\not\in\ot p.$, hence the extension to $\wt{\ot p.}$
 in the first
part of the proof).
So we can choose $r=s$ above in Eqn.~(\ref{integ}), then by continuity,
positivity and by ranging over all $m$, we conclude that
\[
  \b u.(p)-\b e.(p)\big(\b e.(p)\cdot\b u.(p)\big)=0\quad\forall\;
  p\in C_+
\]
and as the second term is just the projection of $\b u.(p)$ in the direction
of $\b p.$, this means $\b u.(p)$ must be proportional to 
$\b p.$ for all $p\in C_+\backslash 0$, i.e. $\b u.(p)=\b p.\,q(p)$,
for some suitable scalar function $q$. 
Since $u_0(p)=\b e.(p)\cdot\b u.(p)=\|\b p.\|\, q(p)=p_0\,q(p)$, $p\in C_+$,
this means $u_\mu(p)=p_\mu \,q(p)$, $p\in C_+$.
By setting $r=0$, we obtain a similar result for $v$, and hence
$f_\mu(p)=p_\mu\, h(p)$, $p\in C_+$. The only restriction on
$h$ is that $f\in\ot Z._{(0)}$, since $f$ is automatically in $\ot p.$ by
its form. Thus by the first part of the proof, $\ot p._0$ consists
of these functions, together with complex multiples of gradients,
and this establishes the theorem.
\end{beweis}
\begin{rem}
\label{Maxsmeared}
\begin{itemize}
\item[(i)]
In the proof above, the fact that $f\in\ot X.$ means that $h$ must be
smooth away from the origin. Since $h$ is undefined at the origin
in the proof, consider the behaviour of $f\in\ot p._0$ at zero.
Let $a\in C_+\backslash 0$, then by continuity of $f$:
\[
\b f.(0)=\lim_{t\to 0^+}\b f.(ta)=\b a.\lim_{t\to 0^+}t\, h(ta)
=\b a.\lim_{t\to 0^+}{\b a.\cdot\b f.(ta)\over\|\b a.\|^2}
=\b e.(a)\big(\b e.(a)\cdot\b f.(0)\big)
\]
which can only be true for all 
$a$ if $f(0)=0$.
\item[(ii)]
Now recall Maxwell's equations ${F_{\mu\nu}}^{,\nu}(x)=0$. In the heuristic
version of Gupta--Bleuler QEM, these need to be imposed as state conditions
to define the physical field. Using the smearing formula for $A(f)$,
${F_{\mu\nu}}^{,\nu}(x)$ corresponds to the space
\[
 \ot f.:=\Big\{f\in\ot X.\mid f_\mu(p)=
        p_\mu p^\nu k_\nu(p),\quad p\in C_+,\; k\in\ot X.\Big\}. 
\]
By Proposition~\ref{Flemma} we observe that
$\ot f.\subset\ot p._0$, and thus enforcing the second stage of constraints
$\wt\al U.=\delta_{\ot p._0}$ 
will also impose the Maxwell equations. Note however that the
inclusion $\ot f.\subset\ot p._0$ is proper, which we see as follows.
Consider a line $t\to ta$, $a\in C_+\backslash 0$ and let $f\in\ot f.$ and
$g\in\ot p._0$, then 
$\lim\limits_{t\to 0^+}{d\over dt}f_\mu(ta)=0$, but
$\lim\limits_{t\to 0^+}{d\over dt}g_\mu(ta)=
a_\mu\,\lim\limits_{t\to 0^+}(h(ta)+t{d\over dt}h(ta))$ if $g_\mu(p)=p_\mu\,h(p)$,
and we can easily choose an $h\in\al S.(\R^4,\,\C)$ which makes the latter 
nonzero. Thus merely imposing the Maxwell equations does not appear to be sufficient
to make the physical algebra simple (contrary to
a claim in \cite{Grundling88b}).
\item[(iii)]
In the next step below, we will factor out $\ot p._0$ from $\ot p..$
Since $\ot p._0\supset{\C\cdot\ot G.\cap\ot Z._{(0)}}$, at this point we factor out
the noncausal fields, and regain the reality condition of $\ot X..$
\item[(iv)] 
From the characterisations of the spaces  $\ot p._0$ and $\ot p.$
above, we notice that the triple of spaces $\ot p._0\subset
\ot p.\subset\ot Y.$ corresponds with the one particle spaces
of the triple of spaces in the
heuristic theory $\al H.''\subset\al H.'\subset\al H.$
hence a Fock--Krein construction on $\ot Y.$ (equipped with
the right indefinite inner product) will reproduce the heuristic
spaces. This is done explicitly in Subsection~\ref{contact}
\end{itemize}
\end{rem}

For completeness we would also like to consider the local structure of the
constraint system $(\al R._0,\,\wt\al U.)$.
Define $\wt\al U.(\B):=\wt\al U.\cap\al R.(\B)=
\delta_{\ot s.(\B)}$, $\B\in\Gamma$,  where
$\ot s.(\B):=\ot p._0\cap\ot p.(\B)$, then it
is clear that $\B\to\big(\al R.(\B),\,\wt\al U.(\B)\big)$ is a system
of local quantum constraints. Since $\wt\al U.\subset Z(\al R._0)$, a 
local T--procedure produces:
\begin{eqnarray*}
 \wt\al D.(\B) &=& 
    [\al R.(\B)(\EINS-\wt\al U.(\B))] 
                      \\[1mm]
\wt\al O.(\B) &=& \al R.(\B) \\[1mm]
 \wt\al R.(\B) &=& \al R.(\B)\big/ 
    [\al R.(\B)(\EINS-\wt\al U.(\B))] \,.
\end{eqnarray*}
The main result of this section is:
\begin{teo}
\label{finalR}
The system of local constraints $\B\to(\al R.(\B),\,
\wt\al U.(\B))$ satisfies reduction isotony, causality and
weak covariance, hence $\B\to\wt\al R.(\B)$ is a HK--QFT.
Moreover 
\[
  \wt\al R.(\B)\cong\ccr\ot p.(\B)\big/\ot s.(\B),
   \wt{B}.\cong\ccr\ot c.(\B),B.
\]
where $\ot c.(\B):=\{f\in\ot p.(\B)\cap\ot X.\mid\b p.\cdot\b f.(p)=0,\;\;
p\in C_+\}$ is the ``Coulomb space'', and 
\[
\wt\al R._0\cong\ccr\ot p.\big/\ot p._0,{\wt{B}}.\cong
\ccr\ot c.,B.\subset\CCR
\]
 where $\wt{B}$ is $B$ factored to $\ot p.\big/\ot p._0$,
and 
$\ot c.:=\{f\in\ot p.\cap\ot X.\mid\b p.\cdot\b f.(p)=0,\;\;
p\in C_+\}$.
\end{teo}
\begin{beweis}
For reduction isotony, since it is obvious that if $\B_1\subseteq\B_2$
then $\wt\al O.(\B_1)=\al R.(\B_1)\subseteq\al R.(\B_2)=\wt\al O.(\B_2)$,
we only need to show that $\wt\al D.(\B_1)=\wt\al D.(\B_2)\cap\al R.(\B_1)$,
which by Lemma~\ref{Lem.3.3} will be the case if every Dirac state on
$\al R.(\B_1)$ extends to a Dirac state on $\al R.(\B_2)$.
We first prove that $\ot p.(\B)=\ot c.(\B)\oplus\ot s.(\B)$
where $\ot c.(\B)$
 is the ``Coulomb space'' above, and 
$\ot s.(\B):=\ot p._0\cap\ot p.(\B)$.
Let $m\in\ot p.(\B)$, so $0=p_\mu m^\mu(p),$ $p\in C_+$ and 
$m=f+n$ where $f\in\ot X.(\B)\cap\ot p.(\B)$ and $n\in\C\cdot\ot G.\cap\ot Z.(\B)
\subset\ot s.(\B).$  Now write
$f_\mu=g_\mu+p_\mu h$ where
  $h(p):=f_0(p)\big/\|\b p.\|=\b p.\cdot\b f.\big/\|\b p.\|^2$
  and $g_\mu(p):=f_\mu(p)-p_\mu h(p)$.
Then obviously $p_\mu h$ is in $\ot s.(\B)$ and 
$\b p.\cdot\b g.(p) =\b p.\cdot\b f.(p)-\|\b p.\|^2h(p)
=0$, so $g\in\ot c.(\B)$. Thus we have a decomposition
$m_\mu=g_\mu + (n_\mu+p_\mu h)$ where $g\in\ot c.(\B)$ and the
function in the bracket is in $\ot s.(\B)$. 
To see that the decomposition is unique,
let $g,\; k\in\ot c.(\B)$ such that $g_\mu-k_\mu=p_\mu h$.
Then $0=\b p.\cdot(\b g.-\b k.)=\|\b p.\|^2h$, i.e. $h=0$.

Since for $\B_1\subseteq\B_2$ we have $\ot s.(\B_1)=
\ot s.(\B_2)\cap\ot p.(\B_1)$ and thus 
${\rm Span}\big(\ot p.(\B_1)\cup\ot s.(\B_2)\big)
=\ot c.(\B_1)\oplus\ot s.(\B_2)$, so
$\al A.:={\rm C}^*\big(\delta_{{\rm Span}(\ot p.(\B_1)
\cup\ot s.(\B_2))}\big)$ is generated by two mutually
commuting C*--algebras  ${\rm C}^*(\delta_{\ot c.(\B_1)})$
and ${\rm C}^*(\delta_{\ot s.(\B_2)})$
where the last one is commutative. Now let $A\in
{\rm C}^*(\delta_{\ot c.(\B_1)})$ and $B\in
{\rm C}^*(\delta_{\ot s.(\B_2)})$
such that $AB=0$. Then we want to show that $A=0$ or $B=0$.
Let 
\begin{eqnarray*}
A_n&:=&\sum\limits_{i=1}^{N_n}\alpha_i^{(n)}\delta_{f_i}
\longrightarrow A=\sum\limits_{i=1}^\infty\alpha_i\delta_{f_i}
\qquad\hbox{where}\qquad f_i\in\ot c.(\B_1),\quad\hbox{and} \\
B_n&:=&\sum\limits_{j=1}^{M_n}\beta_j^{(n)}\delta_{k_j}
\longrightarrow B=\sum\limits_{j=1}^\infty\beta_j\delta_{k_j}
\qquad\hbox{with}\qquad k_j\in\ot s.(\B_2)\,.
\end{eqnarray*}
Then $0=AB=\lim\limits_{n\to\infty}\sum\limits_{i,\, j}^{N_n,\, M_n}
\alpha_i^{(n)}\beta_j^{(n)}\delta_{f_i+k_j}$.
However $f_i+k_j\not=f_{i'}+k_{j'}$ for $i\not=i'$ and
$j\not=j'$ since $\ot c.(\B_1)$ and $\ot s.(\B_2)$
are linear independent spaces intersecting only in $\{0\}$.
Thus the set $\{\delta_{f_i+k_j}\mid i\in\N,\; j\in\N\}$
is linearly independent and so $0=\lim\limits_{n\to\infty}
\alpha_i^{(n)}\beta_j^{(n)}=\alpha_i\beta_j$.
Since this holds for all possible pairs $i,\; j$, there is no
pair $\alpha_i$, $\beta_j$ such that $\alpha_i\beta_j\not=0$
and so either all $\alpha_i=0$ or all $\beta_j=0$,
i.e. $A=0$ or $B=0$. Thus from Takesaki \cite[Exercise~2, p.~220]{Takesaki}
we conclude that $\al A.$ is isomorphic to
${\rm C}^*(\delta_{\ot c.(\B_1)})\otimes
{\rm C}^*(\delta_{\ot s.(\B_2)})$ by the map
$\varphi(A\otimes B):=AB$.

Let $\omega$ be a Dirac state on $\al R.(\B_1)$, i.e. 
$\omega(\delta_{\ot s.(\B_1)})=1$, and then define
a state $\wt\omega$ on $\al A.$ by
$\wt\omega:=(\omega\otimes\widehat\omega)\circ\varphi^{-1}$
where $\widehat\omega$ is the state on 
${\rm C}^*(\delta_{\ot s.(\B_2)})$ satisfying
$\widehat\omega(\delta_{\ot s.(\B_2)})=1$.
Now extend $\wt\omega$ arbitrarily to $\al R.(\B_2)\supset
\al A.$, then it coincides with $\omega$ on $\al R.(\B_1)$
and satisfies
$\widehat\omega(\delta_{\ot s.(\B_2)})=1$
hence is a Dirac state on $\al R.(\B_2)$.
This establishes reduction isotony.

For causality, the fact that $\B\to\al R.(\B)$ is a 
HK--QFT already implies that
$[\al R.(\B_1),\,\al R.(\B_2)]=0$ when $\B_1\perp\B_2$,
so $[\wt\al O.(\B_1),\,\wt\al O.(\B_2)]=0$.

For covariance, we already have that
$\alpha_g(\wt\al O.(\B))=\alpha_g(\al R.(\B))=
\al R.(g\B)=\wt\al O.(g\B)$ for $g\in\al P._+^\uparrow$,
$\B\in\Gamma$. Now 
\[
\alpha_g(\wt\al U.(\B)) = 
           \alpha_g(\delta_{\ot s.(\B)})=\delta_{V_g\ot s.(\B)} 
        \quad\hbox{and}\quad
\wt\al U.(g\B)=\delta_{\ot s.(g\B)}\;.
\]
To see that these are equal, note that $V_g\ot s.(\B)   = 
           V_g(\ot p._0\cap\ot p.(\B))$, $V_g$ is symplectic,
and $V_g\ot p.(\B)\subseteq\ot p.(g\B)$ by
\[
p^\mu(V_gf)_\mu(p)=p^\mu\big(\Lambda f(\Lambda^{-1}p)\big)_\mu
e^{-ip\cdot a}=(\Lambda^{-1}p)^\mu f_\mu(\Lambda^{-1}p)
e^{-ip\cdot a}=0
\]
for $p\in C_+$ and $f\in\ot p.(\B)$. Thus 
$V_g(\ot p._0\cap\ot p.(\B))\subseteq\ot p._0\cap\ot p.(g\B)$.
For the reverse inclusion: $\;{V_{g^{-1}}\big(\ot p._0\cap
\ot p.(g\B)\big)}\subseteq\ot p._0\cap\ot p.(\B)$
implies that $\ot p._0\cap\ot p.(g\B)\subseteq
V_g\big(\ot p._0\cap\ot p.(\B)\big)$.
Thus $\alpha_g(\wt\al U.(\B))=\wt\al U.(g\B)$.

Finally, for the last two isomorphism claims, recall that
$\al R.(\B)={\rm C}^*(\delta_{\ot p.(\B)})
=\ccr\ot p.(\B),B.$ and so since 
 $\wt\al R.(\B) = \al R.(\B)\big/ 
    [\al R.(\B)(\EINS-\wt\al U.(\B))]$
and $\wt\al U.(\B)=\delta_{\ot s.(\B)}$ where $\ot s.(\B)$ is
the degenerate part of $\ot p.(\B)$, we conclude from
Theorem~\ref{Teo.2.14} that
$\wt\al R.(\B)\cong\ccr\ot p.(\B)\big/\ot s.(\B),\wt{B}.$
(providing the symplectic commutant of $\ot s.(\B)$ in
$\ot p.(\B)$, $\ot s.(\B)'=\ot p.(\B)$,
and this is obvious since $\ot s.(\B)=\ot p._0\cap
\ot p.(\B)$ and $\ot p.(\B)\subset\ot p.$). Since
 $\ot p.(\B)=\ot c.(\B)\oplus\ot s.(\B)$,
this is isomorphic to
$\ccr\ot c.(\B),B.$. Since for $\B_1\subseteq\B_2$
the inclusion $\wt\al R.(\B_1)\subseteq\wt\al R.(\B_2)$
comes from $\ot p.(\B_1)\subseteq\ot p.(\B_2)$,
and this inclusion factors through to produce
$\ot p.(\B_1)\subseteq\ot p.(\B_2)$,
hence $\ot c.(\B_1)\subseteq\ot c.(\B_2)$,
so the last isomorphism is clear.
\end{beweis}

Thus the quasi--local algebra $\wt{\al R.}_0$ is simple.
Below we will show that for double cones $\B$ the local algebras
are also simple. For a more general net $\Gamma$
it is not clear whether the local algebras are simple.

\begin{rem}
An apparent puzzle raised by the isomorphisms $\wt\al R.(\B)
\cong\ccr\ot c.(\B),B.$ here, is the noncovariance of the spaces
$\ot c.(\B)$ under $V_g$, $g\in\al P._+^\uparrow$, given that the
net $\wt\al R.(\B)$ is covariant under the isomorphisms
derived from $V_g$. The resolution is that $V_g$  maps an 
equivalence class $f+\ot s.(\B)$ in $\ot p.(\B)$
to the equivalence class $V_gf+\ot s.(g\B)$ in
$\ot p.(g\B)$, and these equivalence classes
correspond to elements $h\in\ot c.(\B)$ and $k\in\ot c.(g\B)$
repectively, but it is not true that $k=V_gh$.
\end{rem}

\begin{teo}
If the sets $\B\in\Gamma$ consist of double cones, then the local algebras
\[
 \wt\al R._0(\B)\cong\ccr\ot p.(\B)\big/\ot s.(\B),{\wt{B}}.
                \cong\ccr\ot c.(\B),B. 
\]
are simple.
\end{teo}
\begin{beweis}
By Theorems~\ref{Teo.2.13} and \ref{finalR} it
suffices to prove that $(\ot c.(\B),B)$ is a 
nondegenerate symplectic space for each double cone $\B$, where we consider
\begin{eqnarray*}
\ot c.(\B) &:=& \Big\{f\in\ot p.(\B)\cap\ot X.
                \mid\b p.\cdot\b f.(p)=0,\;\;p\in C_+ \Big\}       \\
 \ot C.(\B) &:=& \Big\{f\in C_c^\infty(\R^4,\,\R^4)\mid f_0=0\,,\;
                 \sum\partial_\ell f_\ell=0\,,\quad\mathrm{and}\quad
               \mathrm{supp}\,f\subset\B \Big\}                   \\
\ot C.     &:=& \Big\{f\in C_c^\infty(\R^4,\,\R^4)\mid f_0=0\quad
               \mathrm{and}\quad\sum\partial_\ell f_\ell=0    \Big\}\,.
\end{eqnarray*}
Observe that if we define the map
$\rho\colon\ \al S.(\R^4,\,\R^4)\to\ot X.=
\al S.(\R^4,\,\R^4)\wh{}\big/{\rm Ker}(D)$ by $\rho(f)=\wh{f}+{\rm Ker}(D)$,
then $\ot c.(\B)=\rho(\ot C.(\B))$.

%Now to prove the nondegeneracy of the mentioned local symplectic space
We adapt the arguments in Dimock~\cite{Dimock}. 
(Note though that from Proposition~\ref{Flemma} 
we do not need the assumption that the Cauchy
surface is compact, used in~\cite[Proposition~5]{Dimock}). 
For test functions in $\al S.(\R^4,\R^4)$ we have
\begin{eqnarray*}
\widehat{D}(f,\, h)&:=&D(\widehat{f},\,\widehat{h})=
\int\int f_\mu(x)\,h^\mu(y)\, D(x-y)\,d^4x\, d^4y \\
&=&\int f_\mu(x) (Dh)^\mu(x) d^4(x) \\
\hbox{where}\qquad (Dh)_\mu(x)&:=&\int h_\mu(y)\, D(x-y)\, d^4y \\
&=&-i\pi\int_{C_+}{d^3p\over p_0}\Big(e^{ip\cdot x}\widehat{h}_\mu(p)
-e^{-ip\cdot x}\overline{\widehat{h}}_\mu(p)\Big).
\end{eqnarray*}
Note that $D$ is the difference of the retarded and advanced fundamental
solutions of the wave operator $\Box $, hence $Df$ is a solution of the
wave equation (cf.~\cite{ChoquetIn68,Dimock80}).
Henceforth we will only consider test functions in $\ot C.$.
We want to express $D(\widehat{f},\,\widehat{h})$ in terms of 
the corresponding real Cauchy data. 
Given $f\in\ot C.$, we define these by:
\begin{eqnarray*}
Q^f_\ell(\b x.)&:=&\frac{-1}{\pi}(Df)_\ell(0,\b x.)
                   \in C^\infty_c(\R^3,\R) \\
R^f_\ell(\b x.)&:=&\frac{1}{\pi}\big(\partial_0(Df)_\ell\big)(0,\b x.)
                   \in C^\infty_c(\R^3,\R) \,,\quad \ell=1,2,3\,.
\end{eqnarray*}
Then their Fourier transforms are, using the conventions
$\widehat{f}(p)={\int_{\R^4}e^{-ip\cdot x}f(x)d^4x}$ and
$\widehat{h}(\b p.)={(2\pi)^{-3/2}\int_{\R^3}e^{i\b p.\cdot\b  x.}h(\b x.)d^3x}$
for four and three dimensional Fourier transforms:
\def\p{{\|\b p.\|}}
\begin{eqnarray*}
\widehat{Q^f_\ell}(\b p.)&=&{i(2\pi)^{3/2}\over\p}\Big(
\widehat{f}_\ell(\p,\b p.)-\overline{\widehat{f}}_\ell(\p,-\b p.)\Big) 
 \\
\widehat{R^f_\ell}(\b p.)&=&(2\pi)^{3/2}\Big(\widehat{f}_\ell(\p,\b p.)
+\overline{\widehat{f}}_\ell(\p,-\b p.) \Big)\,.
\end{eqnarray*}
If we substitute these into the rhs of the equation
\[
\int_{\R^3}\big(Q^f_\ell R^h_\ell-R^f_\ell Q^h_\ell\big)(\b x.)\, d^3x
=\int_{\R^3}\Big(\widehat{Q^f_\ell}(-\b p.)\widehat{ R^h_\ell}(\b p.)
-\widehat{R^f_\ell}(-\b p.)\widehat{ Q^h_\ell}(\b p.)\Big)d^3p
\]
(summation over $\ell$), then we find
with some algebraic work that
\begin{equation}
\label{Dcdata}
\widehat{D}(f,\, h)={-1\over 16\pi^2}
\int_{\R^3}\big(Q^f_\ell R^h_\ell-R^f_\ell Q^h_\ell\big)(\b x.)\, d^3x \,.
\end{equation}
Now let $\B$ be a double cone, by covariance we can assume it to be
centered at the origin. Let $\Sigma$ be its intersection with
the Cauchy surface $t=0$. Then if $f\in\ot C.(\B)$
we have by the properties of $D$
that ${\rm supp}\,Q^f_\ell\subset\Sigma\supset{\rm supp}\,R^f_\ell$
(cf.~\cite{ChoquetIn68,Dimock80}).
Further from the arguments in the proof of Proposition~2 in \cite{Dimock}
we know that for any pair $(Q,R)\in C^\infty_c(\R^3,\R^3)\times 
C^\infty_c(\R^3,\R^3)$ satisfying $\partial_\ell Q_\ell=0=
\partial_\ell R_\ell$, there exists a unique solution of the wave
equation $\varphi\in C^\infty(\R^4,\R^4)$ with these data and satisfying
$\partial_\ell\varphi_\ell=0=\varphi_0$. Even more by \cite[Proposition~4(c)]{Dimock}
we can always find an $f\in\ot C.(\B)$ such that $Df=\varphi$.

Now take a test function $h\in\ot C.(\B)$ such that
$\rho(h)\in{\rm Ker}(B\rest\ot c.(\B))$, i.e.
\[
0=\int_{\Sigma}\big(Q^f_\ell R^h_\ell-R^f_\ell Q^h_\ell\big)(\b x.)\, d^3x
=\int_{\R^3}\big(Q^f_\ell R^h_\ell-R^f_\ell Q^h_\ell\big)(\b x.)\, d^3x
\]
for all $f\in\ot C.(\B)$. By the arguments above %this implies that
%$(Q^h,R^h)=(0,0)$ (take e.g.~$(Q^f,R^f)=(R^h,0)$ and $(Q^{f'},R^{f'})=(0,-Q^h)$)
we can choose $(Q^f,R^f)=(R^h,0)$ and $(Q^{f'},R^{f'})=(0,-Q^h)$
to conclude that $(Q^h,R^h)=(0,0)$. Then
by uniqueness this implies that $Dh=0$, i.e.~$\rho(h)=0$.
\end{beweis}

In this example we have done our constraint reduction in two stages,
and the question arises as to whether we would have obtained
the same physical algebra from a single reduction by the
full set of constraints. This will be examined in the next main section.

\subsection{Connecting with the indefinite inner product}
\label{contact}

In this subsection we want to connect the C*--algebraic version above of 
Gupta--Bleuler electromagnetism with the usual one on indefinite inner product
space (henceforth abbreviated to IIP--space), sketched in Subsect.~\ref{GBh}.
We will freely use the Fock--Krein construction of Mintchev~\cite{Mintchev}.
Start with the space 
\begin{eqnarray*}
 \ot Y.&=&\al S.(\R^4,\,\C^4)\big/{\rm Ker}(D)\\[1mm]
\hbox{with IIP:}\qquad
   K(f,\, h)&:=&-2\pi\int_{C_+}\overline{{f}^\mu(p)}\,{h}_\mu(p)\,
{d^3p\over p_0}\;,\qquad \forall\, f,\, h\in\ot Y.,
\end{eqnarray*}
which is well--defined on $\ot Y.$ because ${\rm Ker}(D)={\rm Ker}(K)$.
Note that $B={\rm Im}\,K$.
Define now on $\ot Y.$ the operator $J$ by
$(Jf)_0=f_0$, $(Jf)_\ell=-f_\ell,$ $\ell=1,2,3,$ then obviously
$J^2=\EINS$ and ${( f,\, h)}:={K(f,\, Jh)}$
defines a positive definite inner product on $\ot Y.$.
Let $\al N.$ be the Hilbert space completion w.r.t.
this inner product (so in fact, it is just $L^2(C_+,\C^4,\mu_0)$
with $d\mu_0={d^3p/ p_0}$)
 and let $\ot F.(\al N.)$ be the symmetric Fock
space constructed on  $\al N.$.
Below we will use the notation $\ot F._0(\al L.)$ for the finite
particle space with entries taken from a given space $\al L.\subset
\al N.$, and as usual $\ot F._0:=\ot F._0(\al N.)$.
We make  $\ot F.(\al N.)$ into a Krein space with the IIP
${\langle \psi,\,\varphi\rangle}:={\big(\psi,\,\Gamma(J)\varphi\big)}$
where $\Gamma(J)$ is the second quantization of $J$ and the round brackets
indicate the usual Hilbert space inner product. We define creation
and annihilation operators as usual, except for the replacement of the 
inner product by the IIP, i.e. on the n--particle space $\al H.^{(n)}$ they are
\begin{eqnarray*}
a^\d(f)S_nh_1\otimes\cdots\otimes h_n &=& \sqrt{n+1}S_{n+1}f\otimes
h_1\otimes\cdots\otimes h_n\\[1mm]
a(f)S_nh_1\otimes\cdots\otimes h_n &=& {1\over\sqrt{n}}\sum_{i=1}^n
{\langle f,h_i\rangle }S_{n-1}h_1\otimes\cdots\widetilde{h_i}\cdots\otimes h_n
\end{eqnarray*}
where the tilde means omission and $S_n$ is the symmetrisation operator
for $\al H.^{(n)}$.
Note that $a^\d(f)$ is the $\langle\cdot,\cdot\rangle\hbox{--adjoint}$
of $a(f)$, not the Hilbert space adjoint.
The connection with the heuristic creation and annihilation operators in
Subsect.~\ref{GBh} comes from the smearing formula
\[
a(f)=\sqrt{2\pi}\int_{C_+}a_\mu(\b p.)\, f^\mu(p){d^3p\over p_0}
\]
and the Krein adjoint formula for $a^\d(f)$ (which produces a complex
conjugation on the smearing function).
Then the constructed operators have the correct commutation relations,
so that if we define the field operator by
\begin{eqnarray*}
  A(f)&:=&{1\over\sqrt{2}}\big(a^\d(f)+a(f)\big)\\[1mm]
\hbox{then:}\qquad\qquad \big[A(f),\,A(h)\big]&=&iB(f,h)\,.
\end{eqnarray*}
We only need to restrict to $f\in\ot X.$ to make the connection
with the field operators of before. (Note that since $\ot Y.$ is
the complex span of $\ot X.$, the complex span of the set
$\{ A(f)\Omega\,\mid\,f\in\ot X.\}$ is dense in $\ot F.(\al N.)$.)
Following Mintchev~\cite{Mintchev} we now define on the finite particle
space $\ot F._0$ the unbounded  $\langle\cdot,\cdot\rangle\hbox{--unitary}$
operators
\[
  W(f)\psi:=\mathop{\rm lim}_{N\to\infty}\sum_{k=0}^N{[iA(f)]^k\over k!}\psi
\]
which satisfy the Weyl relations, and hence constitute a 
Krein representation $\gamma$  by $\gamma(\delta_f):=W(f)$, of
the dense *--algebra generated by $\delta_{\ot Y.}$ in $\CCRy$, usually
denoted by $\Delta(\ot Y.,B)$. Note that
 $\gamma:\Delta(\ot Y.,B)\to{\rm Op}(\ot F._0)$
takes the C*--involution to the Krein involution.
Moreover, for the constraints we see from the heuristic formula:
$\chi(h):=a(ip_\mu\wh{h}\sqrt\pi)$, so 
the set $\set\chi(h),h\in{\al S.(\R^4,\R)}.$ corresponds to
$\set a(f),{f\in\ot G.}.$. By the commutation relations, we still have
\[
   [\chi(h)^\d\chi(h),\,A(f)]= iA(G_h(f))\,.
\]
We want to extend $\gamma$ so that it also represents the constraint 
unitaries $U_{\al G.}.$
\def\ol #1.{\overline{#1}}
\begin{pro} $\quad$
$T^t_h:\ot Y.\to\ot Y.$ is K--unitary, i.e. $K(f,g)=K(T^t_hf,T^t_hg)$
for all $f,\; g,\; h,\; t.$
\end{pro}
\begin{beweis}
\begin{eqnarray*}
& &K(T^t_hf,T^t_hg)=-{2\pi}\int_{C_+}{d^3p\over p_0}\Big\{\big(
\ol f_\mu.+it\pi p_\mu\ol \wh{h}.\,\ol c(f,h).\big)\cdot
\big(g^\mu-it\pi p^\mu \wh{h}\, c(g,h)\big)\Big\} \\
& &\quad=K(f,g)-2\pi^2 it\int_{C_+}{d^3p\over p_0}p^\mu\big(g_\mu
\,\ol \wh{h}.\,\,\ol c(f,h).-\ol f_\mu.\,\wh{h}\, c(g,h)\big)\\
& & \hbox{and the last integral is:}\\
& &\qquad \Big(\int_{C_+}{d^3p\over p_0}p_\mu g^\mu\ol \wh{h}.(p)\Big)
 \Big(\int_{C_+}{d^3p'\over p'_0}p'_\nu\ol f^\nu.\, \wh{h}(p')\Big)\\
& &\qquad\qquad\qquad- \Big(\int_{C_+}{d^3p\over p_0}p_\mu\ol f^\mu.\,\wh{h}(p)\Big)
 \Big(\int_{C_+}{d^3p'\over p'_0}p'_\nu g^\nu\ol \wh{h}.(p')\Big)\\
& &\qquad=0\;.
\end{eqnarray*}
\end{beweis}
Thus we know from Mintchev~\cite{Mintchev} the second quantized operator
$\Gamma(T^t_h)$ is well-defined on $\ot F._0(\ot Y.)$, it is $\langle\cdot,
\cdot\rangle\hbox{--unitary,}$ and it implements $T^t_h$ on $A(f)$.
By the definition of $\Gamma(T^t_h)$ it is also clear that the set of these
commute, and thus we can extend $\gamma$ to $U_{\al G.}$ by defining
$\gamma(U_{T^t_h}):=\Gamma(T^t_h)$. For the heuristic theory, we would
like to identify this with $\exp(it\chi(h)^\d\chi(h))$, and this is done in the 
next proposition.
\begin{pro}
\[
 {d\over dt}\Gamma(T^t_h)\psi\Big|_{t=0}=i\chi(h)^\d\chi(h)\psi\qquad
\forall\,\psi\in\ot F._0(\ot Y.),\; h\in\al S.(\R^4,\R)\,. 
\]
\end{pro}
\begin{beweis}
Let $\psi=S_nf_1\otimes\cdots\otimes f_n$ with $f_i\in\ot Y.$, and recall that
$T^t_hf=f+tG_h(f)$. Then
\begin{eqnarray}
& &{d\over dt}\Gamma(T^t_h)S_nf_1\otimes\cdots\otimes f_n\Big|_{t=0}
={d\over dt}S_n(T_h^tf_1)\otimes\cdots\otimes(T_h^tf_n)\Big|_{t=0}  
\nonumber \\
&=&\quad S_nG_h(f_1)\otimes f_2\otimes\cdots\otimes f_n
+\cdots+ S_n f_1\otimes\cdots\otimes f_{n-1}\otimes G_h(f_n)\;.\qquad
\label{sum1}
\end{eqnarray}
On the other hand, if we start from the right hand side of the claim
in the proposition, and use
\[
G_h(f_k)=-i\pi p_\mu\wh{h}(p)\int_{C_+}{d^3p'\over p'_0}f_k^\nu p'_\nu
\ol \wh{h}.(p')=\pi  p_\mu\wh{h}(p)\;\langle ip_\nu\wh{h},f_k\rangle
\]
then we see that
\begin{eqnarray*}
& &i\chi(h)^\d\chi(h)S_nf_1\otimes\cdots\otimes f_n
=i\pi a^\d(ip_\mu\wh{h})\, a(ip_\mu\wh{h})\,S_nf_1\otimes\cdots\otimes f_n \\
& &\qquad =i\pi a^\d(ip_\mu\wh{h})\,{1\over\sqrt{n}}\sum_{k=1}^n
\langle ip_\mu\wh{h},\, f_k\rangle S_{n-1}
f_1\otimes\cdots\widetilde{f_k}\cdots\otimes f_n \\
& &\qquad =i\pi\sum_{k=1}^n\langle ip_\mu\wh{h},\, f_k\rangle
S_n\,(ip_\mu\wh{h})\otimes f_1\otimes\cdots\widetilde{f_k}\cdots\otimes f_n \\
& &\qquad =S_n\,G_h(f_1)\otimes f_2\otimes\cdots\otimes f_n
+\cdots+ S_n f_1\otimes\cdots\otimes G_h(f_n)
\end{eqnarray*}
and so comparing this with Eqn.~(\ref{sum1}) establishes the proposition.
\end{beweis}
Thus if $\al E.$ denotes the *--algebra generated by 
$\Delta(\ot Y.,B)\cup U_{\al G.}$ (dense in $\al F._e$), 
and we set $\gamma(U_g):=\Gamma(U_g)$, then we now have a representation
$\gamma:\al E.\to {\rm Op}(\ot F._0(\ot Y.))$ which agrees with the 
Gupta--Bleuler operator theory.

To conclude this section we wish to compare the physical algebra obtained by 
C*--methods with the results of the spatial constraining in the usual theory.
In the latter one defines
\[
\al H.':=\set\psi\in\ot F.(\al N.),{\psi\in{\rm Dom}(\chi(h))
\quad\hbox{and}\quad\chi(h)\psi=0\;
\;\forall\,h\in\al S.(\R^4,\R)}.
\]
so if we take ${\rm Dom}(\chi(h))=\ot F._0(\ot Y.)$, then
\begin{pro} $\qquad$  $\al H.'=\ot F._0(\C\cdot\ot p.)$.
\end{pro}
\begin{beweis}
Since $\chi(h):\al H.^{(n)}\to\al H.^{(n-1)}$ and the n--particle spaces
are linearly independent, it suffices to check the condition $\chi(h)\psi=0$
on each $\al H.^{(n)}$ separately. Write
\begin{equation}
\label{vectorpsi}
\psi\in\al H.'\cap\ot F._0(\ot Y.)\qquad\hbox{in the form:}\qquad
\psi=\sum_{k=1}^{N}\,S_nf_{k1}\otimes\cdots\otimes f_{kn}\,.
\end{equation}
So $\psi\in\al H.'$ means $\chi(h)\,\psi=0$, which implies
${\langle \chi(h)^\d\varphi,\,\psi \rangle}=0$ for all $h\in\al S.(\R^4,\R)$
and $\varphi\in\al H.^{(n-1)}$. Explicitly:
\begin{eqnarray*}
0&=&\langle a^\d(ip_\mu\wh{h})\varphi,\,\psi \rangle \\
&=&\Big\langle a^\d(ip_\mu\wh{h})
\sum_{k=1}^{N}S_{n-1}g_{k1}\otimes\cdots\otimes g_{k(n-1)},\,
\sum_{j=1}^{M}\,S_nf_{j1}\otimes\cdots\otimes f_{jn}
\Big\rangle \\
&=& \sum_{k=1}^{N}\sum_{j=1}^{M}\sqrt{n}
\Big\langle (ip_\mu\wh{h})\otimes g_{k1}\otimes\cdots\otimes g_{k(n-1)},\,
S_nf_{j1}\otimes\cdots\otimes f_{jn}\Big\rangle \\
&=& \sum_{k=1}^{N}\sum_{j=1}^{M}\sum_{\sigma\in P_n}{\sqrt{n}
\over n!}\big\langle ip_\mu\wh{h},f_{j\sigma(1)}\big\rangle
\big\langle g_{k1},f_{j\sigma(2)}\big\rangle\cdots
\big\langle g_{k(n-1)},f_{j\sigma(n)}\big\rangle
\end{eqnarray*}
where $P_n$ denotes the permutation group on $\{1,\ldots,n\}$.
This must hold for all $\varphi$ so if we let the $g_{ki}$ vary
over $\ot Y.$, we get that
$\langle ip_\mu\wh{h},\, f_{ji}\rangle =0$ for all $h$.
Thus $0=\int_{C_+}{d^3p\over p_0}p_\mu f_{ji}^\mu\ol \wh{h}.(p)$
for all $h$, and the choice $\wh{h}=ip_\mu f_{ji}^\mu$ then
implies $p_\mu f_{ji}^\mu\rest C_+=0$ for all $f_{ji}$ i.e. $f_{ji}\in\ot p.$.
Thus $\al H.'\subseteq\ot F._0(\C\cdot\ot p.)$.
The reverse inclusion is obvious.
\end{beweis}
Note that $\al H.'=\ot F._0(\C\cdot\ot p.)$ is of course
preserved by $A(\ot p.\cap\ot X.)$, the generators of $W(\ot p.\cap\ot X.)=
\gamma(\delta_{\ot p.\cap\ot X.})$. Thus by 
the exponential series, $W(\ot p.\cap\ot X.)$ maps $\al H.'$ into its 
Hilbert space closure. Since obviously $\Gamma(T_h^t)\rest\al H.'=\EINS$,
when we restrict the algebra $\gamma(\al E.)$ to $\al H.'$,
the constraints are factored out. We already know that $\ot p.\cap\ot X.$
contains the smearing functions which produce the fields $F_{\mu\nu}$.
We check that the IIP is positive semidefinite on $\al H.'$.
First, the one--particle space.

\begin{pro}
\label{pos-IIP}
We have
$\quad K(f,\, f)\geq 0$ $\forall\, f\in\ot p.\quad$ 
and $\quad\ker(K\restriction
\ot p.)=\ot p._0$.
\end{pro}
\begin{beweis}
Let $f\in\ot p.$, then by Theorem~\ref{Observ} we see
$p_\mu f^\mu(p)=0$ for all $p\in C_+$, and thus by the proof of
Proposition~\ref{Flemma}, $f_0(p)=\b p.\cdot\b f.(p)\big/\|\b p.\|$
for $p\not=0$. Thus
\begin{equation}
\label{(a)}
 K(f,\, f):=-2\pi\int_{C_+}\overline{{f}^\mu(p)}\,{f}_\mu(p)\,{d^3p\over p_0}
           =2\pi\int_{C_+}\big(\overline{\b f.}\cdot\b f.(p)
            -|f_0(p)|^2\big){d^3p\over p_0}\;.
\end{equation}
Now
\[
\overline{\b f.}\cdot\b f.(p) = \|\b f.(p)\|^2
   =\left\|{\b p.\over\|\b p.\|}\right\|^2\|\b f.(p)\|^2
   \geq\left|{\b p.\over\|\b p.\|}\cdot\b f.(p)\right|^2
   =\big|f_0(p)\big|^2
\]
and so $K(f,\, f)\geq 0$ for all $f\in\ot p.$.

Let $K(f,\, f)=0$ for $f\in\ot p.$, then since by the preceding
the integrand in Eqn.~(\ref{(a)}) is positive, we conclude that
$\|\b f.(p)\|=\big|\b p.\cdot\b f.(p)\big|\big/\|\b p.\|$ on $C_+$.
Thus $\b f.(p)$ must be parallel to $\b p.$, i.e.
$\b f.(p)=\b p.\,h(p)$ for some $h$.
Since $f_0(p)=\b p.\cdot\b f.(p)\big/\|\b p.\|
=\|\b p.\|\, h(p)=p_0\, h(p)$ we conclude
$f_\mu(p)=p_\mu\, h(p)$, i.e. $f\in\ot p._0$ by
Proposition~\ref{Flemma}. Thus $\ker(K\restriction\ot p.)\subseteq
\ot p._0$. The reverse inclusion is obvious.
\end{beweis}

This establishes the positivity of the IIP on the one-particle space
of $\al H.'$, and then the positivity on all of
 $\al H.'=\ot F._0(\C\cdot\ot p.)$ follows from the usual arguments
for tensor products.

Next, in the usual theory one factors out the zero norm part of $\al H.'$,
i.e. $\al H.'':=\ker(\langle\cdot,\,\cdot\rangle\rest\al H.')$.

\begin{pro} \chop
(i) $ $  $\al H.''=\set\psi\in\ot F._0(\C\cdot\ot p.),
{\psi^{(n)}\in S_n(\ot p._0\otimes\ot p.\otimes\cdots\otimes\ot p.)}.$\chop
$\quad$ where $\psi^{(n)}$ denotes the n--particle component of $\psi$.\chop
(ii)  $\al H.'\big/\al H.''=\ot F._0(\C\cdot\ot p.\big/\ot p._0)\cong\al H._{\rm phys}$
where the identification is via the factor map.
\end{pro}
\begin{beweis}
%It is obvious from the last lemma that  $\al H.''
%\supseteq\ot F._0(\C\cdot\ot p._0)$, so we show the reverse inclusion.
(i) Since $\langle\cdot,\,\cdot\rangle$ is a positive form on $\al H.'$,
the Cauchy--Schwartz inequality applies, hence $\psi\in\al H.''$
iff $\langle\psi,\,\varphi\rangle=0$ for all $\varphi\in\al H.'$.
Let  $\psi\in\al H.''$ be given by Eqn.~(\ref{vectorpsi}), then
we have
\begin{eqnarray*}
0&=&\Big\langle
\sum_{k=1}^{N}\,S_nf_{k1}\otimes\cdots\otimes f_{kn},\,
S_{n}g_{1}\otimes\cdots\otimes g_{n}\Big\rangle \\
&=&\sum_{k=1}^{N}\sum_{\sigma\in P_n}{1\over n!}
\big\langle f_{k1} ,g_{\sigma(1)}\big\rangle\cdots
\big\langle f_{kn},g_{\sigma(n)}\big\rangle
\end{eqnarray*}
for all $g_i\in\ot p.$. By letting $g_i$  vary over all
$\ot p.$, we conclude for each $k$ there is an $i$ such that
$\big\langle f_{ki} ,g\big\rangle
=0$ for all  $g\in\ot p.$, so $f_{ki}\in\ot p._0$
by Proposition~\ref{pos-IIP}. This establishes the claim in (i).\chop
(ii) It suffices to examine the n--particle spaces independently, and to 
ignore the symmetrisation because it creates symmetric sums in
which we can examine each term independently. 
We first examine elementary tensors where no factor is in $\ot p._0$
(otherwise it is in $\al H.''$ already).
Let $\psi= f_1\otimes\cdots\otimes f_n$.
%By definition of tensor products, we can only add two
%elementary tensors and obtain a different combination of elementary tensors
%(excluding adding combinations which are zero)
%if they are identical in all factors except
%%possibly one. 
Now the factor map comes from the equivalence
$\psi\equiv\phi$ iff $\psi-\phi\in\al H.''$, for $\psi,\;\phi
\in\al H.'$ and so we will show the equivalence class $[\psi]$
depends only on the equivalence classes $[f_i]$ in $\ot p./\ot p._0$.
To generate equivalent elements in the $i\hbox{--th}$ slot, we just add
an $f_1\otimes\cdots\otimes f_{i-1}\otimes g_i\otimes f_{i+1}\cdots f_n
\in\al H.''$ with $g_i\in \ot p._0$. By doing this for all slots,
we have demonstrated for the elementary tensors that the factor map
takes $\psi= f_1\otimes\cdots\otimes f_n\in\al H.'$
to $[\psi]=[f_1]\otimes\cdots\otimes [f_n]$.
Extend by linearity to conclude that 
 $\al H.'\big/\al H.''=\ot F._0(\C\cdot\ot p.\big/\ot p._0)$.
\end{beweis}

Recall from Remark~\ref{Maxsmeared}(ii) that the 
space $\ot f.$ of smearing functions
corresponding to the lhs of the Maxwell equations are in $\ot p._0$
and so as it is obvious that $A(\ot p._0)\al H.'\subset\al H.''$
from the above characterisations, this substantiates the heuristic claim
that the Maxwell equations hold on $\al H.'/\al H.''$.

Considering now the Poincar\'e transformations, recall we have
the symplectic action on $\ot Y.$:
\[
(V_gf)(p):=e^{-ipa}\,\Lambda\,f(\Lambda^{-1}p)
\qquad\forall\; f\in\al S.(\R^4,\C^4),\;
g=(\Lambda,a)\in \al P.^\uparrow_+.
\]
In fact, it is also $K\hbox{--unitary}$ because
\begin{eqnarray*}
K(V_gf,\,V_gh) &=&
  -2\pi \int_{C_+}(\overline{V_gf})^\mu(p)\,(V_gh)_\mu(p)\,
{d^3p\over p_0} \\
   &=&-2\pi \int_{C_+}\big(\Lambda\overline{f}\big)^\mu(\Lambda^{-1}p)\,
\big(\Lambda h\big)_\mu(\Lambda^{-1}p)\;
{d^3p\over p_0} \\
   &=&-2\pi \int_{C_+}\overline{f}^\mu(\Lambda^{-1}p)\,
 h_\mu(\Lambda^{-1}p)\;
{d^3p\over p_0} = K(f,\, h)
\end{eqnarray*}
since the measure $d^3p\big/p_0$ is Lorentz invariant on the light cone.
So, using Mintchev~\cite{Mintchev} again, the second quantized operator
$\Gamma(V_g)$ is well-defined on $\ot F._0(\ot Y.)$, it is $\langle\cdot,
\cdot\rangle\hbox{--unitary,}$ and it implements $V_g$ on $A(f)$.
To see that $\Gamma(V_g)$ preserves $\al H.'$, it suffices 
to note that $V_g\ot p.\subset\ot p.$ because
\[
p^\mu(V_gf)_\mu(p)=p^\mu\big(\Lambda f(\Lambda^{-1}p)\big)_\mu
e^{-ip\cdot a}=(\Lambda^{-1}p)^\mu f_\mu(\Lambda^{-1}p)
e^{-ip\cdot a}=0
\]
for $f\in\ot p.$. Moreover $\Gamma(V_g)$ preserves $\al H.''$ because
it preserves both $\ot p.$ and $\ot p._0$ where the latter follows
from the fact that $\ot p._0$ is the kernel of the symplectic form
on $\ot p.$ and $\Gamma(V_g)$ is a symplectic transformation.
Thus $\Gamma(V_g)$ factors through to 
 $\al H.'\big/\al H.''=\al H._{\rm phys}$
and in fact, since the IIP now is a Hilbert inner product on this space, 
the factored $\Gamma(V_g)$ becomes a unitary operator, which will
extend to the Hilbert closure of $\al H.'\big/\al H.''$.
It obviously will still implement the (factored through) Poincar\'e
transformations $\wt{V}_g$ on the factored field operators 
obtained by restricting $A(\ot p.)$ to $\al H.'$ and then
factoring to $\al H.'\big/\al H.''$.

Returning now to the C*--theory, observe from the characterisations of
$\al H.'$ and $\al H.''$ that $\gamma(\delta_{\ot p.})$
will map $\al H.'$ to its Hilbert space closure in $\ot F.(\ot Y.)$,
and will map $\al H.''$ to its closure. (Also note
 that  $\gamma(\delta_{\ot p._0})-\EINS$ will map  $\al H.'$ 
to the closure of $\al H.''$.) Thus $\gamma(\delta_{\ot p.})$
will lift to operators on the space $\overline{\al H.'}\big/\overline{\al H.''}$
with  $\al H.'\big/{\al H.''}$
in their domains. Equipping $\al H.'\big/\al H.''$ with the inner product coming from
the initial IIP, will make these operators into unitaries which extend to
the closure $\overline{\al H.}_{\rm phys}$ of $\al H.'\big/\al H.''$.
The step of factoring through the operators from  $\al H.'$
to $\al H.'\big/{\al H.''}$ will identify $\gamma(\delta_{\ot p._0})$
with $\EINS$.
Thus we obtain an actual Hilbert space representation $\wt{\gamma}:\ccr
\ot p./\ot p._0,\wt{B}.\to\al B.(\overline{\al H.}_{\rm phys})$.
(Recall this last CCR--algebra was our final quasi--local physical algebra
$\wt{\al R.}_0$ of before).
We know this representation 
is a Fock representation, but this is also clear from 
\begin{eqnarray}
\label{K-state}
\omega_0(\delta_f)&:=&\big\langle\Omega,\,\gamma(\delta_f)\Omega\big\rangle
=\lim_{N\to\infty}\sum_{k=0}^N{i^k\over k!}\big\langle\Omega,\,
[A(f)]^k\Omega\big\rangle \nonumber \\
&=& \exp(-K( f,\, f)\big/4)
\end{eqnarray}
and the fact that $K(\cdot,\, \cdot)$ is positive on $\ot p.$
with kernel $\ot p._0$. (The usual calculation still works for the
last equality).
This state $\omega_0$ thus extends from $\Delta(\ot p./\ot p._0,\wt{B})$
to $\ccr\ot p./\ot p._0,\wt{B}.$. In terms of the original C*--algebra,
note that $\omega_0$ comes from a (nonunique) state $\wt{\omega}_0$
on $\al F._e$, because the formula in Eqn.~(\ref{K-state}) still
defines a state on $C^*(\delta_{\ot p.})$ by positivity of $K\rest\ot p.$
(which obviously becomes $\omega_0$ after constraining out $\delta_{\ot p._0}$)
and it extends by the Hahn--Banach theorem to a state $\wt{\omega}_0$
on $\al F._e$. However,  $\wt{\omega}_0$ must necessarily be nonregular
which we see as follows. We have
$\wt{\omega}_0(\delta_{\ot p._0})=1$, hence for $c\in\ot p._0$
and any $f$ with $B(f,c)\not=0$:
\[
2\wt{\omega}_0(\delta_f)
=\wt{\omega}_0(\delta_f\delta_{tc}+\delta_{tc}\delta_f)
=2\wt{\omega}_0(\delta_{f+tc})\cos[tB(f,c)\big/2]
\]
for all $t\in\R$. This implies $\wt{\omega}_0(\delta_{f})=0$
and thus the map $t\to\wt{\omega}_0(\delta_{tf})$ cannot be continuous at
$t=0$.

This shows there are two ways of obtaining the final physical algebra,
first, we can use Krein representations as studied in this subsection,
but these contain
pathologies (only dense *--subalgebras are represented, and these as unbounded operators),
or second, we can use nonregular representations -- which can still
produce regular representations on the final physical algebra -- 
and now the operator theory is much better understood.
Nonregular representations
avoid the problems spelled out by Strocchi's theorems~\cite{Strocchi67,
Strocchi70} because
due to the nonregularity, one cannot use Stone's theorem to obtain
generators for the one--parameter groups, hence the operators representing
the vector potential do not exist here.
This dichotomy between nonregular representations
and IIP--representations
was pointed out in previous papers,~\cite{Grundling88b, Thirring}.

%%%%%%%%%%%%%%%%%%%%%%%%%%%%%%%%%%%%%%%%%%%%%%%%%%%%%%%%%%%%%%%%%%%%%%%%%%%%%%
\section{Further topics}
\subsection{Global vs local constraining}
\label{GvsL}

For a system of local constraints $\B\to\big(\al F.(\B),\,\al U.(\B)
\big)$ as in Definition~\ref{loc.cons},
a natural question to ask is
the following. What is the relation between the limit algebra
$\al R._0:=\ilim\al R.(\B)$ and the algebra $\al R._e$
obtained from enforcing the full constraint set
$\mathop{\cup}\limits_{\B\in\Gamma}\al U.(\B)=:\al U._e$
in the quasi--local algebra $\al F._0$? In particular,
when will $\al R._0=\al R._e$? In other words, we compare
the local constrainings of the net to
a single global constraining.
(This has bearing on the BRST--constraining algorithm).

Now $\al R._e=\al O._e/\al D._e$ where as usual we have $\al D._e=
[\al F._0(\al U._e-\EINS)]\cap[(\al U._e-\EINS)\al F._0]$
and $\al O._e={\{F\in\al F._0\mid\,[F,\, D]\in\al D._e\;\,\forall\,
D\in\al D._e\}}$. 
\begin{teo}
\label{Loc.glob}
Let the system of local constraints
 $\B\to\big(\al F.(\B),\,\al U.(\B)
\big)$ have reduction isotony, then
$\al R._0:=\ilim\al R.(\B)=\al O._0\big/(\al D._e\cap\al O._0)$
where $\al O._0:=\ilim\al O.(\B)$. Moreover, there is an
injective homomorphism of $\al R._0$ into $\al R._e$.
\end{teo}
\begin{beweis}
First observe that $\al D._e\cap\al O._0=
[\al O._0(\al U._e-\EINS)]\cap[(\al U._e-\EINS)\al O._0]$
because $\al U._e\subset\al O._0$ hence every Dirac state
on $\al O._0$ extends to one on $\al F._0$, and the
$\al D.\hbox{--algebra}$ is characterised as the maximal
C*--algebra in the kernels of all the Dirac states.

Now denote by $\xi_{\B}:\al O.(\B)\to\al R.(\B)$ the 
factoring map by $\al D.(\B)$. Let $\B_1\subseteq\B_2$, 
then the diagram  
\begin{eqnarray*}
\al O.(\B_1)\kern3mm 
&\mathop{-\!\!\!-\!\!\!\longrightarrow}\limits^{{\rm inclusion}}
     & \kern3mm \al O.(\B_2)  \\[3mm]
        \Big\downarrow\, \xi_{\B_1}\kern3mm
                       & & 
       \kern6mm \Big\downarrow\, \xi_{\B_2} \\[3mm]
\al R.(\B_1)\kern3mm 
&\mathop{-\!\!\!-\!\!\!\longrightarrow}\limits^{\iota_{12}}
   & \kern3mm  \al R.(\B_2)  
\end{eqnarray*}
commutes by reduction isotony and the proof of Lemma~\ref{Lem.3.1}.
Thus there exists a surjective homomorphism for the inductive limit algebras:
$\;\xi_0:\al O._0\to\al R._0$, such that $\xi_0\restriction
\al O.(\B)=\xi_{\B}$, $\B\in\Gamma$. Clearly $\al D.(\B)\subset
\ker\xi_0$ for all $\B\in\Gamma$, hence $\al U._e-\EINS\subset
\ker\xi_0$, and so
by the previous paragraph
${\al D._e\cap\al O._0}\subseteq\ker\xi_0$.
Thus $\al D._e\cap\al O.(\B)\subseteq\ker\xi_0\cap\al O.(\B)
=\al D.(\B)$. Since $\al D.(\B)\subset\al D._e$ we conclude
$\al D._e\cap\al O.(\B)=\al D.(\B)$, and so
the global factoring map $\al O._0\to\al O._0\big/(
\al O._0\cap\al D._e)$ coincides on each $\al O.(\B)$ with 
$\xi_{\B}$. Thus it is $\xi_0$, i.e. $\ker\xi_0=\al D._e
\cap\al O._0$, so $\al R._0=\xi_0(\al O._0)
=\al O._0\big/(
\al O._0\cap\al D._e)$.

To prove the last claim, we just apply Lemma~\ref{Lem.3.1}
to the pair $(\al O._0,\,\al U._e)\subset(\al F._0,\,\al U._e)$.
To verify its two conditions, note that we already know by
the first part of the proof that
 $\al D._e\cap\al O._0=
[\al O._0(\al U._e-\EINS)]\cap[(\al U._e-\EINS)\al O._0]$
so we only need to check the second condition.
We also saw above that $\al D._e\cap\al O._0$
is an ideal in $\al O._0$ (the kernel of a homomorphism),
hence the algebra of observables in $\al O._0$ of the
constraints $\al U._e$ is all of $\al O._0$ (using
Theorem~\ref{Teo.2.2}(ii)). Thus we only need to show that
$\al O._0\subset\al O._e$, i.e. that $\al O.(\B)
\subset\al O._e$ for all $\B\in\Gamma$.
By Theorem~\ref{Teo.2.2}(iii) we only need to show that
$[F,\,\al U._e]\subset\al D._e$ $\forall\, F\in\al O.(\B)$,
but this follows immediately from the last paragraph since
for an observable $F\in\al O._0$ we have:
$[F,\,\al U._e]\subset\al D._e\cap\al O._0\subset\al D._e$.
\end{beweis}

We do not as yet have useful general criteria to ensure that
$\al R._0=\al R._e$, though we now verify that it holds for
both stages of constraining in the Gupta--Bleuler example.

\noindent\b Example.. Recall the first stage of constraining
in the previous example. We had a system of local constraints
$\Gamma\ni\B\to\big(\al F.(\B),\,\al U.(\B)\big)$ where
$\al F.(\B)={{\rm C}^*\left(\delta_{\ot X.(\B)}\cup\al U.(\B)\right)}$
and $\al U.(\B)={\big\{U_{T_h}\,\mid\, h\in C_c^\infty(\B,\,\R^4)
\big\}}$.
By Theorem~\ref{Observ} we have $\al R.(\B)\cong{\rm C}^*\big(\delta_{
\ot p.(\B)}\big)$ and so $\al R._0=\ilim\al R.(\B)=\ccr\ot p.,B.$
where $\ot p.=\mathop{\cup}\limits_{\B\in\Gamma}\ot p.(\B)$. 
We need to compare this 
to $\al R._e$ which we obtain from the system
$\big(\al F._0,\,\al U._e\big)$ where $\al F._0=\ilim\al F.(\B)
={\rm C}^*\big(\al U._e\cup\delta_{\ot Z._{(0)}}\big)$, 
 (cf. Remark~\ref{comp.supp}(i)) 
  and $\;\al U._e
=\mathop{\cup}\limits_{\B\in\Gamma}\al U.(\B)=\al U._{(0)}$. 
Now the method in the proof of Theorem~\ref{Observ} did not
use the assumption $\B\in\Gamma$, hence 
it can be transcribed to prove that $\al O._e={\rm C}^*\big(
\delta_{\ot p._e})+\al D._e$ where
$\ot p._e:=\big\{f\in\ot Z._{(0)}\,\mid\, T_h(f)=f\;\forall\,
h\in C_c^\infty(\B,\,\R),\;\B\in\Gamma\big\}$.
 Each $f\in\ot p._e\subset\ot Z._{(0)}$ is in some
$\ot Z.(\B)$, so is in $\ot p.(\B)$ by the defining condition.
Thus $\ot p._e\cap\ot Z.(\B)\subseteq\ot p.(\B)$.
However, by Theorem~\ref{Observ}, these are characterised by
$p_\mu f^\mu\restriction C_+=0$, and by Eqn.~(\ref{gauge})
this implies $T_h(f)=f$ for all $h\in C_c^\infty(\R^4,\,\R)$.
Thus $\ot p._e=\ot p.$, so $\al O._e={\rm C}^*\big(
\delta_{\ot p.})+\al D._e$ and hence by the argument in
the last part of the proof of Theorem~\ref{Observ} we have
$\al R._e\cong{\rm C}^*(\delta_{\ot p.})\cong\al R._0$.

Next we verify for the second stage of constraining that the
local and global constrainings ultimately coincide.
Here we have the system of local constraints:
$\Gamma\ni\B\to\big(\al R.(\B),\,\wt{\al U.}(\B)\big)$ where
$\al R.(\B)=\ccr\ot p.(\B),B.$ and 
$\wt{\al U.}(\B)=\delta_{\ot s.(\B)}$, $\ot s.(\B)=\ot p._0\cap
\ot p.(\B)$. By Theorem~\ref{finalR} we have 
$\wt{\al R.}_0\cong\ccr\ot p./\ot p._0,\wt{B}.$.
We need to compare this to the physical algebra
$\wt{\al R.}_e$ obtained from the system
$\big(\al R._0,\,\wt{\al U.}_e\big)$ where
 $\al R._0=\ilim\al R.(\B)=\ccr\ot p.,B.$
and $\wt{\al U.}_e=\mathop{\cup}\limits_{\B\in\Gamma}\wt{\al U.}(\B)
=\delta_{\ot p._0}$.
By Theorem~5.2 in \cite{Grundling88b} we have
$\wt{\al R.}_e\cong
\ccr\ot p./\ot p._0,\wt{B}.\cong\wt{\al R._0}$,
and this proves the claim.

\subsection{Reduction by stages}

In this subsection we address the problem of reduction by stages, i.e. 
subdivide the initial constraint set, then impose these constraint
sets along an increasing chain (terminating with the full set
of constraints),
and analyse when the final physical algebra of the chain is
the same as that obtained from a single constraining by the full set.
This problem occurred in the Gupta--Bleuler example, and is related
also to the one in the previous subsection.

\begin{defi}
An {\bf n--chain of constraints} consists of a first--class
constraint system $(\al F.,\,\al C.)$ and a chain of subsets
\[
\{0\}\not=\al C._1\subset\al C._2\subset\cdots\subset
\al C._n=\al C.
\]
such that $\al C.\subset\al O._i$ $\forall\; i=1,\,2,\ldots,\, n,$
where we henceforth denote by
${\big({\got S}_{D_i},\,\al D._i,\,\al O._i,\,\al R._i,\,\xi_i\big)}$
the data resulting from application of a T--procedure to
${(\al F.,\,\al C._i)}$. (Recall that $\xi_i:\al O._i\to
\al R._i$ denotes the canonical factorization map).
By convention we will omit the subscript $i$ when $i=n$.
\end{defi}
Note that ${\got S}_D={\got S}_{D_n}\subset{\got S}_{D_{n-1}}
\subset\cdots\subset{\got S}_{D_2}\subset{\got S}_{D_1}$
and $\al D._1\subset\al D._2\subset\cdots\subset
\al D._n=\al D.$. The condition $\al C.\subset\al O._i$
is nontrivial, but necessary for the procedure in the
next theorem. Below we will use subscript notation
$A_i$, $A_{(i)}$ and $A_{\{i\}}$ to distinguish 
between similar objects in different contexts.
\begin{teo}
\label{chain}
Given an n--chain of constraints as above, we 
define inductively the following
cascade of first--class constraint systems
${\big(\al R._{(k-1)},\,\xi_{\{k-1\}}(\al C._k)\big)}$,
$k=1,\ldots,\, n$ with T--procedure data
${\big({\got S}_{D(k)},\,\al D._{(k)}
,\,\al O._{(k)},\,\al R._{(k)},\,\xi_{(k)}\big)}$
and notation $\xi_{\{k\}}:=\xi_{(k)}\circ\xi_{(k-1)}\circ
\cdots\circ\xi_{(1)}$ and conventions
$\xi_{\{0\}}:={\rm id}$, $\al R._{(0)}=\al F.$. Then
\begin{itemize}
\item[(i)] ${\rm Dom}\,\xi_{\{k\}}=\al O._1\cap
\al O._2\cap\cdots\cap\al O._k=:\al O._{\{k\}}$,
${\rm Ker}\,\xi_{\{k\}}=\al O._{\{k\}}\cap\al D._k$,
and ${\rm Ran}\,\xi_{\{k\}}=\al R._{(k)}$
where we use the conventions $\al O._{\{0\}}:=\al F.$
and $\al D._0=\{0\}$.
\item[(ii)]
$\al D._{(k)}=\xi_{\{k-1\}}\big(\al O._{\{k-1\}}\cap
\al D._k\big)$,
\item[(iii)]
$\al O._{(k)}=\xi_{\{k-1\}}(\al O._{\{k\}})$,
\item[(iv)]
$\al R._{(k)}\cong\al O._{\{k\}}\big/
(\al O._{\{k-1\}}\cap\al D._k)\subset\al R._k$,
\item[(v)] the map $\varphi_k:{\got S}_{D_k}\restriction
\al O._{\{k\}}\to{\got S}(\al R._{(k)})$ defined by
$\varphi_k(\omega)\big(\xi_{\{k\}}(A)\big):=
\omega(A)$, $A\in\al O._{\{k\}}$, gives a bijection
${\varphi_k:{\got S}_{D_{k+1}}\restriction
\al O._{\{k\}}}\to{\got S}_{D(k+1)}$.
\end{itemize}
We will call the application of a T--procedure to
${\big(\al R._{(k-1)},\,\xi_{\{k-1\}}(\al C._k)\big)}$,
$k=1,\ldots,\, n$ to produce the  data
${\big({\got S}_{D(k)},\,\al D._{(k)}
,\,\al O._{(k)},\,\al R._{(k)},\,\xi_{(k)}\big)}$
the $k^{\rm th}\hbox{\bf stage reduction}$ of the given
n-chain.
\end{teo}
\begin{beweis}
We apply the second principle of induction, and also
remind the reader that $\al C.\subset\al O._i$ $\forall\, i$.  For $k=1$ we
have by convention that ${(\al R._{(0)},\,\xi_{\{0\}}(\al C._1))}
={(\al F.,\,\al C._1)}$ which is first--class, and
${({\got S}_{D(1)},\,\al D._{(1)},\,\al O._{(1)},\,
\al R._{(1)},\,\xi_{(1)})}=
{({\got S}_{D_1},\,\al D._{1},\,\al O._{1},\,
\al R._{1},\,\xi_{1})}$.
Thus ${\rm Dom}\,\xi_{\{1\}}={\rm Dom}\,\xi_1=\al O._1=\al O._{\{1\}}$,
${\rm Ker}\,\xi_{\{1\}}={\rm Ker}\,\xi_1=\al D._1=\al O._{\{1\}}\cap
\al  D._1$, ${\rm Ran}\,\xi_{\{1\}}={\rm Ran}\,\xi_1
=\al R._1=\al R._{(1)}$. Moreover
$\al D._{(1)}=\al D._1=\xi_{\{0\}}(\al O._{\{0\}}\cap
\al D._1)$, $\al O._{(1)}=\xi_{\{0\}}(\al O._{\{1\}})=\al O._1$
and $\al R._{(1)}=\al R._1=\al O._{\{1\}}/(\al O._{\{0\}}\cap\al D._1)$. 
Now using $\al C._2\subset\al C.\subset\al O._1$,
the bijection $\varphi_1:{\got S}_{D_1}\restriction\al O._1\to
{\got S}(\al R._1)$ given by Theorem~\ref{Teo.2.6},
  produces for  $\omega\in{\got S}_{D_2}$:
\[
  \varphi_1(\omega)\big(\xi_1(C^*C)\big)=\omega(C^*C)=0=
\omega(CC^*)=\varphi_1(\omega)\big(\xi_1(CC^*)\big)
\]
for all $C\in\al C._2$, i.e. $\varphi_1(\omega)\in{\got S}_{D(2)}$.
Conversely, if $\varphi_1(\omega)\in{\got S}_{D(2)}$ then 
$\omega\in{\got S}_{D_2}$.
Thus the theorem holds for $k=1$.

For the induction step, fix an integer $m\geq 1$ and assume the theorem
is true for all $k\leq m$. We prove that it holds for $m+1$.
Now ${\big(\al R._{(m)},\,\xi_{\{m\}}(\al C._{m+1})\big)}$ 
is first--class, because by (v), $\varphi_m({\got S}_{D_{m+1}}
\restriction\al O._{\{m\}})={\got S}_{D(m+1)}\not=\emptyset$
since by $\al C._{m+1}\subset\al C.\subset\al O._i$
we have $\emptyset\not={\got S}_D\restriction\al O._{\{m\}}$
and $\varphi_m$ is a bijection. We first prove (ii).
\begin{eqnarray*}
\al D._{(m+1)}
  \!\!\!&=&\!\!\! 
           \Big\{\xi_{\{m\}}(F)\mid F\in\al O._{\{m\}}\quad\hbox{and}\\
  \!\!\!& &\!\!\! 
           \kern1.7cm \omega(\xi_{\{m\}}(F^*F))=0=\omega(\xi_{\{m\}}(FF^*))
           \;\forall\,\omega\in{\got S}_{D(m+1)}\Big\}\\
  \!\!\!&=& \!\!\!
            \Big\{\xi_{\{m\}}(F)\mid F\in\al O._{\{m\}}\quad\hbox{and}\\
  \!\!\!& &\!\!\!
            \kern1.7cm \widehat\omega(F^*F)=0=\widehat\omega(FF^*)
            \;\forall\,\widehat\omega\in{\got S}_{D_{m+1}}\Big\}\\ 
  \!\!\!& &\!\!\! \hbox{(using (v) of induction assumption)}\\
  \!\!\!&=&\!\!\! \xi_{\{m\}}\big(\al D._{m+1}\cap\al O._{\{m\}}\big)\;.
\end{eqnarray*}
For (iii) we see:
\begin{eqnarray*}
\al O._{(m+1)}
  \!\!\!&=&\!\!\!
           \Big\{\xi_{\{m\}}(F)\mid F\in\al O._{\{m\}}\quad\hbox{and}\\
  \!\!\!& &\!\!\!
        \kern1.2cm\big[\xi_{\{m\}}(F),\,\xi_{\{m\}}(\al C._{m+1})\big]\subset
        \al D._{(m+1)}\Big\}\,\hbox{(by Theorem~\ref{Teo.2.2}(iii))}\\
  \!\!\!&=&\!\!\!\Big\{\xi_{\{m\}}(F)\mid F\in\al O._{\{m\}}\quad\hbox{and}\\
  \!\!\!& &\!\!\!\kern1.7cm\xi_{\{m\}}\big([F,\,\al C._{m+1}]\big)\subset
     \xi_{\{m\}}(\al D._{m+1}\cap\al O._{\{m\}})\Big\}\\
  \!\!\!&=&\!\!\!\Big\{\xi_{\{m\}}(F)\mid F\in\al O._{\{m\}}\quad\hbox{and}\\
  \!\!\!& &\!\!\!\kern1.7cm [F,\,\al C._{m+1}]\subset
     \al D._{m+1}\cap\al O._{\{m\}}+\al O._{\{m\}}\cap\al D._m\Big\}\;.
\end{eqnarray*}
 Now since $\al D._m\subset\al D._{m+1}$, we have
$\al D._{m+1}\cap\al O._{\{m\}}+
\al O._{\{m\}}\cap\al D._m=\al D._{m+1}\cap\al O._{\{m\}}$ and
$\al C._{m+1}\subset\al O._{\{m\}}$ and so
$[F,\,\al C._{m+1}]\subset\al O._{\{m\}}$ for all
$F\in\al O._{\{m\}}$. Thus
\begin{eqnarray*}
\al O._{(m+1)}
  &=& \Big\{\xi_{\{m\}}(F)\;\mid\; F\in\al O._{\{m\}}\quad\hbox{and}\quad
[F,\,\al C._{m+1}]\subset\al D._{m+1}\Big\}\\
  &=& \xi_{\{m\}}(\al O._{\{m\}}\cap\al O._{m+1})
  = \xi_{\{m\}}(\al O._{\{m+1\}})\;.
\end{eqnarray*}
For (iv), note that 
\[
  \al R._{(m+1)}=\al O._{(m+1)}\big/\al D._{(m+1)}
  =\xi_{\{m\}}(\al O._{\{m+1\}})\Big/\xi_{\{m\}}(
  \al D._{m+1}\cap\al O._{\{m\}})\;.
\]
Define a map $\psi:\al R._{(m+1)}\to\al R._{m+1}$ by
\[
  \psi\left(\xi_{\{m\}}(A)+\xi_{\{m\}}(\al D._{m+1}
\cap\al O._{\{m\}})\right):=
   A+\al D._{m+1}\;\,\quad A\in\al O._{\{m+1\}}\;.
\]
To see that it is well--defined, let 
$B\in\al O._{\{m+1\}}$ be such that
  $\xi_{\{m\}}(A)-\xi_{\{m\}}(B)\in
\xi_{\{m\}}(\al D._{m+1}\cap\al O._{\{m\}})$, i.e.
$A-B\in\al D._{m+1}\cap\al O._{\{m\}}+
\al O._{\{m\}}\cap\al D._m=\al D._{m+1}\cap
\al O._{\{m\}}$, and so
\[
  \psi\left(\xi_{\{m\}}(B)+\xi_{\{m\}}(\al D._{m+1}
\cap\al O._{\{m\}})\right)=
   B+\al D._{m+1}=
   A+\al D._{m+1}\;.
\]   
It is easy to see that $\psi$ is a *--homomorphism onto the subalgebra
$\al O._{\{m+1\}}\big/\al D._{m+1}\cap\al O._{\{m\}}
\subset\al R._{m+1}$ and since ${\rm Ker}\,\psi
=\xi_{\{m\}}(\al D._{m+1}\cap\al O._{\{m\}})=\al D._{(m+1)}$
which is the zero of $\al R._{(m+1)}$, $\psi$ is a monomorphism.

To prove (i), recall that $\xi_{\{m+1\}}=\xi_{(m+1)}\circ\xi_{\{m\}}$, so
\begin{eqnarray*}
{\rm Dom}\,\xi_{\{m+1\}}
  &=& \Big\{F\in{\rm Dom}\,\xi_{\{m\}}\;\mid\; 
\xi_{\{m\}}(F)\in{\rm Dom}\,\xi_{(m+1)}=\al O._{(m+1)}\Big\}\\
  &=& \Big\{F\in\al O._{\{m\}}\;\mid\; 
\xi_{\{m\}}(F)\in\xi_{\{m\}}(\al O._{\{m+1\}})\Big\}\\
  &=& \Big\{F\in\al O._{\{m\}}\;\mid\; 
F\in\al O._{\{m+1\}}+\al O._{\{m\}}\cap\al D._m=
\al O._{\{m+1\}}\Big\}
\end{eqnarray*}
because $\al D._m\subset\al D._{m+1}\subset\al O._{m+1}$.
Thus ${\rm Dom}\,\xi_{\{m+1\}}=\al O._{\{m+1\}}$. Now
\begin{eqnarray*}
{\rm Ker}\,\xi_{\{m+1\}}
  &=& \Big\{F\in{\rm Dom}\,\xi_{\{m+1\}}\;\mid\; 
\xi_{\{m\}}(F)\in{\rm Ker}\,\xi_{(m+1)}=\al D._{(m+1)}\Big\}\\
  &=& \Big\{F\in\al O._{\{m+1\}}\;\mid\; 
\xi_{\{m\}}(F)\in\xi_{\{m\}}(\al D._{m+1}\cap\al O._{\{m\}})\Big\}\\
  &=& \Big\{F\in\al O._{\{m+1\}}\;\mid \\
  & &\kern.5cm F\in\al D._{m+1}\cap\al O._{\{m\}}+\al O._{\{m\}}\cap\al D._m
     =\al D._{m+1}\cap\al O._{\{m\}}\Big\}\\
  &=& \al O._{\{m+1\}}\cap\al D._{m+1}\;.\\
{\rm Ran}\,\xi_{\{m+1\}}
  &=& \xi_{(m+1)}\left(\xi_{\{m\}}({\rm Dom}\,\xi_{\{m+1\}})\right)
  = \xi_{(m+1)}\left(\xi_{\{m\}}(\al O._{\{m+1\}})\right)\\
  &=& \xi_{(m+1)}(\al O._{\{m+1\}})=\al R._{(m+1)}\;.
\end{eqnarray*}
Finally, to prove (v), since $\varphi_{m+1}$ is a surjection,
each $\omega\in{\got S}_{D(m+2)}$ is of the form
$\omega=\varphi_{m+1}(\widehat\omega)$ for some
$\widehat\omega\in{\got S}_{D_{m+1}}\restriction\al O._{\{m+1\}}$.
Now $\omega\in{\got S}_{D(m+2)}$ iff
$\omega\big(\xi_{\{m+1\}}(C^*C)\big)=0=
\omega\big(\xi_{\{m+1\}}(CC^*)\big)$ for all $C\in\al C._{m+2}$
iff $\widehat\omega(C^*C)=0=\widehat\omega(CC^*)$ for all 
$C\in\al C._{m+2}$ iff $\widehat\omega\in{\got S}_{D_{m+2}}
\restriction\al O._{\{m+1\}}$.
\end{beweis}

%\begin{cor}
%\label{stages}
%Given an n--chain of constraints as above, a sufficient
%condition to obtain $\al R._n\cong\al R._{(n)}$ is that
%$\al O._n\backslash\al O._{\{n\}}\subset\al D._n$.
%\end{cor}
%\begin{beweis}
%\begin{eqnarray*}
%\al R._n &=& \al O._n\big/\al D._n
%= \al O._{\{n\}}\big/\big(\al D._n\cap\al O._{\{n\}}\big)
%\qquad\quad\hbox{(since
%$\al O._n\backslash\al O._{\{n\}}\subset\al D._n$)}\\
%&=& \al O._{\{n\}}\big/\big(\al D._n\cap\al O._{\{n-1\}}\big)
%\cong\al R._{(n)}
%\end{eqnarray*}
%where we used Theorem~\ref{chain}(iv) for the last isomorphism.
%\end{beweis}
\bigskip\noindent{\bf Example.}
The 
Gupta--Bleuler model of the previous section
 provide examples of 2--chains of constraints
both at the local and the global levels.
We will only consider the global level, and refer freely
to the example of the last section where both global constrainings were
done. 
Let $\al C._1:=
 \{U_{T_h}-\EINS\mid h\in C_c^\infty(\R^4,\, \R)\}$
and let the total constraint set in $\al F._0$
be $\al C.=\al C._2:=\al C._1\cup\wt\al C.$ where
$\wt\al C.:=\EINS-\wt{\al U.}_e=
\{\EINS-\delta_f\mid f\in\ot p._0\}$.
\begin{claim} 
\label{GB2Chain}
$\quad\al C._1\subset\al C._2=\al C.$ is a 2--chain of constraints
in $\al F._0={\rm C}^*\big(\delta_{\ot Z._{(0)}}\cup\al U._{(0)}\big)$.
(Notation as in Remark~\ref{comp.supp}(i)).
\end{claim}
\begin{beweis} To see that $\al C.$ is first--class,
define a state $\widehat\omega$ on $\ccr\ot Z._{(0)},B.\subset\al F._0$
by $\widehat\omega(\delta_f)=1$ if $f\in\ot p._0 $, and 
otherwise $\widehat\omega(\delta_f)=0$ (that this defines a state is 
easy to check).
Since by Theorem~\ref{Observ} the space $\ot p.$ is pointwise invariant
under $T_h$ for $h\in C_c^\infty(\R^4,\,\R)$, 
(also using Eqn.~(\ref{gauge})) so is $\ot p._0$,
hence $\widehat\omega$ is invariant under $\al G._{(0)}=$
the group generated by $\al G.(\B)$, $\B\in\Gamma$.
Thus $\widehat\omega$ extends (uniquely) to a Dirac state on $\al F._0$
(by a trivial application of Theorem~\ref{Outer}).
Thus $\al C.$ is first class.

It is obvious that $\al C.\subset\al D._2\subset\al O._2$,
so we only need to show that $\al C.\subset\al O._1$.
By Theorem~\ref{Observ} and the last subsection
 we have $\al O._1={\rm C}^*(\delta_{
\ot p.})+\al D._1$ and as $\al C._1\subset\al D._1$
and $\delta_{\ot p._0}\subset{\rm C}^*(\delta_{\ot p.})$
it follows that $\al C.\subset\al O._1$.
\end{beweis}
%We now show that $\al O._2\big\backslash\al O._{\{2\}}
%\subset\al D._2$, i.e. since $\al O._{\{2\}}=\al O._1\cap
%\al O._2$, that $\al O._2\backslash\al O._1\subset\al D._2$.
%Then by Corollary~\ref{stages} we have $\al R._2=\al R._{(2)}$, i.e. 
%the two--step reduction by stages produces the same physical 
%algebra as a single reduction by the full constraint set $\al C.$.
%This, we will do as follows. We know that 
%from the last subsection that $\al O._1={\rm C}^*(\delta_{
%\ot p.})+\al D._1$ and in the next two claims we prove
% that $\al O._2={\rm C}^*(\delta_{
%\ot p.})+\al D._2$. Then since $\al D._1\subset\al D._2$
%it follows that $\al O._2\backslash\al O._1\subset\al D._2$. [MISTAKE!!!]

Now we want to show that $\al R._2=\al R._{(2)}$, i.e.
the two--step reduction by stages produces the same physical
algebra as a single reduction by the full constraint set $\al C.$.
Recall that by Theorem~\ref{chain}(iv), we have a monomorphic 
imbedding:
\[
\al R._{(2)}\cong\al O._{\{2\}}\big/(\al O._{\{1\}}\cap\al D._2)
\subset\al R._2=\al O._2\big/\al D._2
\]
where $\al O._{\{2\}}=\al O._1\cap\al O._2$, $\al O._1=\al O._{\{1\}}$.
So we will have the desired isomorphism  $\al R._2=\al R._{(2)}$,
if we can show that this imbedding is surjective.
Now we know from the last subsection that $\al O._1=C^*(\delta_{\ot p.})
+\al D._1$, and below in the next two claims we prove that
$\al O._2=C^*(\delta_{\ot p.})+\al D._2$.
Then since $\al D._1\subset\al D._2$ we have $\al O._1\subset\al O._2$,
so
\[
\al R._{(2)}\cong\al O._1\big/(\al O._1\cap\al D._2)
\subset\al R._2=\al O._2\big/\al D._2\;.
\]
Now note that each equivalence class corresponding to an element of $\al R._2$
is of the form $A+\al D._2$ with $A\in C^*(\delta_{\ot p.})$,
and this contains the equivalence class of  $A+\al D._1$ from
$\al O._1\big/(\al O._1\cap\al D._2)$. So the imbedding is surjective.

It remains to prove that $\al O._2=C^*(\delta_{\ot p.})+\al D._2$.
We first prove:
\begin{claim}   
\chop $\al O._2={\rm C}^*(\delta_{
\ot p._0'})+\al D._2\qquad\hbox{where}\qquad\ot p._0'={\{f\in\ot Z._{(0)}\,\mid\,
B(f,\, s)=0\quad\forall\, s\in\ot p._0\}}\;$.
\end{claim}
\begin{beweis}
We adapt the proof of Theorem~\ref{Observ}.
Since $\al O._2\subset\al F._0={\rm C}^*\big(\delta_{\ot Z._{(0)}}
\cup\al U._{(0)}\big)$, for a general $A\in\al O._2$ we can
write
\begin{equation}
\label{(1)}
A=\lim_{n\to\infty}A_n\qquad\hbox{where}\qquad
A_n=\sum_{i=1}^{N_n}\delta_{f_i}\sum_{j=1}^{L_n}
\lambda_{ij}^{(n)}U_{\gamma_{ij}^{(n)}}
\end{equation}
where $f_i\not= f_j$ if $i\not= j$, $f_i\in\ot Z._{(0)}$,
$\gamma_{ij}^{(n)}\in\al G._{(0)}$ and 
$\lambda_{ij}^{(n)}\in\C$. Consider the
equivalence classes of $\ot Z._{(0)}\big/\ot p._0$.
If $f_i-f_j=:s\in\ot p._0$, then
$\delta_{f_i}=\delta_{f_j}\cdot\delta_s\exp\big(iB(s,\, f_j)/2\big)$
 and $\delta_s\in\wt{\al U.}_e$. Thus we can write 
Eqn.~(\ref{(1)}) in the form
\begin{equation}
\label{(2)}
A_n=\sum_{i=1}^{N_n}\delta_{f_i}\sum_{j=1}^{L_n}
\sum_{k=1}^{K_n}
\lambda_{ijk}^{(n)}\delta_{s_{ij}}U_{\gamma_{ik}^{(n)}}
\end{equation}
where $f_i-f_j\not\in\ot p._0$ if $i\not=j$ and 
$s_{ij}\in\ot p._0$. Let $\omega\in\ot S._{D_2}$, then
\begin{eqnarray}
   \pi_\omega(A_n)\Omega_\omega   &=&
  \sum_{i=1}^{N_n}\zeta_i^{(n)}\,\pi_\omega(\delta_{f_i})\,\Omega_\omega
   \label{(3)}\\
\hbox{where}\qquad\quad
  \zeta_i^{(n)}  &=&
\sum_{j=1}^{L_n}
\sum_{k=1}^{K_n}
\lambda_{ijk}^{(n)}\in\C\;.\nonumber
\end{eqnarray}
Let $h\in\ot p._0$, then from $\delta_h\in\wt{\al U.}_e$ and 
$A\in\al O._2$, we get, using Eqn.~(\ref{(3)}):
\begin{eqnarray}
0 &=& \omega\big(A^*(\delta_h-\EINS)^*(\delta_h-\EINS)A\big)
=\lim_{n\to\infty}\omega\big(A_n^*(2\EINS-\delta_h-\delta_{-h})A_n\big)
\nonumber\\
&=&\lim_{n\to\infty}\Big\{2\omega(A_n^*A_n)-
\sum_{i,\,j=1}^{N_n}\overline{\zeta}_i^{(n)}
\zeta_j^{(n)}\,\omega\big(\delta_{-f_i}(\delta_h+\delta_{-h})
\delta_{f_j}\big)\Big\}
\label{(4)}
\end{eqnarray}
The state $\ww$ on $\al F._0$
defined by $\ww(\delta_f):=\chi_{\ot p._0}(f)$
and $\ww(U_g)=1$ $\forall\, g\in\al G._{(0)}$
(encountered in
 the proof of Claim~\ref{GB2Chain}) 
is in $\ot S._{D_2}$, so Eqn.~(\ref{(4)}) becomes for it:
\begin{eqnarray}
0 &=& \lim_{n\to\infty}\Big\{2\sum_{j=1}^{N_n}|\zeta_j^{(n)}|^2
-\sum_{j=1}^{N_n}|\zeta_j^{(n)}|^2\big(e^{iB(h,\,f_j)}+
e^{-iB(h,\,f_j)}\big)\Big\}\nonumber\\
 &=& 2\lim_{n\to\infty}\sum_{j=1}^{N_n}|\zeta_j^{(n)}|^2
\Big(1-\cos\,B(h,\, f_j)\Big)\label{(5)}
\end{eqnarray}
where we made use of $f_i-f_j\not\in\ot p._0$ if
$i\not=j$ and the equation
\begin{eqnarray*}
\pi_\omega\big(\delta_{-f_i}\delta_h\delta_{f_j}\big)\,\Omega_\omega
  &=& \pi_\omega\big(\delta_{-f_i}e^{iB(h,\,f_j)}
      \delta_{f_j}\delta_h\big)\,\Omega_\omega \\
  &=& e^{iB(h,\,f_j)}e^{-iB(f_i,\, f_j)/2}\;
      \pi_\omega\big(\delta_{f_j-f_i}\big)\,\Omega_\omega\;.
\end{eqnarray*}
Now the terms in the sum of Eqn.~(\ref{(5)}) are all positive
so in the limit these must individually vanish, i.e.
\[
\lim_{n\to\infty}|\zeta_j^{(n)}|^2\big(1-\cos\,B(h,\, f_j)\big)=0\qquad
\forall\; h\in\ot p._0\;.
\]
Thus since the first factor is independent of $h$ and the
second is independent of $n$, either
$\lim\limits_{n\to\infty}|\zeta_j^{(n)}|^2=0$ or
$B(h,\, f_j)=0$ $\;\forall\, h\in\ot p._0$
(i.e. $f_j\in\ot p._0'$). Now we can rewrite the argument
around Eqn.~(\ref{later}) almost verbatim to obtain
$\al O._2={\rm C}^*(\delta_{\ot p._0'})+\al D._2$.
\end{beweis}
\begin{claim} $\ot p._0'=\ot p.$.
\end{claim}
\begin{beweis}
Clearly $\ot p._0'\supseteq\ot p.$ by definition.
For the converse inclusion, let $f\in C_c^\infty(\R^4,\,\R^4)$
such that $\rho(f):=\wh{f}\restriction C_+\in\ot p._0'$,
i.e. we have $B(\rho(f),\,\rho(k))=0$
$\forall\,\rho(k)\in\ot p._0$. Now by Proposition~\ref{Flemma},
if $\rho(k)$ is in $\ot p._0$, then $\rho(k)_\mu(p)=
ip_\mu h(p)$ for some $h$ such that $\rho(k)\in\ot Z._{(0)}$.
In particular, we can take $h=\rho(r)$ for $r\in C^\infty_c(\R^4,\, \R)$,
in which case $\rho(k)_\mu=\rho(\partial_\mu r)$, and so
\begin{eqnarray*}
0 &=& B(\rho(f),\,\rho(k))=\widehat{D}(f,\, k) \\
&=& \int\int f_\mu(x)\, k^\mu(y)\,D(x-y)\, d^4x\, d^4y \\
&=& \int\int f_\mu(z+y)\, k^\mu(y)\,D(z)\, d^4z\, d^4y\\
&=& \int\Big(\int f_\mu(z+y)\, {\partial r(y)\over\partial y_\mu}
\, d^4y\Big)\,D(z)\, d^4z\\
&=& -\int\Big(\int (\partial^\mu f_\mu)(z+y)\,
r(y)\, d^4y\Big)\,D(z)\, d^4z\\
&=& -\int\int (\partial^\mu f_\mu)(x)\,
r(y)\, D(x-y)\, d^4y\, d^4x\\
&=& -\wt{D}\big(\rho(\partial^\mu f_\mu),\,\rho(r)\big)
\end{eqnarray*}
for all $r\in C_c^\infty(\R^4,\, \R)$ and where $D$
is the Pauli--Jordan distribution and
 $\wt{D}$ is the symplectic form for the free neutral scalar
bosonic field. It is well--known that $\wt{D}$ is nondegenerate
on $\rho\big(C_c^\infty(\R^4,\, \R)\big)$,
(to see this, use the Schwartz density argument in the proof of
Proposition~\ref{Flemma})
hence since $\rho(\partial^\mu f_\mu)$ is also in this space, we
conclude $\rho(\partial^\mu f_\mu)=0$, i.e.
$p^\mu\rho(f_\mu)=0$, i.e. by Eqn.~(\ref{gauge})
$\rho(f)\in\ot p.$. Hence we have the reverse inclusion, so
$\ot p._0'=\ot p.$.
\end{beweis}
Thus $\al O._2={\rm C}^*(\delta_{\ot p.})+\al D._2$
and so $\al R._2\cong\al R._{(2)}$.

\subsection{The weak spectral condition}

A very important additional property which is used in
the analysis of algebraic QFT, is that of the spectral condition.
\begin{defi}
\label{spectralC}
An action $\alpha:\al P._+^\uparrow\to\aut\al F._0$
on a C*--algebra $\al F._0$ satisfies the {\bf spectral condition}
if there is a state $\omega\in\ot S.(\al F._0)$ such that
\begin{itemize}
\item[(i)] $\omega$ is translation--invariant, i.e. 
$\omega\circ\alpha_g=\omega$ $\forall\, g\in\R^4\subset\al P._+^\uparrow$,
\item[(ii)] the spectrum of the generators of translations
in $\pi_\omega$ is in the forward light cone $V_+$.
\end{itemize}
\end{defi}
Let now $\B\to\big(\al F.(\B),\,\al U.(\B)\big)$ be a system of local constraints
with reduction isotony and weak covariance. We want to find the
weakest requirement on 
 $\B\to\big(\al F.(\B),\,\al U.(\B)\big)$  to ensure that
$\wt\alpha:\al P._+^\uparrow\to\aut\al R._0$ (cf. Theorem~\ref{Teo.4.2}(iii))
satisfies the spectral condition. We propose:
\begin{defi}
The given action $\alpha:\al P._+^\uparrow\to\aut\al F._0$
on $\al F._0=\ilim\al F.(\B)$ satisfies the {\bf weak spectral condition}
iff the set
\[
  \al C.:=(\al U._e-\EINS)\cup\big\{\alpha_f(A)\,\mid\, A\in\al O._0,\;
   f\in F(V_+)\big\}\;\subset\al O._0:=\ilim\al O.(\B)
\]
is a first--class constraint set, where we used the notation
\begin{eqnarray*}
\al U._e&:=&\bigcup_{\B\in\Gamma}\al U.(\B),\qquad\quad
\alpha_f(A):=\int_{\R^4}\alpha_{t}(A)\,f(t)\,d^4t\;, \\
F(V_+) &:=&\big\{f\in L^1(\R^4)\,\mid\,{\rm supp}\,\wh{f}
\subset\R^4\backslash V_+\quad\hbox{and
$\;{\rm supp}\,\wh{f}\;$ is compact}\big\}
\end{eqnarray*}
where $\wh{f}$ denotes the Fourier transform of $f$.
\end{defi}
\begin{teo}
Let  $\B\to\big(\al F.(\B),\,\al U.(\B)\big)$  be a local system
of contraints with reduction isotony and weak covariance.
Then $\wt\alpha:\al P._+^\uparrow\to\aut\al R._0$ satisfies the
spectral condition iff
 $\alpha:\al P._+^\uparrow\to\aut\al F._0$
satisfies the weak spectral condition.
\end{teo}
\begin{beweis}
We will use the notation in Subsection~\ref{GvsL}
By the last definition, $\alpha$ satisfies the weak spectral condition
iff $\al C.$ is first--class iff the left ideal $[\al O._0\al C.]$
in $\al O._0$ is proper iff the left ideal
$\xi_0([\al O._0\al C.])=[\al R._0\xi_0(\al C.)]$
is proper in $\al R._0$ where $\xi_0:\al O._0\to\al O._0\big/(\al D._e
\cap \al O._0)=\al R._0$ is the canonical factoring map,
and we used $\ker\xi_0=\al D._e\cap\al O._0\subset
[\al O._0(\al U._e-\EINS)]\subset[\al O._0\al C.]$.
Now $\xi_0(\al C.)=
  {\big\{\xi_0(\alpha_f(A))\,\mid\, A\in\al O._0,\;
   f\in F(V_+)\big\}}$ because $\xi_0(\al U._e-\EINS)=0$.
Let $A\in\al O._0$ and $f\in F(V_+)$, then
\begin{eqnarray*}
\xi_0(\alpha_f(A)) &=&  \xi_0\Big(\int_{\R^4}\alpha_{t}(A)\,
f(t)\,d^4t\Big) =
\int_{\R^4}\xi_0\big(\alpha_{t}(A)\big)\,
f(t)\,d^4t\\
&=&  \int_{\R^4}\wt\alpha_{t}(\xi_0(A))\,
f(t)\,d^4t = \wt\alpha_f(\xi_0(A))\;.
\end{eqnarray*}
Thus $[\al R._0\xi_0(\al C.)]$ is the left ideal generated in
$\al R._0$ by
  ${\big\{\wt\alpha_f(B))\,\mid\, B\in\al R._0,\;
   f\in F(V_+)\big\}}$  and this is precisely Doplicher's left ideal
(cf.~\cite{Dop,Sakai}), which is proper iff $\wt\alpha$
satisfies the spectral condition by 2.7.2 in Sakai~\cite{Sakai}.
\end{beweis}
In general the weak spectral condition seems very difficult to
verify directly. However, for the Gupta--Bleuler example in this paper, 
it is easily verified via the last theorem:\medskip\hfill\break
{\bf Example.} 
We show that the final HK--QFT of the Gupta--Bleuler example
(as expressed in Theorem~\ref{finalR}) satisfies the
spectral condition, and hence the initial system must 
satisfy the weak spectral condition.
Thus we need to
show the existence of a state on $\wt{\al R.}_0:=\ilim\wt{\al R.}(\B)$
which satisfies the two conditions in Definition~\ref{spectralC}
for the action $\wt\alpha:\al P._+^\uparrow\to\aut\wt{\al R.}_0$.
Now the usual Gupta--Bleuler theory studied in Subsection~\ref{contact}
produced a Fock state $\omega_0$ on  $\wt{\al R.}_0=\ccr\ot p./\ot p._0,\wt{B}.$
given by the formula
\[
  \omega_0(\delta_{\zeta(f)}):=\exp\big(-K(f,\, f)\big/4\big)\;,
\qquad f\in\ot p.
\]
where  $\zeta:\ot p.\to\ot p./\ot p._0$
is the usual factoring map, and we want to show that it satisfies the
spectral condition. We already know that $K$ produces a Hilbert inner
product on $\ot p./\ot p._0$, and that it is Poincar\'e invariant
w.r.t.  $\wt{V}_g$ which denotes the factoring of
$V_g$ to $\ot p./\ot p._0$. Thus
the implementer of a Poincar\'e transformation $g$ is just the 
second quantization  of  $\wt{V}_g$,
i.e. $U_g:=\Gamma(\wt{V}_g)$.
We need to verify the spectral condition for the generators of the
translations.
Recall for $g=(\Lambda,\, a)$ we have $(V_gf)(p)=e^{-ipa}\,\Lambda
f(\Lambda^{-1}p)$. Translation by $a$ therefore acts by
multiplication operators $(V_af)(p)=e^{-ipa}f(p)$ with infinitesimal
generators $\wt{P}_\mu$ of $\wt{V}_a$ being the factoring
to $\ot p./\ot p._0$ of the multiplication operators $ f(p)\to p_\mu f(p)$.
Now for $f\in\ot p.$:
\[
  K\big(\zeta(f),\,\wt{P}_0\,\zeta(f)\big)
   =-2\pi\int_{C_+}\overline{{f}^\mu(p)}\,p_0\,{f}_\mu(p)\,
{d^3p\over p_0}
   =-2\pi\int_{C_+}\overline{{f}^\mu(p)}\,{f}_\mu(p)\,
d^3p   \geq 0
\]
since we have shown in the proof of Proposition~\ref{pos-IIP} that
   $\overline{{f}^\mu(p)}\,{f}_\mu(p)\leq 0$ $\;\forall
f\in\ot p.$, $p\in C_+$. So $\wt{P}_0\geq 0$. Since $U_a=\Gamma(\wt{V}_a)
=\exp\big(-ia_\mu\, d\Gamma(\wt{P}^\mu)\big)$, the generators for translation
for $\omega$ are $d\Gamma(\wt{P}^\mu)$, and so
 $\wt{P}_0\geq 0$ implies
 $d\Gamma(\wt{P}_0)\geq 0$.

To conclude, notice from the fact that $\wt{P}_\mu$ acts
on $\ot p./\ot p._0$ and in $\ot p.$ we have restriction
to $C_+$, that the spectrum of $\wt{P}_\mu$ must be in $C_+$.
Now in second quantization on an n--particle space:
\[
  d\Gamma(\wt{P}_\mu) = \wt{P}_\mu\otimes\EINS\otimes\cdots\otimes\EINS
  +\EINS\otimes \wt{P}_\mu\otimes\EINS\otimes\cdots\otimes\EINS
  +\cdots+
  \EINS\otimes\cdots\otimes\EINS\otimes\wt{P}_\mu
\]
and the spectrum for this will be all possible sums of $n$ vectors
in $C_+$, and this will always be in $V_+$. Since the spectrum for
the full $d\Gamma(\wt{P}_\mu)$ is the sum over all those
on the n--particle spaces, this will still be in $V_+$ since
$V_+$ is a cone. Thus we have verified the spectral condition as claimed.
 
%%%%%%%%%%%%%%%%%%%%%%%%%%%%%%%%%%%%%%%%%%%%%%%%%%%%%%%%%%%%%%%%%%
\section{Conclusions.}
In this paper we introduced the concept of a system of
local quantum constraints and we obtained a ``weak'' version
of each of the Haag--Kastler axioms of isotony, causality, covariance
and spectrality in such a way that after a local constraining
procedure the resulting system of physical algebras satisfies
the usual version of these axioms. We analyzed Gupta--Bleuler electromagnetism
in detail and showed that it satisfies these weak axioms, but that it
violates the causality axiom. This example was particularly
satisfying, in that we obtained by pure C*--algebra techniques
the correct physical algebra and positive energy Fock--representation
without having to pass through an indefinite metric representation.
We did however also point out the precise connection with the usual
indefinite metric representation.

There are some further
aspects of our Gupta--Bleuler example which are of independent
interest. These are:
\begin{itemize}
\item[(1)] the use of nonlinear constraints $\chi(h)^\d\chi(h),$ which we realised
as automorphisms on the field algebra (outer constraint situation),
\item[(2)] a nonstandard extension of our smearing formulii to complex--valued functions
(cf. Remark~\ref{comp.supp}(ii)),
which implied noncausal behaviour on nonphysical objects, where the 
latter were eliminated in the final theory (hence the need for
weak causality),
\item[(3)] the use of nonregular representations, but as in the last point 
the nonregularity was restricted to nonphysical objects.
\end{itemize}

There are many future directions of development for this project,
and a few of the more evident ones are:
\begin{itemize}
\item{} Find an example of a realistic constrained local field theory which
satisfies the weak Haag--Kastler axioms, but violates the usual
covariance axiom (a variant of the Coulomb gauge may work).
\item{} Continue the analysis here for the rest of the Haag--Kastler
axioms, i.e. find the appropriate weak versions of e.g.
the axioms of additivity, local normality, local definiteness etc.,
as well as examples which satisfy the weak axioms but not the usual ones.
\item{} In the present paper we assumed a system of local constraints 
which is first--class. Now a reduction procedure at the C*--level
exists also for second class constraints (cf.~\cite{Grundling88a})
and so one can therefore ask what the appropriate weakened form of the
Haag--Kastler axioms should be for such a system.
A possible example for such an analysis is electromagnetism in the 
Coulomb gauge.
\item{} Develop a QFT example with nonlinear constraints.
This is related to the difficulty of abstractly defining the C*--algebra
of a QFT with nontrivial interaction.
Our Gupta--Bleuler example has several similarities with Dimock's version
of a Yang--Mills theory on the cylinder~\cite{Dimock3}, and so this
seems to be a possible candidate for further development.
\end{itemize}
%%%%%%%%%%%%%%%%%%%%%%%%%%%%%%%%%%%%%%%%%%%%%%%%%%%%%%%%%%%%%%%%%%%%%%%%%%%%%%
\section{Appendix 1}
\label{Hereditary}

Next we wish to gain further understanding of the
algebras $\al D.,\;\al O.,\;\al R.$ by
exploiting the hereditary property of $\al D.$. 
Denote by $\pi_u$ the universal representation of $\al F.$ on the
universal Hilbert space $\al H._u$ \cite[Section~3.7]{bPedersen89}.
$\al F.''$ is the strong closure of $\pi_u(\al F.)$ and since $\pi_u$
is faithful we make the usual identification of $\al F.$
with a subalgebra of $\al F.''$, i.e.~generally omit explicit
indication of $\pi_u$. If $\omega\in{\got S}(\al F.)$, we will
use the same symbol for the unique extension of $\omega$ from $\al F.$ to 
$\al F.''$. 

\begin{teo}
\label{Teo.2.7}
For a constrained system $(\al F.,\al C.)$ there exists a 
 projection%\footnote{Such a projection is called {\em open} in
%\cite{bPedersen89}.} 
 $P\in\al F.''$ such that
\begin{itemize}
 \item[{\rm (i)}] $\al N.=\al F.''\,P\cap \al F.$,
 \item[{\rm (ii)}] $\al D.=P\,\al F.''\,P \cap \al F.$ and 
 \item[{\rm (iii)}] ${\got S}_D=\{\omega\in{\got S}(\al F.)\mid\omega(P)=0\}$.
\end{itemize}
\end{teo}
\begin{beweis}
{}From Theorem~\ref{Teo.2.2}~(i) $\al D.$ is a hereditary C$^*$--subalgebra
of $\al F.$ and by 3.11.10 and 3.11.9 in \cite{bPedersen89} there exists 
a projection
 $P\in\al F.''$ such that $\al D.=P\al F.''P\cap \al F.$. 
Further by the proof of Theorem~\ref{Teo.2.1}~(iii) as well as 3.10.7
and 3.11.9 in \cite{bPedersen89} we obtain that
$\al N. =\al F.''\,P\cap \al F.$ and
\[
 {\got S}_D=\{ \omega\in{\got S}(\al F.)\mid \omega(P)=0 \}\,,
\]
which concludes this proof.
\end{beweis}
 A projection satisfying the conditions of Theorem~\ref{Teo.2.7}
is called {\em open} in
\cite{bPedersen89}. 

\begin{teo}
\label{Teo.2.10}
Let $P$ be the open projection in Theorem~$\ref{Teo.2.7}$. Then: 
\begin{itemize}
\item[{\rm(i)}] $\al O.=\{ A\in\al F. \mid PA(\EINS-P)=0=
                        (\EINS-P)AP \}=P'\cap \al F.$, and
\item[{\rm(ii)}] $\al C.' \cap \al F. \subset \al O.$.
\end{itemize}
\end{teo}
\begin{beweis}
(i) Recall that $\al O.=M_{\al F.}(\al D.)$, and let $A\in\al F.$ and 
$D\in \al D.$. Then by Theorem~\ref{Teo.2.7} there exists an 
$F\in \al F.''$ such that $D=PFP$ and so
\begin{eqnarray*}
 AD &=& (PAP+ (\EINS-P)AP+PA(\EINS -P)+(\EINS-P)A(\EINS-P) )\, PFP\\[1mm]
    &=& PAPFP + (\EINS-P)APFP \\[1mm]
    &=& PAPD + (\EINS-P)APD\,.
\end{eqnarray*}
Therefore using Theorem~\ref{Teo.2.7} again we have
$AD\in\al D.$ for all $D\in\al D.$ iff $(\EINS-P)APD=0$
for all $D\in\al D.$. But from 3.11.9 in \cite{bPedersen89} $P$ is in 
the strong closure of $\al D.$ in $\al F.''$ so that $AD\in
\al D.\;\forall\, D\in\al D.$
iff $(\EINS-P)AP=0$. Taking adjoints we get also the condition
$PA(\EINS-P)=0$ iff $\al D.A\subset\al D.$.

(ii) Let $D\in\al D.=[ \al FC.]\; \cap\; [ \al CF.]$ and 
$A\in \al C.' \cap \al F.$. Then $AD\in {[ {\al FC.}]\; \cap\; 
[A\,{\al CF.}]}=[ {\al FC.}]\; \cap\; [ \al C.\,A\,{\al F.}]\subset \al D.$.
Similarly, $DA\in\al D.$ so that by definition we have $A\in\al O.$.
\end{beweis}

What these two last theorems mean, is that with respect to
the decomposition 
\[
 \al H._u=P\,\al H._u\oplus (\EINS-P)\,\al H._u
\] 
we may rewrite 
\begin{eqnarray*}
 \al D.&=& \Big\{ F\in\al F.\;\Big|\; F=
          \left(\kern-1.5mm\begin{array}{cc}
              D \kern-1.6mm & 0 \\ 0 \kern-1.6mm & 0
          \end{array} \kern-1.5mm\right),\;
       D\in P\al F.P \Big\}\;\;{\rm and}  \\
 \al O.&=& \Big\{ F\in\al F.\;\Big|\; F=
          \left(\kern-1.5mm\begin{array}{cc}
              A \kern-1.6mm & 0 \\ 0 \kern-1.6mm & B
          \end{array} \kern-1.5mm\right),\;
      A\in P\al F.P,\; B\in(\EINS-P)\al F.(\EINS-P) \Big\}\,.
\end{eqnarray*}
It is clear that in general $\al O.$ can be much greater than the
traditional observables $\al C.'\cap\al F.$. Next we show how to identify the 
final algebra of physical observables $\al R.$ with a subalgebra of
$\al F.''$.

\begin{teo}
\label{Teo.2.11}
For $P$ as above we have:
\[
 \al R.\cong\Big\{ F\in\al F.\;\Big|\; F=
          \left(\kern-1.5mm\begin{array}{cc}
              0 \kern-1.6mm & 0 \\ 0 \kern-1.6mm & A
          \end{array} \kern-1.5mm\right) \Big\}=
 (\EINS-P)\,(P'\cap\al F.) \subset \al F.''.
\]
\end{teo}
\begin{beweis}
The homomorphism $\Phi\colon\ \al O.\rightarrow (\EINS-P)\,(P'\cap\al F.)$
defined by $\Phi(A):=(\EINS-P)A$, $A\in\al O.=P'\cap\al F.$, will establish an
isomorphism with ${\al R.}:={\al O.}/{\al D.}$ if we can show that 
${\rm Ker}\,\Phi=\al D.$, i.e.~$(\EINS-P)A=0$ iff $A\in\al D.$. Clearly, if
$A\in\al D.=P\,\al F.''\,P \cap \al F.$ (cf.~Theorem~\ref{Teo.2.7}),
then $(\EINS-P)A=0$. Conversely, assume that $A\in\al O.=P'\cap\al F.$
satisfies $(\EINS-P)A=0$, i.e.~$A=PA$. Then, $A\in P\,\al F.'' \cap \al F.$
and so since $A\in P'\cap\al F.$, we have 
$A\in P\,\al F.''\,P \cap \al F.=\al D.$, which ends the proof.
\end{beweis}

\begin{rem}
\label{Rem.2.12}
\begin{itemize}
\item[(i)] 
With the preceding result we may interpret the projection $P$ in
Theorem~\ref{Teo.2.7} as being equivalent to the set $\al C.$ if we 
are willing to enlarge $\al F.$ to ${\rm C}^*(\al F.\cup \{P\})$.
This can be partially justified by the fact that
\[
 \al R.\cong (\EINS-P)\,(P'\cap\al F.)\subset {\rm C}^*(\al F.\cup \{P\})\,.
\]
\item[(ii)]
The projection $P$ can also be used to
make contact with the original heuristic picture. 
Given a Dirac state $\omega\in{\got S}_D$ we see from 
Theorem~\ref{Teo.2.7}~(iii) that $\EINS-\pi_\omega(P)$ is 
the projection onto the physical subspace
\[
 \al H._\omega^{(p)}:=\{ \psi\in\al H._\omega\mid\pi_\omega(\al C.)\psi=0\}.
\]
Since $\pi_\omega(\al O.)\subset \pi_\omega(P)'
\cap\pi_\omega(\al F.)$, then $\pi_\omega(\al O.)$ is a subalgebra of
the algebra of observables in the field algebra $\pi_\omega(\al F.)$
which preserves the physical subspace. (In fact $\al O.=\{F\in\al F.\,\mid\,
\pi_\omega(F)\al H._\omega^{(p)}\subseteq\al H._\omega^{(p)}
\quad\forall\,\omega\in{\got S}(\al F.)\}$). If $\pi_\omega$ is faithful, then 
$\pi_\omega(\al O.)$ contains the traditional observables 
$\pi_\omega(\al C.)'\cap\pi_\omega(\al F.)$. Now restricting 
$\pi_\omega(\al O.)$ to the subspace 
$\al H._\omega^{(p)}={(\EINS-\pi_\omega(P))\,\al H._\omega}$
we have for the final constrained system:
\[
 \pi_\omega(\al O.)\restriction\al H._\omega^{(p)}=
 (\EINS-\pi_\omega(P))\pi_\omega(\al O.)=
 \pi_\omega((\EINS-P)(P'\cap\al F.)),
\]
which by Theorem~\ref{Teo.2.11} produces a representation of $\al R.$.
\end{itemize}
\end{rem}

%%%%%%%%%%%%%%%%%%%%%%%%%%%%%%%%%%%%%%%%%%%%%%%%%%%%%%%%%%%%%%%%%%%%%%%%%%%%%%
\section*{Appendix 2}

We will give in this appendix the proof of Theorem~\ref{FixingBlemish}.
Recall the notation and results of Subsection~\ref{Manuceau}.
\begin{teo}
Given nondegenerate $({\got X},\,B)$ and ${\frak s}\subset {\got X}$ 
as in Subsection~\ref{Manuceau},
where ${\frak s}\subset{\frak s}'$ and 
${\frak s}={\frak s}''$, then
\[
 \al O.= {\rm C}^*(\delta_{{\frak s}'}\cup \al D.)=\big[ 
          {\rm C}^*(\delta_{{\frak s}'}) \cup \al D. \big]\,.
\]
\end{teo}
\begin{beweis}
{}From Theorem~\ref{Teo.2.2}~(ii) and (v) it is clear that 
$\al O.\supset {\rm C}^*(\delta_{{\frak s}'}\cup \al D.)$ and we only 
have to prove the converse inclusion. 
We first show that for any Dirac state $\omega$ of $\CCR$ we have that
$\pi_\omega(\delta_f)\Omega_\omega\perp\al H._\omega^{(p)}:=\{\psi
\in\al H._\omega\mid\pi_\omega(\delta_{\ot s.})\psi=\psi\}$
for all $f\not\in\ot s.'$.
For any $f\not\in\ot s.'$, choose a $k\in\ot s.$ such that
$B(f,\, k)\not\in 2\pi\Z$. Then 
\[
  \pi_\omega(\delta_k)\big(\pi_\omega(\delta_f)\Omega_\omega\big)
=e^{iB(k,\, f)}\pi_\omega(\delta_f\delta_k)\Omega_\omega
=e^{iB(k,\, f)}\pi_\omega(\delta_f)\Omega_\omega
\]
and so $\pi_\omega(\delta_f)\Omega_\omega$ is in a different
eigenspace of $\pi_\omega(\delta_k)$ than $\al H._\omega^{(p)}$,
hence must be orthogonal to it.

Now recall (cf. Remark~\ref{Rem.2.12}~(ii)) that if $A\in\al O.$,
then $\pi_\omega(A)$ preserves $\al H._\omega^{(p)}$ (here $\omega$
is a Dirac state). Thus $\pi_\omega(A)\Omega_\omega\perp
\pi_\omega(\delta_f)\Omega_\omega$ for $f\not\in\ot s.'$,
and so $\omega(\delta_fA)=0$ for $f\not\in\ot s.'$.
Now let $A_n=\sum\limits_{i=1}^{N_n}\lambda_i^{(n)}\delta_{h_i}
\subset\Delta(\ot X.,\, B)$ be a sequence converging to $A$.
We furthermore partition the set ${\{h_i\mid i\in\N\}}$ into
$\ot s.\hbox{--equivalence}$ classes and choose one representative
in each, so that we can write
\[
  A_n=\sum_j\delta_{h_j}\sum_i\beta_{ij}^{(n)}\delta_{c_{ij}}
\qquad\hbox{where}\quad c_{ij}\in\ot s.\quad\hbox{and}\quad
h_j-h_k\not\in\ot s.\quad\hbox{if}\quad j\not=k
\]
i.e. the first sum is over representatives in different equivalence classes.
Thus we get for $A\in\al O.$ that for all $f\not\in\ot s.'$:
\begin{equation}
\label{star}
  0=\lim_{n\to\infty}\omega(\delta_fA_n)
=\lim_{n\to\infty}\sum_{j,\, i}\beta_{ij}^{(n)}
\omega(\delta_f\delta_{h_j}\delta_{c_{ij}})\;.
\end{equation}
Now we make a particular choice for $\omega$, by  
setting  $\omega(\delta_f)=1$ if $f\in\ot s.$ and zero otherwise.
(To see that this indeed defines a state, note that if we factors
the central state on $\overline{\Delta(\ot s.'/\ot s.'',\,\wt{B})}$
to ${\rm C}^*(\delta_{\ot s.'})$, then by \cite[p.~387]{Grundling85} 
it extends uniquely to $\CCR$ and will coincide with $\omega$).
Then
\begin{eqnarray*}
\omega(\delta_f\delta_{h_j}\delta_{c_{ij}})
  &=& \exp\big[i\big(B(f,\, h_j)+B(f+h_j,\, c_{ij})\big)\big/2\big]
       \;\omega(\delta_{f+h_j+c_{ij}})\\
 &=& e^{{i\over 2}B(f,\, h_j)}\chi_{\ot s.}(f+h_j)
\end{eqnarray*}
where $\chi_{\ot s.}$ denotes the characteristic function of the set $\ot s.$.
Thus Eqn.~(\ref{star}) becomes:
\[
  \lim_{n\to\infty}\sum_{j,\,i}\beta_{ij}^{(n)}e^{{i\over 2}
B(f,\,h_j)}\chi_{\ot s.}(f+h_j)=0\qquad\forall\; f\not\in\ot s.'\;.
\]
If there is a $k$ such that $h_k\not\in\ot s.'$, we can choose
$f=-h_k$ in the previous equation, then since
$h_j-h_k\not\in\ot s.$ for $j\not=k$, we get $\chi_{\ot s.}(h_j-h_k)
=\delta_{jk}$ and thus ${\lim\limits_{n\to\infty}
\sum_i\beta^{(n)}_{ik}=0}$.

Consider now the universal representation
$\pi_u:\al A.\to\al B.(\al H._u)$ and let $\psi\in
\al H._u^{(p)}:=\{\psi\in\al H._u^{(p)}\mid\;\pi_u(\al D.)\psi=0\}$.
Then for each $k$ such that $h_k\not\in\ot s.'$ we have
\[
  \lim_n\pi_u\Big(\delta_{h_k}\sum_i\beta^{(n)}_{ik}
\delta_{c_{ik}}\Big)\,\psi=\pi_u(\delta_{h_k})\lim_n
\sum_i\beta_{ik}^{(n)}\psi=0\;.
\]
Hence in the sums involved in $\pi_u(A_n)\psi$ we can drop all
terms where $h_j\not\in\ot s.'$, i.e.
\begin{eqnarray*}
\pi_u(A)\psi &=&\lim_n\pi_u(A_n)\psi=
  \lim_n\sum_{h_j\in\ot s.'}\pi_u(\delta_{h_j})\sum_i\beta^{(n)}_{ik}
\pi_u(\delta_{c_{ik}})\,\psi\\
 &=&
  \lim_n\sum_{h_j\in\ot s.'}\pi_u(\delta_{h_j})\sum_i\beta^{(n)}_{ik}
\psi\;\in\;\pi_u\big({\rm C}^*(\delta_{\ot s.'})\big)\,\psi
\qquad\forall\psi\in\al H._u^{(p)}\;.
\end{eqnarray*}
Thus the restriction of $\al O.=\pi_u(\al O.)$ to $\al H._u^{(p)}$
is the same as the restriction of 
${\rm C}^*(\delta_{\ot s.'})$ to $\al H._u^{(p)}$
(given that 
${\rm C}^*(\delta_{\ot s.'})\subset\al O.$).
However, recall from Theorem~\ref{Teo.2.11} and the preceding remarks,
that $\al O./\al D.=\al R.\cong\al O.\restriction\al H._u^{(p)}
={\rm C}^*(\delta_{\ot s.'})\restriction\al H._u^{(p)}$
(with respect to the open projection $P$, we have
$\al H._u^{(p)}=(\EINS-P)\al H._u$).
Thus $\al O.=
{\rm C}^*(\delta_{\ot s.'})+\al D.$.
\end{beweis}

\section*{Acknowledgements.}

We are both very grateful to Prof. D. Buchholz for the
friendly interest which he took in this paper, and for pointing out
an important mistake in an earlier version of it.
One of us (H.G.) would like to thank the Erwin Schr\"odinger
Institute for Mathematical Physics
in Vienna (where a substantial part of this work
was done) for their generous assistance, as well as
Prof.~H.~Baumg\"artel for his warm hospitality at
the University of Potsdam, where this project was started.
We thank the sfb 288 for support during this visit.
We also benefitted from an ARC grant which funded a visit of
F.Ll. to the University of New South Wales. Finally, F.Ll. would
like to thank Hanno Gottschalk and Wolfgang Junker
for helpful conversations,  as well
as Sergio Doplicher for his kind hospitality at the 
 `Dipartamento di Matematica dell' Universit\`a di Roma `La
Sapienza'' in March 2000, when the final version of this paper
was prepared. The visit was supported by a 
EU TMR network ``Implementation of concept and methods from
Non--Commutative Geometry to Operator Algebras and its applications'', 
contract no.~ERB FMRX-CT 96-0073.

%\bigskip
\vfill\eject

\providecommand{\bysame}{\leavevmode\hbox to3em{\hrulefill}\thinspace}

\end{document}